\documentclass[12pt]{article}
\usepackage{epsfig,amsfonts,amssymb}
\usepackage{hyperref}

\usepackage{cite}
\usepackage{cite}

\usepackage{comment}

\def\ZZZ{{\hbox{ Z\kern-1.6mm Z}}}
\def\RRR{{\hbox{ R\kern-2.4mm R}}}
\def\CCC{{\hbox{ C\kern-2.0mm C}}}
\def\zzz{{\hbox{z\kern-1mm z}}}

\newcommand{\qeq}{{\hbox{=\kern-2.3mm ? \kern.5mm }}}
\renewcommand{\qeq}{=}

\newcommand{\eps}{\epsilon}

\newcommand{\ve}{\varepsilon}

\newcommand{\AAA}{{\cal A}}

\newcommand{\JJ}{{\cal J}}

\newcommand{\CC}{{\cal C}}

\newcommand{\OO}{{\cal O}}

\newcommand{\wt}{\widetilde}

\newcommand{\NN}{{\cal N}}

\newcommand{\be}{\begin{equation}}
\newcommand{\ee}{\end{equation}}
\newcommand{\ben}{\begin{eqnarray}\displaystyle}
\newcommand{\een}{\end{eqnarray}}

\newcommand{\refb}[1]{(\ref{#1})}
\newcommand{\p}{\partial}
\newcommand{\sectiono}[1]{\section{#1}\setcounter{equation}{0}}

\def\one{{\hbox{ 1\kern-.8mm l}}}
\def\zero{{\hbox{ 0\kern-1.5mm 0}}}

\newcommand{\bea}[1]{\begin{eqnarray}\label{#1} }
\newcommand{\eea}{\end{eqnarray}}

\newcommand{\eqref}{\refb}

\usepackage{bm}
\usepackage[table]{xcolor}

\def\figone{

\def\JPicScale{0.6}
\ifx\JPicScale\undefined\def\JPicScale{1}\fi
\unitlength \JPicScale mm
\begin{picture}(236.25,70)(0,0)
\linethickness{0.3mm}
\put(45,40){\circle{12.5}}

\linethickness{0.3mm}
\put(120,40){\circle{12.5}}

\linethickness{0.3mm}
\put(85,25){\circle{12.5}}

\linethickness{0.3mm}
\put(85,55){\circle{12.5}}

\linethickness{0.7mm}
\multiput(45,40)(0.32,0.12){125}{\line(1,0){0.32}}
\linethickness{0.7mm}
\multiput(85,55)(0.28,-0.12){125}{\line(1,0){0.28}}
\linethickness{0.7mm}
\multiput(85,25)(0.28,0.12){125}{\line(1,0){0.28}}
\linethickness{0.7mm}
\multiput(45,40)(0.32,-0.12){125}{\line(1,0){0.32}}

\end{picture}

}

\def\figtwo{

\def\JPicScale{0.8}
\ifx\JPicScale\undefined\def\JPicScale{1}\fi
\unitlength \JPicScale mm
\begin{picture}(70,60)(0,0)
\linethickness{0.7mm}
\multiput(50,40)(0.12,0.12){167}{\line(1,0){0.12}}
\linethickness{0.7mm}
\multiput(30,60)(0.12,-0.12){167}{\line(1,0){0.12}}
\linethickness{0.7mm}
\multiput(30,30)(0.24,0.12){83}{\line(1,0){0.24}}
\linethickness{0.3mm}
\multiput(50,40)(0.12,-0.24){83}{\line(0,-1){0.24}}
\linethickness{0.3mm}
\put(50,40){\line(1,0){5}}
\linethickness{0.3mm}
\put(50,40){\line(0,1){5}}
\linethickness{0.3mm}
\put(45,40){\line(1,0){5}}
\put(58,40){\makebox(0,0)[cc]{$i_1$}}

\put(50,48){\makebox(0,0)[cc]{$i_2$}}

\put(42,40){\makebox(0,0)[cc]{$i_3$}}

\end{picture}

}

\def\figthree{

\def\JPicScale{0.8}
\ifx\JPicScale\undefined\def\JPicScale{1}\fi
\unitlength \JPicScale mm
\begin{picture}(70,80)(0,0)
\linethickness{0.7mm}
\put(20,40){\line(1,0){30}}
\linethickness{0.7mm}
\multiput(50,40)(0.12,0.12){167}{\line(1,0){0.12}}
\linethickness{0.7mm}
\multiput(50,40)(0.12,-0.12){167}{\line(1,0){0.12}}
\linethickness{0.7mm}
\put(20,70){\line(1,0){50}}
\put(20,70){\makebox(0,0)[cc]{$\times$}}

\put(70,60){\makebox(0,0)[cc]{$\times$}}

\put(70,20){\makebox(0,0)[cc]{$\times$}}

\put(75,70){\makebox(0,0)[cc]{$\omega_2$}}

\put(15,40){\makebox(0,0)[cc]{$\tilde\omega_1$}}

\put(68,52){\makebox(0,0)[cc]{$\omega_3$}}

\put(70,25){\makebox(0,0)[cc]{$\omega_4$}}

\linethickness{0.2mm}
\qbezier(10,80)(9.58,79.22)(8.32,71.87)
\qbezier(8.32,71.87)(7.07,64.53)(10,60)
\qbezier(10,60)(18.21,53.02)(31.34,56.28)
\qbezier(31.34,56.28)(44.46,59.53)(55,55)
\qbezier(55,55)(58.91,52.52)(61.3,48.52)
\qbezier(61.3,48.52)(63.69,44.53)(65,40)
\qbezier(65,40)(66.96,32.62)(66.12,25.18)
\qbezier(66.12,25.18)(65.28,17.74)(65,10)
\end{picture}

}

\def\figthreenew{

\def\JPicScale{0.8}
\ifx\JPicScale\undefined\def\JPicScale{1}\fi
\unitlength \JPicScale mm
\begin{picture}(70,80)(0,0)
\linethickness{0.7mm}
\put(20,40){\line(1,0){30}}
\linethickness{0.7mm}
\multiput(50,40)(0.12,0.12){167}{\line(1,0){0.12}}
\linethickness{0.7mm}
\multiput(50,40)(0.12,-0.12){167}{\line(1,0){0.12}}
\linethickness{0.7mm}
\put(20,70){\line(1,0){50}}
\put(20,70){\makebox(0,0)[cc]{$\times$}}

\put(70,60){\makebox(0,0)[cc]{$\times$}}

\put(70,20){\makebox(0,0)[cc]{$\times$}}

\put(75,70){\makebox(0,0)[cc]{$\omega_2$}}

\put(15,40){\makebox(0,0)[cc]{$\tilde\omega_1$}}

\put(68,52){\makebox(0,0)[cc]{$\omega_3$}}

\put(70,25){\makebox(0,0)[cc]{$\omega_4$}}

\end{picture}

}

\def\figfour{

\def\JPicScale{0.8}
\ifx\JPicScale\undefined\def\JPicScale{1}\fi
\unitlength \JPicScale mm
\begin{picture}(150,65.51)(0,0)
\linethickness{0.7mm}
\multiput(20,60)(0.12,-0.12){167}{\line(1,0){0.12}}
\linethickness{0.7mm}
\multiput(20,20)(0.12,0.12){167}{\line(1,0){0.12}}
\linethickness{0.7mm}
\multiput(70,40)(0.12,0.12){167}{\line(1,0){0.12}}
\linethickness{0.7mm}
\multiput(70,40)(0.12,-0.12){167}{\line(1,0){0.12}}
\linethickness{0.7mm}
\qbezier(40,40)(40.21,40.45)(43.32,44.31)
\qbezier(43.32,44.31)(46.44,48.16)(50,50)
\qbezier(50,50)(52.4,50.98)(55,50.98)
\qbezier(55,50.98)(57.6,50.98)(60,50)
\qbezier(60,50)(63.56,48.16)(66.68,44.31)
\qbezier(66.68,44.31)(69.79,40.45)(70,40)
\linethickness{0.7mm}
\qbezier(40,40)(40.21,39.55)(43.32,35.69)
\qbezier(43.32,35.69)(46.44,31.84)(50,30)
\qbezier(50,30)(52.4,29.02)(55,29.02)
\qbezier(55,29.02)(57.6,29.02)(60,30)
\qbezier(60,30)(63.56,31.84)(66.68,35.69)
\qbezier(66.68,35.69)(69.79,39.55)(70,40)
\linethickness{0.7mm}
\multiput(110,60)(0.12,-0.12){167}{\line(1,0){0.12}}
\linethickness{0.7mm}
\multiput(130,40)(0.12,0.12){167}{\line(1,0){0.12}}
\linethickness{0.7mm}
\multiput(130,40)(0.12,-0.12){167}{\line(1,0){0.12}}
\linethickness{0.7mm}
\multiput(110,20)(0.12,0.12){167}{\line(1,0){0.12}}
\linethickness{0.7mm}
\qbezier(130,40)(129.02,40.67)(123.57,47.69)
\qbezier(123.57,47.69)(118.13,54.71)(120,60)
\qbezier(120,60)(123.16,64.13)(130,64.13)
\qbezier(130,64.13)(136.84,64.13)(140,60)
\qbezier(140,60)(141.87,54.71)(136.43,47.69)
\qbezier(136.43,47.69)(130.98,40.67)(130,40)
\linethickness{0.7mm}
\qbezier(130,40)(129.02,39.33)(123.57,32.31)
\qbezier(123.57,32.31)(118.13,25.29)(120,20)
\qbezier(120,20)(123.16,15.87)(130,15.87)
\qbezier(130,15.87)(136.84,15.87)(140,20)
\qbezier(140,20)(141.87,25.29)(136.43,32.31)
\qbezier(136.43,32.31)(130.98,39.33)(130,40)
\put(100,40){\makebox(0,0)[cc]{$\Rightarrow$}}

\put(130,40){\makebox(0,0)[cc]{\Large $\bullet$}}

\put(40,40){\makebox(0,0)[cc]{\Large $+$}}

\put(70.3,40){\makebox(0,0)[cc]{\Large $+$}}

\end{picture}

}

\def\figfive{

\def\JPicScale{0.8}
\ifx\JPicScale\undefined\def\JPicScale{1}\fi
\unitlength \JPicScale mm
\begin{picture}(140,70)(0,0)
\linethickness{0.7mm}
\multiput(10,70)(0.12,-0.12){167}{\line(1,0){0.12}}
\linethickness{0.7mm}
\multiput(10,30)(0.12,0.12){167}{\line(1,0){0.12}}
\linethickness{0.2mm}
\put(30,50){\line(1,0){20}}
\linethickness{0.7mm}
\multiput(50,50)(0.12,0.12){167}{\line(1,0){0.12}}
\linethickness{0.7mm}
\multiput(50,50)(0.12,-0.12){167}{\line(1,0){0.12}}
\linethickness{0.7mm}
\multiput(85,70)(0.2,-0.12){125}{\line(1,0){0.2}}
\linethickness{0.7mm}
\multiput(110,55)(0.2,0.12){125}{\line(1,0){0.2}}
\linethickness{0.7mm}
\put(110,30){\line(0,1){25}}
\linethickness{0.7mm}
\put(140,30){\line(0,1){40}}
\put(29.2,50){\makebox(0,0)[cc]{{\Large $\times$}}}

\put(51,50){\makebox(0,0)[cc]{{\Large $\times$}}}

\put(110,55.5){\makebox(0,0)[cc]{{\Large $\times$}}}

\put(140,30){\makebox(0,0)[cc]{{\Large $\times$}}}

\put(35,20){\makebox(0,0)[cc]{(a)}}

\put(125,20){\makebox(0,0)[cc]{(b)}}

\end{picture}

}

\def\figseven{

\def\JPicScale{0.8}
\ifx\JPicScale\undefined\def\JPicScale{1}\fi
\unitlength \JPicScale mm
\begin{picture}(70,40)(0,0)
\linethickness{0.7mm}
\put(30,40){\line(1,0){40}}
\put(30,40){\makebox(0,0)[cc]{$\times$}}

\put(70,40){\makebox(0,0)[cc]{$\times$}}

\put(30,35){\makebox(0,0)[cc]{1}}

\put(70,35){\makebox(0,0)[cc]{2}}

\end{picture}

}

\def\figeight{

\def\JPicScale{0.6}
\ifx\JPicScale\undefined\def\JPicScale{1}\fi
\unitlength \JPicScale mm
\begin{picture}(75,60)(0,0)
\linethickness{0.7mm}
\put(30,60){\line(1,0){30}}
\linethickness{0.7mm}
\put(30,40){\line(1,0){30}}
\put(30,60){\makebox(0,0)[cc]{$\times$}}

\put(60,40){\makebox(0,0)[cc]{$\times$}}

\put(65,40){\makebox(0,0)[cc]{1}}

\put(25,60){\makebox(0,0)[cc]{2}}

\put(25,40){\makebox(0,0)[cc]{$\tilde\omega_1$}}

\put(65,60){\makebox(0,0)[cc]{$\omega_2$}}

\put(75,50){\makebox(0,0)[cc]{$\qquad \times \ \exp[\AAA_{12}]$}}

\put(45,30){\makebox(0,0)[cc]{(a)}}

\end{picture}

}

\def\fignine{

\def\JPicScale{0.6}
\ifx\JPicScale\undefined\def\JPicScale{1}\fi
\unitlength \JPicScale mm
\begin{picture}(75,60)(0,0)
\linethickness{0.7mm}
\put(30,60){\line(1,0){30}}
\linethickness{0.7mm}
\put(30,40){\line(1,0){30}}
\put(30,60){\makebox(0,0)[cc]{$\times$}}

\put(60,40){\makebox(0,0)[cc]{$\times$}}

\put(65,40){\makebox(0,0)[cc]{1}}

\put(25,60){\makebox(0,0)[cc]{1}}

\put(75,50){\makebox(0,0)[cc]{$\qquad \times \ \left(\exp[\AAA_{12}]-1\right)$}}

\put(45,30){\makebox(0,0)[cc]{(b)}}

\put(25,40){\makebox(0,0)[cc]{$\tilde\omega_1$}}

\put(65,60){\makebox(0,0)[cc]{$\omega_2$}}

\end{picture}

}

\def\figten{

\def\JPicScale{0.6}
\ifx\JPicScale\undefined\def\JPicScale{1}\fi
\unitlength \JPicScale mm
\begin{picture}(75,60)(0,0)
\linethickness{0.7mm}
\put(30,60){\line(1,0){30}}
\linethickness{0.7mm}
\put(30,40){\line(1,0){30}}
\put(30,60){\makebox(0,0)[cc]{$\times$}}

\put(60,40){\makebox(0,0)[cc]{$\times$}}

\put(65,40){\makebox(0,0)[cc]{2}}

\put(25,60){\makebox(0,0)[cc]{2}}

\put(75,50){\makebox(0,0)[cc]{$\qquad \times \ \left(\exp[\AAA_{12}]-1\right)$}}

\put(45,30){\makebox(0,0)[cc]{(c)}}

\put(25,40){\makebox(0,0)[cc]{$\tilde\omega_1$}}

\put(65,60){\makebox(0,0)[cc]{$\omega_2$}}

\end{picture}

}

\def\figeleven{

\def\JPicScale{1}
\ifx\JPicScale\undefined\def\JPicScale{1}\fi
\unitlength \JPicScale mm
\begin{picture}(70,70)(0,0)
\linethickness{0.7mm}
\put(10,60){\line(1,0){20}}
\linethickness{0.7mm}
\put(30,50){\line(1,0){20}}
\linethickness{0.7mm}
\put(30,40){\line(1,0){20}}
\linethickness{0.7mm}
\put(30,30){\line(1,0){20}}
\linethickness{0.7mm}
\put(50,60){\line(1,0){20}}
\put(30,60){\makebox(0,0)[cc]{$\times$}}

\put(50,60){\makebox(0,0)[cc]{$\times$}}

\put(30,50){\makebox(0,0)[cc]{$\times$}}

\put(50,50){\makebox(0,0)[cc]{$\times$}}

\put(30,40){\makebox(0,0)[cc]{$\times$}}

\put(50,40){\makebox(0,0)[cc]{$\times$}}

\put(30,30){\makebox(0,0)[cc]{$\times$}}

\put(50,30){\makebox(0,0)[cc]{$\times$}}

\put(30,63){\makebox(0,0)[cc]{1}}

\put(50,63){\makebox(0,0)[cc]{2}}

\put(30,53){\makebox(0,0)[cc]{1}}

\put(50,53){\makebox(0,0)[cc]{2}}

\put(30,43){\makebox(0,0)[cc]{1}}

\put(50,43){\makebox(0,0)[cc]{2}}

\put(30,33){\makebox(0,0)[cc]{1}}

\put(50,33){\makebox(0,0)[cc]{2}}

\put(44,33){\makebox(0,0)[cc]{$\cdot$}}

\put(44,37){\makebox(0,0)[cc]{$\cdot$}}

\put(44,35){\makebox(0,0)[cc]{$\cdot$}}

\put(12,65){\makebox(0,0)[cc]{$\tilde\omega_1$}}

\put(70,65){\makebox(0,0)[cc]{$\omega_2$}}

\linethickness{0.1mm}
\put(40,20){\line(0,1){50}}
\end{picture}

}

\def\figthirteen{

\def\JPicScale{1.1}
\ifx\JPicScale\undefined\def\JPicScale{1}\fi
\unitlength \JPicScale mm
\begin{picture}(70,50)(0,0)
\linethickness{0.7mm}
\put(10,40){\line(1,0){20}}
\linethickness{0.7mm}
\put(50,40){\line(1,0){20}}
\linethickness{0.7mm}
\qbezier(30,40)(35.19,45.25)(40,45.25)
\qbezier(40,45.25)(44.81,45.25)(50,40)
\linethickness{0.7mm}
\qbezier(30,40)(35.19,34.75)(40,34.75)
\qbezier(40,34.75)(44.81,34.75)(50,40)
\put(50,40){\makebox(0,0)[cc]{$\bullet$}}

\put(10,45){\makebox(0,0)[cc]{$\tilde\omega_1$}}

\put(70,45){\makebox(0,0)[cc]{$\omega_2$}}

\linethickness{0.1mm}
\put(40,30){\line(0,1){20}}
\linethickness{0.7mm}
\put(30,40){\line(1,0){20}}
\put(30,40){\makebox(0,0)[cc]{$\bullet$}}

\put(42,38){\makebox(0,0)[cc]{$\cdot$}}

\put(42,36.5){\makebox(0,0)[cc]{$\cdot$}}

\end{picture}

}

\def\figelevennew{

\def\JPicScale{1}
\ifx\JPicScale\undefined\def\JPicScale{1}\fi
\unitlength \JPicScale mm
\begin{picture}(70,70)(0,0)
\linethickness{0.7mm}
\put(10,60){\line(1,0){20}}
\linethickness{0.7mm}
\put(30,50){\line(1,0){20}}
\linethickness{0.7mm}
\put(30,40){\line(1,0){20}}
\linethickness{0.7mm}
\put(30,30){\line(1,0){20}}
\linethickness{0.7mm}
\put(50,60){\line(1,0){20}}
\put(30,60){\makebox(0,0)[cc]{$\times$}}

\put(50,60){\makebox(0,0)[cc]{$\times$}}

\put(30,50){\makebox(0,0)[cc]{$\times$}}

\put(50,50){\makebox(0,0)[cc]{$\times$}}

\put(30,40){\makebox(0,0)[cc]{$\times$}}

\put(50,40){\makebox(0,0)[cc]{$\times$}}

\put(30,30){\makebox(0,0)[cc]{$\times$}}

\put(50,30){\makebox(0,0)[cc]{$\times$}}

\put(30,63){\makebox(0,0)[cc]{1}}

\put(50,63){\makebox(0,0)[cc]{2}}

\put(30,53){\makebox(0,0)[cc]{1}}

\put(50,53){\makebox(0,0)[cc]{2}}

\put(30,43){\makebox(0,0)[cc]{1}}

\put(50,43){\makebox(0,0)[cc]{2}}

\put(30,33){\makebox(0,0)[cc]{1}}

\put(50,33){\makebox(0,0)[cc]{2}}

\put(44,33){\makebox(0,0)[cc]{$\cdot$}}

\put(44,37){\makebox(0,0)[cc]{$\cdot$}}

\put(44,35){\makebox(0,0)[cc]{$\cdot$}}

\put(12,65){\makebox(0,0)[cc]{$\tilde\omega_1$}}

\put(70,65){\makebox(0,0)[cc]{$\omega_2$}}

\linethickness{0.1mm}
\end{picture}

}

\def\figthirteennew{

\def\JPicScale{1.1}
\ifx\JPicScale\undefined\def\JPicScale{1}\fi
\unitlength \JPicScale mm
\begin{picture}(70,50)(0,0)
\linethickness{0.7mm}
\put(10,40){\line(1,0){20}}
\linethickness{0.7mm}
\put(50,40){\line(1,0){20}}
\linethickness{0.7mm}
\qbezier(30,40)(35.19,45.25)(40,45.25)
\qbezier(40,45.25)(44.81,45.25)(50,40)
\linethickness{0.7mm}
\qbezier(30,40)(35.19,34.75)(40,34.75)
\qbezier(40,34.75)(44.81,34.75)(50,40)
\put(50,40){\makebox(0,0)[cc]{$\bullet$}}

\put(10,45){\makebox(0,0)[cc]{$\tilde\omega_1$}}

\put(70,45){\makebox(0,0)[cc]{$\omega_2$}}

\linethickness{0.1mm}
\linethickness{0.7mm}
\put(30,40){\line(1,0){20}}
\put(30,40){\makebox(0,0)[cc]{$\bullet$}}

\put(42,38){\makebox(0,0)[cc]{$\cdot$}}

\put(42,36.5){\makebox(0,0)[cc]{$\cdot$}}

\end{picture}

}

\def\figtwelve{

\def\JPicScale{0.8}
\ifx\JPicScale\undefined\def\JPicScale{1}\fi
\unitlength \JPicScale mm
\begin{picture}(70,50)(0,0)
\linethickness{0.5mm}
\put(10,40){\line(1,0){20}}
\linethickness{0.5mm}
\put(50,40){\line(1,0){20}}
\linethickness{0.5mm}
\qbezier(30,40)(35.19,45.25)(40,45.25)
\qbezier(40,45.25)(44.81,45.25)(50,40)
\linethickness{0.5mm}
\qbezier(30,40)(35.19,34.75)(40,34.75)
\qbezier(40,34.75)(44.81,34.75)(50,40)
\put(50,40){\makebox(0,0)[cc]{$\bullet$}}

\put(10,45){\makebox(0,0)[cc]{$\tilde\omega_1$}}

\put(70,45){\makebox(0,0)[cc]{$\omega_2$}}

\linethickness{0.1mm}
\put(40,30){\line(0,1){20}}
\end{picture}

}

\def\figtwelvea{

\def\JPicScale{0.8}
\ifx\JPicScale\undefined\def\JPicScale{1}\fi
\unitlength \JPicScale mm
\begin{picture}(70,50)(0,0)
\linethickness{0.5mm}
\put(10,40){\line(1,0){20}}
\linethickness{0.5mm}
\put(50,40){\line(1,0){20}}
\linethickness{0.5mm}
\qbezier(30,40)(35.19,45.25)(40,45.25)
\qbezier(40,45.25)(44.81,45.25)(50,40)
\linethickness{0.5mm}
\qbezier(30,40)(35.19,34.75)(40,34.75)
\qbezier(40,34.75)(44.81,34.75)(50,40)
\put(50,40){\makebox(0,0)[cc]{$\bullet$}}

\put(10,45){\makebox(0,0)[cc]{$\tilde\omega_1$}}

\put(70,45){\makebox(0,0)[cc]{$\omega_2$}}

\end{picture}

}

\def\figfourteen{

\def\JPicScale{.6}
\ifx\JPicScale\undefined\def\JPicScale{1}\fi
\unitlength \JPicScale mm
\begin{picture}(70,70)(0,0)
\linethickness{0.7mm}
\put(10,60){\line(1,0){20}}
\linethickness{0.7mm}
\put(30,50){\line(1,0){20}}
\linethickness{0.7mm}
\put(30,40){\line(1,0){20}}
\linethickness{0.7mm}
\put(30,30){\line(1,0){20}}
\linethickness{0.7mm}
\put(50,60){\line(1,0){20}}
\put(30,60){\makebox(0,0)[cc]{$\times$}}

\put(50,60){\makebox(0,0)[cc]{$\times$}}

\put(30,50){\makebox(0,0)[cc]{$\times$}}

\put(50,50){\makebox(0,0)[cc]{$\times$}}

\put(30,40){\makebox(0,0)[cc]{$\times$}}

\put(50,40){\makebox(0,0)[cc]{$\times$}}

\put(30,30){\makebox(0,0)[cc]{$\times$}}

\put(50,30){\makebox(0,0)[cc]{$\times$}}

\put(30,63){\makebox(0,0)[cc]{1,2}}

\put(50,63){\makebox(0,0)[cc]{1,2}}

\put(30,53){\makebox(0,0)[cc]{1}}

\put(50,53){\makebox(0,0)[cc]{2}}

\put(30,43){\makebox(0,0)[cc]{1}}

\put(50,43){\makebox(0,0)[cc]{2}}

\put(30,33){\makebox(0,0)[cc]{1}}

\put(50,33){\makebox(0,0)[cc]{2}}

\put(44,33){\makebox(0,0)[cc]{$\cdot$}}

\put(44,37){\makebox(0,0)[cc]{$\cdot$}}

\put(44,35){\makebox(0,0)[cc]{$\cdot$}}

\put(12,65){\makebox(0,0)[cc]{$\tilde\omega_1$}}

\put(70,65){\makebox(0,0)[cc]{$\omega_2$}}

\put(99,53){\makebox(0,0)[cc]{$\times(\exp[A_{12}]-1)$}}

\linethickness{0.1mm}
\end{picture}

}

\def\figfifteen{

\def\JPicScale{0.8}
\ifx\JPicScale\undefined\def\JPicScale{1}\fi
\unitlength \JPicScale mm
\begin{picture}(50,40)(0,0)
\linethickness{0.7mm}
\put(10,40){\line(1,0){20}}
\linethickness{0.7mm}
\put(30,40){\line(1,0){20}}
\put(30,40){\makebox(0,0)[cc]{$\bullet$}}

\put(30,40){\makebox(0,0)[cc]{${\Large\times}$}}

\put(10,43){\makebox(0,0)[cc]{$\tilde\omega_1$}}

\put(50,43){\makebox(0,0)[cc]{$\omega_2$}}

\linethickness{0.7mm}
\put(30,35){\circle{10}}

\linethickness{0.7mm}
\put(40.83,28.92){\line(0,1){0.5}}
\multiput(40.81,29.92)(0.02,-0.5){1}{\line(0,-1){0.5}}
\multiput(40.76,30.42)(0.05,-0.5){1}{\line(0,-1){0.5}}
\multiput(40.69,30.91)(0.07,-0.5){1}{\line(0,-1){0.5}}
\multiput(40.6,31.4)(0.09,-0.49){1}{\line(0,-1){0.49}}
\multiput(40.49,31.89)(0.11,-0.49){1}{\line(0,-1){0.49}}
\multiput(40.35,32.37)(0.14,-0.48){1}{\line(0,-1){0.48}}
\multiput(40.19,32.85)(0.16,-0.47){1}{\line(0,-1){0.47}}
\multiput(40.01,33.31)(0.09,-0.23){2}{\line(0,-1){0.23}}
\multiput(39.81,33.77)(0.1,-0.23){2}{\line(0,-1){0.23}}
\multiput(39.58,34.22)(0.11,-0.22){2}{\line(0,-1){0.22}}
\multiput(39.34,34.66)(0.12,-0.22){2}{\line(0,-1){0.22}}
\multiput(39.08,35.08)(0.13,-0.21){2}{\line(0,-1){0.21}}
\multiput(38.79,35.49)(0.14,-0.21){2}{\line(0,-1){0.21}}
\multiput(38.49,35.89)(0.1,-0.13){3}{\line(0,-1){0.13}}
\multiput(38.17,36.28)(0.11,-0.13){3}{\line(0,-1){0.13}}
\multiput(37.84,36.65)(0.11,-0.12){3}{\line(0,-1){0.12}}
\multiput(37.48,37)(0.12,-0.12){3}{\line(1,0){0.12}}
\multiput(37.11,37.34)(0.12,-0.11){3}{\line(1,0){0.12}}
\multiput(36.73,37.66)(0.13,-0.11){3}{\line(1,0){0.13}}
\multiput(36.33,37.96)(0.13,-0.1){3}{\line(1,0){0.13}}
\multiput(35.91,38.24)(0.21,-0.14){2}{\line(1,0){0.21}}
\multiput(35.49,38.51)(0.21,-0.13){2}{\line(1,0){0.21}}
\multiput(35.05,38.75)(0.22,-0.12){2}{\line(1,0){0.22}}
\multiput(34.6,38.97)(0.22,-0.11){2}{\line(1,0){0.22}}
\multiput(34.15,39.18)(0.23,-0.1){2}{\line(1,0){0.23}}
\multiput(33.68,39.36)(0.23,-0.09){2}{\line(1,0){0.23}}
\multiput(33.2,39.52)(0.47,-0.16){1}{\line(1,0){0.47}}
\multiput(32.72,39.65)(0.48,-0.14){1}{\line(1,0){0.48}}
\multiput(32.24,39.77)(0.49,-0.11){1}{\line(1,0){0.49}}
\multiput(31.74,39.86)(0.49,-0.09){1}{\line(1,0){0.49}}
\multiput(31.25,39.93)(0.5,-0.07){1}{\line(1,0){0.5}}
\multiput(30.75,39.97)(0.5,-0.05){1}{\line(1,0){0.5}}
\multiput(30.25,40)(0.5,-0.02){1}{\line(1,0){0.5}}
\put(29.75,40){\line(1,0){0.5}}
\multiput(29.25,39.97)(0.5,0.02){1}{\line(1,0){0.5}}
\multiput(28.75,39.93)(0.5,0.05){1}{\line(1,0){0.5}}
\multiput(28.26,39.86)(0.5,0.07){1}{\line(1,0){0.5}}
\multiput(27.76,39.77)(0.49,0.09){1}{\line(1,0){0.49}}
\multiput(27.28,39.65)(0.49,0.11){1}{\line(1,0){0.49}}
\multiput(26.8,39.52)(0.48,0.14){1}{\line(1,0){0.48}}
\multiput(26.32,39.36)(0.47,0.16){1}{\line(1,0){0.47}}
\multiput(25.85,39.18)(0.23,0.09){2}{\line(1,0){0.23}}
\multiput(25.4,38.97)(0.23,0.1){2}{\line(1,0){0.23}}
\multiput(24.95,38.75)(0.22,0.11){2}{\line(1,0){0.22}}
\multiput(24.51,38.51)(0.22,0.12){2}{\line(1,0){0.22}}
\multiput(24.09,38.24)(0.21,0.13){2}{\line(1,0){0.21}}
\multiput(23.67,37.96)(0.21,0.14){2}{\line(1,0){0.21}}
\multiput(23.27,37.66)(0.13,0.1){3}{\line(1,0){0.13}}
\multiput(22.89,37.34)(0.13,0.11){3}{\line(1,0){0.13}}
\multiput(22.52,37)(0.12,0.11){3}{\line(1,0){0.12}}
\multiput(22.16,36.65)(0.12,0.12){3}{\line(0,1){0.12}}
\multiput(21.83,36.28)(0.11,0.12){3}{\line(0,1){0.12}}
\multiput(21.51,35.89)(0.11,0.13){3}{\line(0,1){0.13}}
\multiput(21.21,35.49)(0.1,0.13){3}{\line(0,1){0.13}}
\multiput(20.92,35.08)(0.14,0.21){2}{\line(0,1){0.21}}
\multiput(20.66,34.66)(0.13,0.21){2}{\line(0,1){0.21}}
\multiput(20.42,34.22)(0.12,0.22){2}{\line(0,1){0.22}}
\multiput(20.19,33.77)(0.11,0.22){2}{\line(0,1){0.22}}
\multiput(19.99,33.31)(0.1,0.23){2}{\line(0,1){0.23}}
\multiput(19.81,32.85)(0.09,0.23){2}{\line(0,1){0.23}}
\multiput(19.65,32.37)(0.16,0.47){1}{\line(0,1){0.47}}
\multiput(19.51,31.89)(0.14,0.48){1}{\line(0,1){0.48}}
\multiput(19.4,31.4)(0.11,0.49){1}{\line(0,1){0.49}}
\multiput(19.31,30.91)(0.09,0.49){1}{\line(0,1){0.49}}
\multiput(19.24,30.42)(0.07,0.5){1}{\line(0,1){0.5}}
\multiput(19.19,29.92)(0.05,0.5){1}{\line(0,1){0.5}}
\multiput(19.17,29.42)(0.02,0.5){1}{\line(0,1){0.5}}
\put(19.17,28.92){\line(0,1){0.5}}
\multiput(19.17,28.92)(0.02,-0.5){1}{\line(0,-1){0.5}}
\multiput(19.19,28.42)(0.05,-0.5){1}{\line(0,-1){0.5}}
\multiput(19.24,27.92)(0.07,-0.5){1}{\line(0,-1){0.5}}
\multiput(19.31,27.42)(0.09,-0.49){1}{\line(0,-1){0.49}}
\multiput(19.4,26.93)(0.11,-0.49){1}{\line(0,-1){0.49}}
\multiput(19.51,26.44)(0.14,-0.48){1}{\line(0,-1){0.48}}
\multiput(19.65,25.96)(0.16,-0.47){1}{\line(0,-1){0.47}}
\multiput(19.81,25.49)(0.09,-0.23){2}{\line(0,-1){0.23}}
\multiput(19.99,25.02)(0.1,-0.23){2}{\line(0,-1){0.23}}
\multiput(20.19,24.56)(0.11,-0.22){2}{\line(0,-1){0.22}}
\multiput(20.42,24.12)(0.12,-0.22){2}{\line(0,-1){0.22}}
\multiput(20.66,23.68)(0.13,-0.21){2}{\line(0,-1){0.21}}
\multiput(20.92,23.25)(0.14,-0.21){2}{\line(0,-1){0.21}}
\multiput(21.21,22.84)(0.1,-0.13){3}{\line(0,-1){0.13}}
\multiput(21.51,22.44)(0.11,-0.13){3}{\line(0,-1){0.13}}
\multiput(21.83,22.06)(0.11,-0.12){3}{\line(0,-1){0.12}}
\multiput(22.16,21.69)(0.12,-0.12){3}{\line(0,-1){0.12}}
\multiput(22.52,21.33)(0.12,-0.11){3}{\line(1,0){0.12}}
\multiput(22.89,20.99)(0.13,-0.11){3}{\line(1,0){0.13}}
\multiput(23.27,20.67)(0.13,-0.1){3}{\line(1,0){0.13}}
\multiput(23.67,20.37)(0.21,-0.14){2}{\line(1,0){0.21}}
\multiput(24.09,20.09)(0.21,-0.13){2}{\line(1,0){0.21}}
\multiput(24.51,19.83)(0.22,-0.12){2}{\line(1,0){0.22}}
\multiput(24.95,19.58)(0.22,-0.11){2}{\line(1,0){0.22}}
\multiput(25.4,19.36)(0.23,-0.1){2}{\line(1,0){0.23}}
\multiput(25.85,19.16)(0.23,-0.09){2}{\line(1,0){0.23}}
\multiput(26.32,18.98)(0.47,-0.16){1}{\line(1,0){0.47}}
\multiput(26.8,18.82)(0.48,-0.14){1}{\line(1,0){0.48}}
\multiput(27.28,18.68)(0.49,-0.11){1}{\line(1,0){0.49}}
\multiput(27.76,18.57)(0.49,-0.09){1}{\line(1,0){0.49}}
\multiput(28.26,18.47)(0.5,-0.07){1}{\line(1,0){0.5}}
\multiput(28.75,18.41)(0.5,-0.05){1}{\line(1,0){0.5}}
\multiput(29.25,18.36)(0.5,-0.02){1}{\line(1,0){0.5}}
\put(29.75,18.34){\line(1,0){0.5}}
\multiput(30.25,18.34)(0.5,0.02){1}{\line(1,0){0.5}}
\multiput(30.75,18.36)(0.5,0.05){1}{\line(1,0){0.5}}
\multiput(31.25,18.41)(0.5,0.07){1}{\line(1,0){0.5}}
\multiput(31.74,18.47)(0.49,0.09){1}{\line(1,0){0.49}}
\multiput(32.24,18.57)(0.49,0.11){1}{\line(1,0){0.49}}
\multiput(32.72,18.68)(0.48,0.14){1}{\line(1,0){0.48}}
\multiput(33.2,18.82)(0.47,0.16){1}{\line(1,0){0.47}}
\multiput(33.68,18.98)(0.23,0.09){2}{\line(1,0){0.23}}
\multiput(34.15,19.16)(0.23,0.1){2}{\line(1,0){0.23}}
\multiput(34.6,19.36)(0.22,0.11){2}{\line(1,0){0.22}}
\multiput(35.05,19.58)(0.22,0.12){2}{\line(1,0){0.22}}
\multiput(35.49,19.83)(0.21,0.13){2}{\line(1,0){0.21}}
\multiput(35.91,20.09)(0.21,0.14){2}{\line(1,0){0.21}}
\multiput(36.33,20.37)(0.13,0.1){3}{\line(1,0){0.13}}
\multiput(36.73,20.67)(0.13,0.11){3}{\line(1,0){0.13}}
\multiput(37.11,20.99)(0.12,0.11){3}{\line(1,0){0.12}}
\multiput(37.48,21.33)(0.12,0.12){3}{\line(1,0){0.12}}
\multiput(37.84,21.69)(0.11,0.12){3}{\line(0,1){0.12}}
\multiput(38.17,22.06)(0.11,0.13){3}{\line(0,1){0.13}}
\multiput(38.49,22.44)(0.1,0.13){3}{\line(0,1){0.13}}
\multiput(38.79,22.84)(0.14,0.21){2}{\line(0,1){0.21}}
\multiput(39.08,23.25)(0.13,0.21){2}{\line(0,1){0.21}}
\multiput(39.34,23.68)(0.12,0.22){2}{\line(0,1){0.22}}
\multiput(39.58,24.12)(0.11,0.22){2}{\line(0,1){0.22}}
\multiput(39.81,24.56)(0.1,0.23){2}{\line(0,1){0.23}}
\multiput(40.01,25.02)(0.09,0.23){2}{\line(0,1){0.23}}
\multiput(40.19,25.49)(0.16,0.47){1}{\line(0,1){0.47}}
\multiput(40.35,25.96)(0.14,0.48){1}{\line(0,1){0.48}}
\multiput(40.49,26.44)(0.11,0.49){1}{\line(0,1){0.49}}
\multiput(40.6,26.93)(0.09,0.49){1}{\line(0,1){0.49}}
\multiput(40.69,27.42)(0.07,0.5){1}{\line(0,1){0.5}}
\multiput(40.76,27.92)(0.05,0.5){1}{\line(0,1){0.5}}
\multiput(40.81,28.42)(0.02,0.5){1}{\line(0,1){0.5}}

\put(30,25){\makebox(0,0)[cc]{$\vdots$}}

\end{picture}
}

\def\figseventeen{

\def\JPicScale{0.8}
\ifx\JPicScale\undefined\def\JPicScale{1}\fi
\unitlength \JPicScale mm
\begin{picture}(130,45)(0,0)
\linethickness{0.7mm}
\put(10,40){\line(1,0){20}}
\linethickness{0.7mm}
\put(40,40){\line(1,0){20}}
\put(10,45){\makebox(0,0)[cc]{$\tilde\omega_1$}}

\put(60,45){\makebox(0,0)[cc]{$\omega_2$}}

\put(30,40){\makebox(0,0)[cc]{$\times$}}

\put(40,40){\makebox(0,0)[cc]{$\times$}}

\put(30,45){\makebox(0,0)[cc]{1,2}}

\put(40,45){\makebox(0,0)[cc]{1,2}}

\put(80,40){\makebox(0,0)[cc]{$\times (e^{A_{12}}-1)$}}

\linethickness{0.7mm}
\put(110,40){\line(1,0){20}}
\linethickness{0.7mm}
\put(130,40){\line(1,0){20}}
\put(130,40){\makebox(0,0)[cc]{$\bullet$}}

\put(130,40){\makebox(0,0)[cc]{${\Large\times}$}}

\put(110,45){\makebox(0,0)[cc]{$\tilde\omega_1$}}

\put(150,45){\makebox(0,0)[cc]{$\omega_2$}}

\put(35,30){\makebox(0,0)[cc]{(a)}}

\put(130,30){\makebox(0,0)[cc]{(b)}}

\end{picture}

}

\def\fignew{

\def\JPicScale{1}
\ifx\JPicScale\undefined\def\JPicScale{1}\fi
\unitlength \JPicScale mm
\begin{picture}(70,70)(0,0)
\linethickness{0.7mm}
\put(10,60){\line(1,0){20}}
\linethickness{0.7mm}
\put(30,50){\line(1,0){20}}
\linethickness{0.7mm}
\put(30,40){\line(1,0){20}}
\linethickness{0.7mm}
\put(30,30){\line(1,0){20}}
\linethickness{0.7mm}
\put(50,60){\line(1,0){20}}
\put(30,60){\makebox(0,0)[cc]{$\times$}}

\put(50,60){\makebox(0,0)[cc]{$\times$}}

\put(30,50){\makebox(0,0)[cc]{$\times$}}

\put(50,50){\makebox(0,0)[cc]{$\times$}}

\put(30,40){\makebox(0,0)[cc]{$\times$}}

\put(50,40){\makebox(0,0)[cc]{$\times$}}

\put(30,30){\makebox(0,0)[cc]{$\times$}}

\put(50,30){\makebox(0,0)[cc]{$\times$}}

\put(30,63){\makebox(0,0)[cc]{1}}

\put(50,63){\makebox(0,0)[cc]{1}}

\put(30,53){\makebox(0,0)[cc]{1}}

\put(50,53){\makebox(0,0)[cc]{2}}

\put(30,43){\makebox(0,0)[cc]{1}}

\put(50,43){\makebox(0,0)[cc]{2}}

\put(30,33){\makebox(0,0)[cc]{1}}

\put(50,33){\makebox(0,0)[cc]{2}}

\put(44,33){\makebox(0,0)[cc]{$\cdot$}}

\put(44,37){\makebox(0,0)[cc]{$\cdot$}}

\put(44,35){\makebox(0,0)[cc]{$\cdot$}}

\put(12,65){\makebox(0,0)[cc]{$\tilde\omega_1$}}

\put(70,65){\makebox(0,0)[cc]{$\omega_2$}}

\put(40,20){\makebox(0,0)[cc]{(a)}}

\linethickness{0.1mm}
\end{picture}

}

\def\figK{
\def\JPicScale{0.8}
\ifx\JPicScale\undefined\def\JPicScale{1}\fi
\unitlength \JPicScale mm
\begin{picture}(110,60)(0,0)
\linethickness{0.7mm}
\put(70,50){\line(1,0){20}}
\linethickness{0.7mm}
\put(90,50){\line(1,0){20}}
\linethickness{0.7mm}
\qbezier(90,50)(84.75,44.81)(84.75,40)
\qbezier(84.75,40)(84.75,35.19)(90,30)
\linethickness{0.7mm}
\qbezier(90,50)(95.25,44.81)(95.25,40)
\qbezier(95.25,40)(95.25,35.19)(90,30)
\put(90,50){\makebox(0,0)[cc]{$\bullet$}}

\put(90,30){\makebox(0,0)[cc]{$\bullet$}}

\put(35,20){\makebox(0,0)[cc]{(a)}}

\put(90,20){\makebox(0,0)[cc]{(b)}}

\linethickness{0.7mm}
\put(0,50){\line(1,0){25}}
\linethickness{0.7mm}
\put(25,50){\line(1,0){15}}
\linethickness{0.7mm}
\put(40,50){\line(1,0){20}}
\linethickness{0.7mm}
\put(31.01,39.25){\line(0,1){0.5}}
\multiput(30.99,40.25)(0.02,-0.5){1}{\line(0,-1){0.5}}
\multiput(30.94,40.75)(0.05,-0.5){1}{\line(0,-1){0.5}}
\multiput(30.87,41.24)(0.07,-0.5){1}{\line(0,-1){0.5}}
\multiput(30.77,41.73)(0.09,-0.49){1}{\line(0,-1){0.49}}
\multiput(30.65,42.22)(0.12,-0.49){1}{\line(0,-1){0.49}}
\multiput(30.51,42.7)(0.14,-0.48){1}{\line(0,-1){0.48}}
\multiput(30.35,43.17)(0.16,-0.47){1}{\line(0,-1){0.47}}
\multiput(30.16,43.64)(0.09,-0.23){2}{\line(0,-1){0.23}}
\multiput(29.96,44.09)(0.1,-0.23){2}{\line(0,-1){0.23}}
\multiput(29.73,44.54)(0.11,-0.22){2}{\line(0,-1){0.22}}
\multiput(29.48,44.97)(0.13,-0.22){2}{\line(0,-1){0.22}}
\multiput(29.21,45.39)(0.14,-0.21){2}{\line(0,-1){0.21}}
\multiput(28.92,45.8)(0.15,-0.2){2}{\line(0,-1){0.2}}
\multiput(28.61,46.19)(0.1,-0.13){3}{\line(0,-1){0.13}}
\multiput(28.28,46.57)(0.11,-0.13){3}{\line(0,-1){0.13}}
\multiput(27.93,46.93)(0.12,-0.12){3}{\line(0,-1){0.12}}
\multiput(27.57,47.28)(0.12,-0.12){3}{\line(1,0){0.12}}
\multiput(27.19,47.61)(0.13,-0.11){3}{\line(1,0){0.13}}
\multiput(26.8,47.92)(0.13,-0.1){3}{\line(1,0){0.13}}
\multiput(26.39,48.21)(0.2,-0.15){2}{\line(1,0){0.2}}
\multiput(25.97,48.48)(0.21,-0.14){2}{\line(1,0){0.21}}
\multiput(25.54,48.73)(0.22,-0.13){2}{\line(1,0){0.22}}
\multiput(25.09,48.96)(0.22,-0.11){2}{\line(1,0){0.22}}
\multiput(24.64,49.16)(0.23,-0.1){2}{\line(1,0){0.23}}
\multiput(24.17,49.35)(0.23,-0.09){2}{\line(1,0){0.23}}
\multiput(23.7,49.51)(0.47,-0.16){1}{\line(1,0){0.47}}
\multiput(23.22,49.65)(0.48,-0.14){1}{\line(1,0){0.48}}
\multiput(22.73,49.77)(0.49,-0.12){1}{\line(1,0){0.49}}
\multiput(22.24,49.87)(0.49,-0.09){1}{\line(1,0){0.49}}
\multiput(21.75,49.94)(0.5,-0.07){1}{\line(1,0){0.5}}
\multiput(21.25,49.99)(0.5,-0.05){1}{\line(1,0){0.5}}
\multiput(20.75,50.01)(0.5,-0.02){1}{\line(1,0){0.5}}
\put(20.25,50.01){\line(1,0){0.5}}
\multiput(19.75,49.99)(0.5,0.02){1}{\line(1,0){0.5}}
\multiput(19.25,49.94)(0.5,0.05){1}{\line(1,0){0.5}}
\multiput(18.76,49.87)(0.5,0.07){1}{\line(1,0){0.5}}
\multiput(18.27,49.77)(0.49,0.09){1}{\line(1,0){0.49}}
\multiput(17.78,49.65)(0.49,0.12){1}{\line(1,0){0.49}}
\multiput(17.3,49.51)(0.48,0.14){1}{\line(1,0){0.48}}
\multiput(16.83,49.35)(0.47,0.16){1}{\line(1,0){0.47}}
\multiput(16.36,49.16)(0.23,0.09){2}{\line(1,0){0.23}}
\multiput(15.91,48.96)(0.23,0.1){2}{\line(1,0){0.23}}
\multiput(15.46,48.73)(0.22,0.11){2}{\line(1,0){0.22}}
\multiput(15.03,48.48)(0.22,0.13){2}{\line(1,0){0.22}}
\multiput(14.61,48.21)(0.21,0.14){2}{\line(1,0){0.21}}
\multiput(14.2,47.92)(0.2,0.15){2}{\line(1,0){0.2}}
\multiput(13.81,47.61)(0.13,0.1){3}{\line(1,0){0.13}}
\multiput(13.43,47.28)(0.13,0.11){3}{\line(1,0){0.13}}
\multiput(13.07,46.93)(0.12,0.12){3}{\line(1,0){0.12}}
\multiput(12.72,46.57)(0.12,0.12){3}{\line(0,1){0.12}}
\multiput(12.39,46.19)(0.11,0.13){3}{\line(0,1){0.13}}
\multiput(12.08,45.8)(0.1,0.13){3}{\line(0,1){0.13}}
\multiput(11.79,45.39)(0.15,0.2){2}{\line(0,1){0.2}}
\multiput(11.52,44.97)(0.14,0.21){2}{\line(0,1){0.21}}
\multiput(11.27,44.54)(0.13,0.22){2}{\line(0,1){0.22}}
\multiput(11.04,44.09)(0.11,0.22){2}{\line(0,1){0.22}}
\multiput(10.84,43.64)(0.1,0.23){2}{\line(0,1){0.23}}
\multiput(10.65,43.17)(0.09,0.23){2}{\line(0,1){0.23}}
\multiput(10.49,42.7)(0.16,0.47){1}{\line(0,1){0.47}}
\multiput(10.35,42.22)(0.14,0.48){1}{\line(0,1){0.48}}
\multiput(10.23,41.73)(0.12,0.49){1}{\line(0,1){0.49}}
\multiput(10.13,41.24)(0.09,0.49){1}{\line(0,1){0.49}}
\multiput(10.06,40.75)(0.07,0.5){1}{\line(0,1){0.5}}
\multiput(10.01,40.25)(0.05,0.5){1}{\line(0,1){0.5}}
\multiput(9.99,39.75)(0.02,0.5){1}{\line(0,1){0.5}}
\put(9.99,39.25){\line(0,1){0.5}}
\multiput(9.99,39.25)(0.02,-0.5){1}{\line(0,-1){0.5}}
\multiput(10.01,38.75)(0.05,-0.5){1}{\line(0,-1){0.5}}
\multiput(10.06,38.25)(0.07,-0.5){1}{\line(0,-1){0.5}}
\multiput(10.13,37.76)(0.09,-0.49){1}{\line(0,-1){0.49}}
\multiput(10.23,37.27)(0.12,-0.49){1}{\line(0,-1){0.49}}
\multiput(10.35,36.78)(0.14,-0.48){1}{\line(0,-1){0.48}}
\multiput(10.49,36.3)(0.16,-0.47){1}{\line(0,-1){0.47}}
\multiput(10.65,35.83)(0.09,-0.23){2}{\line(0,-1){0.23}}
\multiput(10.84,35.36)(0.1,-0.23){2}{\line(0,-1){0.23}}
\multiput(11.04,34.91)(0.11,-0.22){2}{\line(0,-1){0.22}}
\multiput(11.27,34.46)(0.13,-0.22){2}{\line(0,-1){0.22}}
\multiput(11.52,34.03)(0.14,-0.21){2}{\line(0,-1){0.21}}
\multiput(11.79,33.61)(0.15,-0.2){2}{\line(0,-1){0.2}}
\multiput(12.08,33.2)(0.1,-0.13){3}{\line(0,-1){0.13}}
\multiput(12.39,32.81)(0.11,-0.13){3}{\line(0,-1){0.13}}
\multiput(12.72,32.43)(0.12,-0.12){3}{\line(0,-1){0.12}}
\multiput(13.07,32.07)(0.12,-0.12){3}{\line(1,0){0.12}}
\multiput(13.43,31.72)(0.13,-0.11){3}{\line(1,0){0.13}}
\multiput(13.81,31.39)(0.13,-0.1){3}{\line(1,0){0.13}}
\multiput(14.2,31.08)(0.2,-0.15){2}{\line(1,0){0.2}}
\multiput(14.61,30.79)(0.21,-0.14){2}{\line(1,0){0.21}}
\multiput(15.03,30.52)(0.22,-0.13){2}{\line(1,0){0.22}}
\multiput(15.46,30.27)(0.22,-0.11){2}{\line(1,0){0.22}}
\multiput(15.91,30.04)(0.23,-0.1){2}{\line(1,0){0.23}}
\multiput(16.36,29.84)(0.23,-0.09){2}{\line(1,0){0.23}}
\multiput(16.83,29.65)(0.47,-0.16){1}{\line(1,0){0.47}}
\multiput(17.3,29.49)(0.48,-0.14){1}{\line(1,0){0.48}}
\multiput(17.78,29.35)(0.49,-0.12){1}{\line(1,0){0.49}}
\multiput(18.27,29.23)(0.49,-0.09){1}{\line(1,0){0.49}}
\multiput(18.76,29.13)(0.5,-0.07){1}{\line(1,0){0.5}}
\multiput(19.25,29.06)(0.5,-0.05){1}{\line(1,0){0.5}}
\multiput(19.75,29.01)(0.5,-0.02){1}{\line(1,0){0.5}}
\put(20.25,28.99){\line(1,0){0.5}}
\multiput(20.75,28.99)(0.5,0.02){1}{\line(1,0){0.5}}
\multiput(21.25,29.01)(0.5,0.05){1}{\line(1,0){0.5}}
\multiput(21.75,29.06)(0.5,0.07){1}{\line(1,0){0.5}}
\multiput(22.24,29.13)(0.49,0.09){1}{\line(1,0){0.49}}
\multiput(22.73,29.23)(0.49,0.12){1}{\line(1,0){0.49}}
\multiput(23.22,29.35)(0.48,0.14){1}{\line(1,0){0.48}}
\multiput(23.7,29.49)(0.47,0.16){1}{\line(1,0){0.47}}
\multiput(24.17,29.65)(0.23,0.09){2}{\line(1,0){0.23}}
\multiput(24.64,29.84)(0.23,0.1){2}{\line(1,0){0.23}}
\multiput(25.09,30.04)(0.22,0.11){2}{\line(1,0){0.22}}
\multiput(25.54,30.27)(0.22,0.13){2}{\line(1,0){0.22}}
\multiput(25.97,30.52)(0.21,0.14){2}{\line(1,0){0.21}}
\multiput(26.39,30.79)(0.2,0.15){2}{\line(1,0){0.2}}
\multiput(26.8,31.08)(0.13,0.1){3}{\line(1,0){0.13}}
\multiput(27.19,31.39)(0.13,0.11){3}{\line(1,0){0.13}}
\multiput(27.57,31.72)(0.12,0.12){3}{\line(1,0){0.12}}
\multiput(27.93,32.07)(0.12,0.12){3}{\line(0,1){0.12}}
\multiput(28.28,32.43)(0.11,0.13){3}{\line(0,1){0.13}}
\multiput(28.61,32.81)(0.1,0.13){3}{\line(0,1){0.13}}
\multiput(28.92,33.2)(0.15,0.2){2}{\line(0,1){0.2}}
\multiput(29.21,33.61)(0.14,0.21){2}{\line(0,1){0.21}}
\multiput(29.48,34.03)(0.13,0.22){2}{\line(0,1){0.22}}
\multiput(29.73,34.46)(0.11,0.22){2}{\line(0,1){0.22}}
\multiput(29.96,34.91)(0.1,0.23){2}{\line(0,1){0.23}}
\multiput(30.16,35.36)(0.09,0.23){2}{\line(0,1){0.23}}
\multiput(30.35,35.83)(0.16,0.47){1}{\line(0,1){0.47}}
\multiput(30.51,36.3)(0.14,0.48){1}{\line(0,1){0.48}}
\multiput(30.65,36.78)(0.12,0.49){1}{\line(0,1){0.49}}
\multiput(30.77,37.27)(0.09,0.49){1}{\line(0,1){0.49}}
\multiput(30.87,37.76)(0.07,0.5){1}{\line(0,1){0.5}}
\multiput(30.94,38.25)(0.05,0.5){1}{\line(0,1){0.5}}
\multiput(30.99,38.75)(0.02,0.5){1}{\line(0,1){0.5}}

\linethickness{0.7mm}
\qbezier(20,50)(25.19,55.25)(30,55.25)
\qbezier(30,55.25)(34.81,55.25)(40,50)
\put(20,50){\makebox(0,0)[cc]{$\bullet$}}

\put(40,50){\makebox(0,0)[cc]{$\bullet$}}

\linethickness{0.7mm}
\put(90,30){\line(0,1){20}}
\end{picture}
}

\def\figKnew{
\def\JPicScale{1.2}
\ifx\JPicScale\undefined\def\JPicScale{1}\fi
\unitlength \JPicScale mm
\begin{picture}(110,60)(0,0)
\linethickness{0.7mm}
\put(70,50){\line(1,0){20}}
\linethickness{0.7mm}
\put(90,50){\line(1,0){20}}
\linethickness{0.7mm}
\qbezier(90,50)(84.75,44.81)(84.75,40)
\qbezier(84.75,40)(84.75,35.19)(90,30)
\linethickness{0.7mm}
\qbezier(90,50)(95.25,44.81)(95.25,40)
\qbezier(95.25,40)(95.25,35.19)(90,30)
\put(90,50){\makebox(0,0)[cc]{$\bullet$}}

\put(90,30){\makebox(0,0)[cc]{$\bullet$}}

\put(93,40){\makebox(0,0)[cc]{$\cdots$}}

\put(90,20){\makebox(0,0)[cc]{(b)}}

\linethickness{0.7mm}
\put(90,30){\line(0,1){20}}
\end{picture}
}

\def\figstar{

\def\JPicScale{0.8}
\ifx\JPicScale\undefined\def\JPicScale{1}\fi
\unitlength \JPicScale mm
\begin{picture}(70,60)(0,0)
\linethickness{0.3mm}
\multiput(30,60)(0.12,-0.12){167}{\line(1,0){0.12}}
\linethickness{0.3mm}
\put(30,40){\line(1,0){20}}
\linethickness{0.3mm}
\multiput(30,20)(0.12,0.12){167}{\line(1,0){0.12}}
\linethickness{0.3mm}
\multiput(50,40)(0.12,0.12){167}{\line(1,0){0.12}}
\linethickness{0.3mm}
\put(50,40){\line(1,0){20}}
\linethickness{0.3mm}
\multiput(50,40)(0.12,-0.12){167}{\line(1,0){0.12}}
\end{picture}

}

\def\figsixteen{

\def\JPicScale{0.8}
\ifx\JPicScale\undefined\def\JPicScale{1}\fi
\unitlength \JPicScale mm
\begin{picture}(70,45)(0,0)
\linethickness{0.7mm}
\put(30,40){\line(1,0){40}}
\put(30,40){\makebox(0,0)[cc]{$\times$}}

\put(70,40){\makebox(0,0)[cc]{$\times$}}

\put(30,45){\makebox(0,0)[cc]{1}}

\put(70,45){\makebox(0,0)[cc]{2}}

\end{picture}

}

\def\figcontour{

\def\JPicScale{0.8}
\ifx\JPicScale\undefined\def\JPicScale{1}\fi
\unitlength \JPicScale mm
\begin{picture}(130,80)(0,0)
\linethickness{0.1mm}
\put(10,40){\line(1,0){120}}
\linethickness{0.1mm}
\put(60,0){\line(0,1){80}}
\put(40,45){\makebox(0,0)[cc]{$\times$}}

\put(80,35){\makebox(0,0)[cc]{$\times$}}

\put(90,45){\makebox(0,0)[cc]{$\times$}}

\put(110,35){\makebox(0,0)[cc]{$\times$}}

\linethickness{0.7mm}
\qbezier(30,40)(35.16,32.19)(42.38,27.38)
\qbezier(42.38,27.38)(49.59,22.56)(60,20)
\linethickness{0.7mm}
\qbezier(110,40)(99.66,53.05)(87.62,59.06)
\qbezier(87.62,59.06)(75.59,65.08)(60,65)
\linethickness{0.7mm}
\put(60,0){\line(0,1){20}}
\linethickness{0.7mm}
\put(60,65){\line(0,1){15}}
\put(40,50){\makebox(0,0)[cc]{{\small $-2P_2+i\ve$}}}

\linethickness{0.7mm}
\put(30,40){\line(1,0){80}}

\put(80,30){\makebox(0,0)[cc]{{\small $2P_2-i\ve$}}}

\put(85,50){\makebox(0,0)[cc]{\small{$\tilde\omega_1-2P_1+i\ve$}}}

\put(110,30){\makebox(0,0)[cc]{{\small $\tilde\omega_1+2P_1-i\ve$}}}

\end{picture}

}

\def\figpicard{

\def\JPicScale{0.8}
\ifx\JPicScale\undefined\def\JPicScale{1}\fi
\unitlength \JPicScale mm
\begin{picture}(75,70)(0,0)
\linethickness{0.1mm}
\put(20,40){\line(1,0){50}}
\linethickness{0.3mm}
\put(20,40){\line(0,1){30}}
\linethickness{0.3mm}
\put(20,10){\line(0,1){30}}
\linethickness{0.3mm}
\qbezier(70,40)(54.37,40)(45.34,40)
\qbezier(45.34,40)(36.32,40)(32.5,40)
\qbezier(32.5,40)(28.59,39.98)(26.78,41.19)
\qbezier(26.78,41.19)(24.98,42.39)(25,45)
\qbezier(25,45)(25,47.58)(25,51.19)
\qbezier(25,51.19)(25,54.8)(25,60)
\qbezier(25,60)(25,65.22)(25,67.62)
\qbezier(25,67.62)(25,70.03)(25,70)
\put(10,40){\makebox(0,0)[cc]{$\theta=0$}}

\put(80,40){\makebox(0,0)[cc]{$\theta=\pi$}}

\put(20,55){\makebox(0,0)[cc]{$\downarrow$}}

\put(20,25){\makebox(0,0)[cc]{$\uparrow$}}

\put(25,50){\makebox(0,0)[cc]{$\downarrow$}}

\put(50,40){\makebox(0,0)[cc]{$\rightarrow$}}

\put(10,65){\makebox(0,0)[cc]{$\JJ^+_{\rm inst}$}}

\put(10,15){\makebox(0,0)[cc]{$\JJ^-_{\rm inst}$}}

\put(50,45){\makebox(0,0)[cc]{$\JJ_{\rm pert}$}}

\end{picture}

}

\def\figpicard{

\def\JPicScale{0.8}
\ifx\JPicScale\undefined\def\JPicScale{1}\fi
\unitlength \JPicScale mm
\begin{picture}(75,70)(0,0)
\linethickness{0.3mm}
\put(20,40){\line(1,0){50}}
\linethickness{0.3mm}
\put(20,40){\line(0,1){30}}
\linethickness{0.3mm}
\put(20,10){\line(0,1){30}}
\linethickness{0.3mm}
\qbezier(70,40)(54.37,40)(45.34,40)
\qbezier(45.34,40)(36.32,40)(32.5,40)
\qbezier(32.5,40)(28.59,39.98)(26.78,41.19)
\qbezier(26.78,41.19)(24.98,42.39)(25,45)
\qbezier(25,45)(25,47.58)(25,51.19)
\qbezier(25,51.19)(25,54.8)(25,60)
\qbezier(25,60)(25,65.22)(25,67.62)
\qbezier(25,67.62)(25,70.03)(25,70)
\put(10,40){\makebox(0,0)[cc]{$\theta=0$}}

\put(80,40){\makebox(0,0)[cc]{$\theta=\pi$}}

\put(20,55){\makebox(0,0)[cc]{$\downarrow$}}

\put(20,25){\makebox(0,0)[cc]{$\uparrow$}}

\put(25,50){\makebox(0,0)[cc]{$\downarrow$}}

\put(50,40){\makebox(0,0)[cc]{$\rightarrow$}}

\put(10,65){\makebox(0,0)[cc]{$\JJ^+_{\rm inst}$}}

\put(10,15){\makebox(0,0)[cc]{$\JJ^-_{\rm inst}$}}

\put(32,55){\makebox(0,0)[cc]{$\JJ^+_{\rm pert}$}}

\put(32,28){\makebox(0,0)[cc]{$\JJ^-_{\rm pert}$}}

\linethickness{0.3mm}
\qbezier(70,40)(54.37,40)(45.34,40)
\qbezier(45.34,40)(36.32,40)(32.5,40)
\qbezier(32.5,40)(28.59,40.02)(26.78,38.81)
\qbezier(26.78,38.81)(24.98,37.61)(25,35)
\qbezier(25,35)(25,32.42)(25,28.81)
\qbezier(25,28.81)(25,25.2)(25,20)
\qbezier(25,20)(25,14.78)(25,12.38)
\qbezier(25,12.38)(25,9.97)(25,10)
\put(25,20){\makebox(0,0)[cc]{$\uparrow$}}

\end{picture}

}

\def\figpicardtwo{

\def\JPicScale{0.8}
\ifx\JPicScale\undefined\def\JPicScale{1}\fi
\unitlength \JPicScale mm
\begin{picture}(75,70)(0,0)
\linethickness{0.3mm}
\put(20,40){\line(1,0){50}}
\linethickness{0.3mm}
\put(20,40){\line(0,1){30}}
\linethickness{0.3mm}
\put(20,10){\line(0,1){30}}
\linethickness{0.3mm}
\qbezier(20,40)(35.63,40)(44.66,40)
\qbezier(44.66,40)(53.68,40)(57.5,40)
\qbezier(57.5,40)(61.41,39.98)(63.22,41.19)
\qbezier(63.22,41.19)(65.02,42.39)(65,45)
\qbezier(65,45)(65,47.58)(65,51.19)
\qbezier(65,51.19)(65,54.8)(65,60)
\qbezier(65,60)(65,65.22)(65,67.62)
\qbezier(65,67.62)(65,70.03)(65,70)
\put(10,40){\makebox(0,0)[cc]{$\theta=0$}}

\put(80,40){\makebox(0,0)[cc]{$\theta=\pi$}}

\put(70,55){\makebox(0,0)[cc]{$\downarrow$}}

\put(70,25){\makebox(0,0)[cc]{$\uparrow$}}

\put(65,50){\makebox(0,0)[cc]{$\downarrow$}}

\put(50,40){\makebox(0,0)[cc]{$\leftarrow$}}

\put(57,65){\makebox(0,0)[cc]{$\JJ^+_{\rm inst}$}}

\put(57,15){\makebox(0,0)[cc]{$\JJ^-_{\rm inst}$}}

\put(80,65){\makebox(0,0)[cc]{$\JJ^+_{\rm pert}$}}

\put(80,15){\makebox(0,0)[cc]{$\JJ^-_{\rm pert}$}}

\linethickness{0.3mm}
\qbezier(20,40)(35.63,40)(44.66,40)
\qbezier(44.66,40)(53.68,40)(57.5,40)
\qbezier(57.5,40)(61.41,40.02)(63.22,38.81)
\qbezier(63.22,38.81)(65.02,37.61)(65,35)
\qbezier(65,35)(65,32.42)(65,28.81)
\qbezier(65,28.81)(65,25.2)(65,20)
\qbezier(65,20)(65,14.78)(65,12.38)
\qbezier(65,12.38)(65,9.97)(65,10)
\put(65,20){\makebox(0,0)[cc]{$\uparrow$}}

\linethickness{0.3mm}
\put(70,10){\line(0,1){60}}
\end{picture}

}

\def\figpicardthree{

\def\JPicScale{0.8}
\ifx\JPicScale\undefined\def\JPicScale{1}\fi
\unitlength \JPicScale mm
\begin{picture}(130,80)(0,0)
\linethickness{0.1mm}
\put(70,0){\line(0,1){80}}
\linethickness{0.1mm}
\put(10,40){\line(1,0){120}}
\linethickness{0.3mm}
\linethickness{0.3mm}
\linethickness{0.3mm}
\linethickness{0.3mm}
\linethickness{0.3mm}
\linethickness{0.3mm}
\linethickness{0.5mm}
\qbezier(70,40)(80.38,40)(90,40)
\qbezier(90,40)(99.62,40)(110,40)
\qbezier(110,40)(120.44,40)(125.25,40)
\qbezier(125.25,40)(130.06,40)(130,40)
\linethickness{0.5mm}
\qbezier(70,40)(70,37.4)(70,35.59)
\qbezier(70,35.59)(70,33.79)(70,32.5)
\qbezier(70,32.5)(70.02,31.23)(68.81,27.62)
\qbezier(68.81,27.62)(67.61,24.02)(65,17.5)
\qbezier(65,17.5)(62.41,10.99)(60,6.78)
\qbezier(60,6.78)(57.59,2.57)(55,0)

\put(95,40){\makebox(0,0)[cc]{$\times$}}

\put(95,45){\makebox(0,0)[cc]{$T=1$}}

\put(60,45){\makebox(0,0)[cc]{$T=0$}}

\put(80,47){\makebox(0,0)[cc]{$\JJ_{\rm pert}^-$}}

\put(120,47){\makebox(0,0)[cc]{$\JJ_{\rm pert}^+$}}

\put(80,39.8){\makebox(0,0)[cc]{$\Rightarrow$}}

\put(120,39.8){\makebox(0,0)[cc]{$\Leftarrow$}}

\put(70,40){\makebox(0,0)[cc]{$\times$}}

\put(70,35){\makebox(0,0)[cc]{$\Uparrow$}}

\put(58,20){\makebox(0,0)[cc]{$\JJ_{\rm inst}^-$}}

\end{picture}

}

\newcommand{\NNN}{{\cal N}}

\begin{document}

\baselineskip 24pt

\begin{center}

{\Large \bf Cutkosky Rules and Unitarity (Violation) in D-instanton Amplitudes}


\end{center}

\vskip .6cm
\medskip

\vspace*{4.0ex}

\baselineskip=18pt

\centerline{\large \rm Ashoke Sen}

\vspace*{4.0ex}

\centerline{\large \it Harish-Chandra Research Institute, HBNI}
\centerline{\large \it  Chhatnag Road, Jhusi,
Allahabad 211019, India}


\vspace*{1.0ex}
\centerline{\small E-mail:  sen@hri.res.in}

\vspace*{5.0ex}

\centerline{\bf Abstract} \bigskip

In perturbative amplitudes in quantum field theory and string field theory, Cutkosky 
rule expresses
the anti-hermitian 
part of a Feynman diagram in terms of sum over all its cut diagrams, and this
in turn is used to prove unitarity of the theory. For D-instanton contribution to a string theory
amplitude, the cutting rule needed for the proof of unitarity is somewhat different; we need 
to sum over only those cut diagrams for which all 
the world-sheet boundaries ending on some
particular D-instanton lie on the same side of the cut. 
By working with the closed string effective action,
obtained after integrating out the open string modes, we
prove that
the D-instanton 
amplitudes actually satisfy these cutting rules, provided the effective action  is real.
The  violation of unitarity in the closed string sector of
two dimensional string theory can be traced to the failure of this reality condition.
In the critical superstring theory, multi-instanton and
multi anti-instanton amplitudes satisfy  the reality condition. 
Contribution to the amplitudes from the  instanton anti-instanton sector 
satisfies the reality condition if we make a specific choice of integration cycle over the
configuration space of string fields, whereas contribution due to the non-BPS
D-instantons will need to either vanish or have an overall real normalization 
in order for it to give real contribution. We use 
Picard-Lefschetz theory to argue that these conditions are indeed satisfied in superstring
theories.

\vfill \eject

\renewcommand{\NN}{{\bf N}}

\tableofcontents

\sectiono{Introduction and summary}

Cutkosky rule in quantum field theory expresses the anti-hermitian part of a Feynman
diagram in terms
of sum of its cut diagrams, and is used to prove perturbative unitarity of the 
theory\cite{Cutkosky, fowler,veltman,diagrammar,1512.01705}.
In local quantum field theory the Cutkosky rule is proved using a particular property of
the position space Green's function known as the largest 
time equation\cite{veltman,diagrammar}. In string
(field) theory the interaction vertices in the momentum space representation
have exponential dependence on
the quadratic functions of momenta, and there is no simple local
position space description
of these vertices. For this reason the proof based on the largest time equation is not
available. Nevertheless it is possible to prove the Cutkosky rules directly by analyzing the
momentum space Feynman diagrams of string field theory\cite{1604.01783}.

The results described above hold for perturbative amplitudes in quantum field theory and
string theory to all orders in the coupling constant. However in string theory, certain
non-perturbative contributions to the amplitudes may be computed using perturbative 
techniques. These refer to amplitudes associated to D-instantons\cite{9407031,9701093}, 
and the perturbative
technique involves the quantum (field) theory of open and closed strings. Therefore it is
natural to ask if one can extend the proof of unitarity of string theory 
perturbative amplitudes based on  Cutkosky rules also to D-instanton induced amplitudes.

Even if string theory is a consistent quantum theory, the answer to this question is not
obvious. For example, let us suppose that we consider the scattering of a pair of closed string
states at high energy in type IIA string theory. 
At sufficiently high energy, this process may produce a D0-$\bar{\rm D}0$ brane pair.
Therefore if we consider the scattering amplitudes in the closed string sector alone, 
where
we allow the external states to be only closed strings, we expect the 
Cutkosky rules to fail.
Formally this can be stated by saying that if we integrate out all the degrees of freedom
other than closed strings, then the effective action for the closed strings will acquire an
imaginary part at energies above the threshold of production of the 
D0-$\bar{\rm D}0$ brane pair.

Now suppose that we consider scattering at energies below the threshold of production
of the D0-$\bar{\rm D}0$ brane pair. Naively one would expect that the unitarity in the
closed string sector will be restored, and the closed string effective action below this
energy will be real. However, in the open string theory on the D0-$\bar{\rm D}0$ brane pair,
there are classical solutions with energy below this threshold -- these are the rolling tachyon
solutions\cite{0203211,0203265}. 
The energies of these solutions can go all
the way down to zero, and therefore if these are to be counted as independent states in
type IIA string theory then we would expect violation of unitarity in the closed string sector
even at low energy. 
In type IIB string theory there are similar low energy configurations
associated with rolling tachyons on non-BPS 
D0-branes to which the closed string states could make transitions.
Indeed something similar happens in two dimensional
bosonic string theory, where after taking into account the D-instanton corrections to the
S-matrix, one finds violation of unitarity in the closed string sector by itself, and this can
be traced to possible transition to low energy rolling tachyon configuration on 
non-BPS
D0-branes\cite{1907.07688,1912.07170}. 
There is reason to believe however that in critical string theory the 
rolling tachyon configurations correspond to some configurations 
of closed string states, and should not be counted as independent states\cite{0303139}. 
In particular, if the rolling tachyon
system, after quantization, 
represents new kinds of low mass particles with standard coupling to gravity, they would
affect the perturbative amplitudes by appearing as intermediate states in the
graviton exchange diagrams, in contradiction to
the observation that D-instanton effects are of order $e^{-C/g_s}$ for some constant $C$.
Therefore we would expect that unitarity is present in the closed string sector itself.
Nevertheless it is clearly important to analyze in which theories this happens, and
in particular the difference between the two dimensional bosonic string theory and the
critical superstring theories that renders the former non-unitary and the latter
unitary within the closed
string sector.

This is the problem we shall address in this paper. 
Since the excitations around a D-instanton background involve both open and closed
strings,  the amplitudes will be
expressed as integrals over the moduli spaces of Riemann surfaces with boundaries,
with the integrand constructed from the correlation functions of vertex operators
on the Riemann surfaces.
However even for the usual string amplitudes, this is not an effective approach to 
proving the cutting rules, since the integrands that appear 
in these integrals
are always real, and the imaginary parts predicted by cutting rules can only arise from
the integrals encountering divergences from the boundaries of the moduli spaces,
requiring us to define the integrals via analytic continuation, or deforming the integration
contour into the complexified moduli 
space\cite{sundborg,amano,marcus,sundborg1,
9302003,9404128,9410152,berera2,1307.5124}. 
This makes the analysis of the imaginary
parts in the world-sheet formalism difficult. For this reason, the natural framework for
studying the cutting rules in string theory is string field theory, where we
can try to apply the usual method of quantum field theory to try to prove the
cutting rules\cite{1604.01783}.
In the presence of D-instantons, the relevant string field theory is open closed
string field theory, describing the dynamics of open and closed strings.

It turns out however, that one cannot directly apply the analysis of \cite{1604.01783}
to the current problem. One source of the difficulty 
is that the individual interaction vertices of open closed string
field theory in the D-instanton background
do not satisfy the usual momentum conservation laws that were used in the
analysis of \cite{1604.01783}. 
Another difficulty is that for a given D-instanton system, each
amplitude is accompanied by an overall normalization factor independently of
the number of vertices and propagators that the Feynman diagram has, and as
a result, the cutting rules, that give  non-linear relations between the amplitudes,
do not hold in their usual form. A third difficulty is that a general cut of a Feynman
diagram will have cut running through both open and closed string propagators, but
since open strings are not asymptotic states, we should not have cut open
string propagators.

We shall 
circumvent these problems by  integrating out the open string modes and
defining an effective field theory of closed
strings whose interaction vertices satisfy the usual momentum conservation
rules. This is somewhat different from the conventional procedure for integrating out
a set of fields in a quantum field theory or string field 
theory\cite{0112228,0306332,1609.00459,2006.16270}, 
since here we need to sum over both connected
and disconnected diagrams for constructing the interaction vertices of the effective
field theory and include the overall normalization factor mentioned in the last paragraph
in the definition of the interaction vertices. Nevertheless we find that the action
still has the gauge invariance required for proving
the decoupling of
the unphysical states. We can then
derive the cutting rules directly in this closed string effective field theory, leading to
a proof of unitarity within the closed string sector, {\em provided
the action of this effective field theory is real}. 

In conventional quantum field theory, when we integrate out a set of heavy modes
to construct an effective action for the light modes, the effective action is real at low
energy, but above the threshold of production of the heavy modes, it acquires an
imaginary part, indicating the possibility of transition of the light modes into heavy
modes. In contrast, here we are integrating out the open string modes on the D-instanton
which are not allowed asymptotic states. Therefore, unitarity requires the effective 
action to be exactly real.
We shall find however that sometime the reality condition may fail to hold.
This can arise from various sources that we shall now describe.
\begin{enumerate}
\item  The boundary 
state\cite{callan1,callan2,cai,ishibashi,9604091,9707068,9912275} describing the
D-instanton may be complex, leading to complex contribution to the effective action.
The D-instantons in type IIB string theory are of this kind\cite{9403040,9701093}. 
However in such cases,
one usually finds a complex conjugate boundary state describing the anti-D-instanton,
and when one adds the contribution to the effective action from the D-instanton and the
anti-D-instanton, one gets real result. In our analysis we shall assume that this is always
so, i.e.\ there is no violation of reality of the effective action from having complex boundary
states. In the two dimensional bosonic string theory that we shall discuss in detail, 
the boundary states of the D-instantons
are all real.
\item
A D-instanton amplitude is accompanied by
an overall normalization factor $\NN$ that measures the ratio of the integration 
measure over the string fields in the D-instanton sector and the vacuum sector, 
including the ratio of one loop
determinants of the string fields in these sectors. 
If this factor is complex, then the corresponding contribution to the closed string
effective action is complex, leading to violation of unitarity. 
Although
it should be possible in principle to determine this constant explicitly by regarding both
the D-instanton and the perturbative vacuum as different classical solutions of some
string field theory, {\it e.g.} the ones involving unstable D-branes, at present we do not
have sufficient computational tools to determine this. 

In some cases, however, $\NN$ may
be determined by making use of known duality relations. For example, in two
dimensional bosonic string theory, $\NN$ can be determined by comparison with the
dual matrix model\cite{1907.07688} and turns out to be imaginary, 
causing violation of cutting rules and
unitarity in the purely closed string sector of the theory. On the other hand in type IIB
string theory, application of S-duality leads to real D-instanton induced term in the
effective action\cite{9701093}. This implies that the constant $\NN$ for the 
D-instanton and the constant $\bar\NN$ for the 
anti-D-instanton are complex conjugates of each other, leading
to real effective action. In this case the
actual phase of $\NN$ is ambiguous since it can be absorbed into a shift of the RR scalar
field, but we can set $\NN$ and $\bar\NN$ to be real and equal
by expanding the action around the 
vanishing RR scalar background. 

Since in the presence of a tachyon the steepest descent contour has to run along the
imaginary axis, 
it is reasonable
to conjecture that whether $\NN$ is real or complex is determined by the absence or
presence of tachyons on the D-instanton system. The examples described above
are
consistent with this. 
\item
The third source of violation of the reality condition is more directly related to tachyons
-- those arising from open strings that connect two different D-instantons. 
If we denote the mass$^2$ of this tachyon for coincident D-instantons
as $-M^2$ and denote by $\chi^2$ the squared distance between the D-instantons in
the Euclidean space-time, then the mass$^2$ of the tachyon in the $\alpha'=1$
unit is given by $-M^2+\chi^2/(4\pi^2)$.
Furthermore these tachyons typically arise in pairs related by the reversal in orientation
of the open strings. Then the one loop determinant of these fields  
is given by a term
proportional to:
\be \label{eint11}
(\chi^2 - 4\pi^2 M^2)^{-1}\, .
\ee
At higher orders, there will also be such terms from internal open string
tachyon propagators.
Since $\chi$ represents a mode of the open string and needs to be integrated over to
define the closed string effective action, we need a prescription for dealing with the
singularities at $\chi=\pm 2\pi M$. 
The prescription of \cite{1912.07170,talk}, called the Lorentzian prescription,
is to replace \refb{eint11} by:
\be \label{eint13}
(\chi^2 - 4\pi^2 M^2+i\eps)^{-1}\, .
\ee
However, this typically generates imaginary contribution to the closed string effective
action and leads to violation of unitarity. 

\end{enumerate}

The violation of unitarity associated with the Lorentzian prescription
\refb{eint13} was not a problem in the analysis
of \cite{1912.07170}, where the underlying theory is not unitary when restricted to
purely closed string sector and the lack of unitarity signals possible transition to
rolling tachyon configuration on the D0-brane. 
However, if we use the same prescription for the
instanton anti-instanton sector of type IIB string theory we shall get unitarity violating
amplitudes in the closed string sector, which, as we have discussed earlier, does
not seem likely. 
For this reason, we suggest a different prescription, where we deal with the
singularities of the type given in \refb{eint11} by 
averaging over the Lorentzian prescription of \cite{1912.07170} and its complex
conjugate. This means that we replace all the $i\eps$'s in the Lorentzian 
prescription by $-i\eps$ and then take the average of the original result and
the new result. 
This leads to real closed string effective action, restoring unitarity. 
We shall call this the unitary prescription.
Physically, these two prescriptions can be
regarded as different choices of integration cycles for the variables
$\chi$. Since the modes $\chi$ are particular
modes of the open 
string field, the choice of integration cycle refers to the choice of integration
cycle in the space of string fields.
The unitary prescription also seems to be needed in two dimensional type 0B
string theory\cite{private}.

For a general amplitude with $k$ instantons the singularity structure is quite
complicated. For example if we denote by $\phi_i$ the Euclidean 
space-time coordinate of the
$i$-th D-instanton, then in general the singular factor in the Lorentzian 
prescription takes the form,
\be\label{esingone}
\prod_{i,j=1\atop i<j}^k \left\{(\phi_i-\phi_j)^2 - 4\pi^2 M_{ij}^2 +i\eps\right\}^{-a_{ij}}\, ,
\ee
for some integers $a_{ij}$. In the unitary prescription one needs to average between
\refb{esingone} and its complex conjugate. Despite 
the  proliferation of singularities of the
kind given in \refb{esingone} for large number of instantons,
one can check that
both the Lorentzian and the unitary prescriptions give non-singular results, i.e.\ 
the integration 
contour is not pinched between a pair of poles from the
opposite side. For the Lorentzian prescription this follows from the fact that deforming
each $\phi_i$ integration contour to $\phi_i = e^{i\theta} u_i$ with real $u_i$ and 
$\theta>0$, 
we can
move away from the singularities, since in the complex $u_i$ plane the poles are now
at $(u_i-u_j)^2=(4\pi^2 M_{ij}^2 - i\eps)e^{-2i\theta}$, i.e.\ away from the real axis,
and furthermore,  as we vary $\theta$ from 0 to some positive value, 
the poles do not cross the real $u_i$ axis along which the integration
contour lies. For
this argument we must use the same $\theta$ for all components of all 
the $\phi_i$'s, so that the
difference between any pair of $\phi_i$'s also has the same phase $e^{i\theta}$.
For the complex conjugate contour, needed for the
unitary prescription, we can similarly deform the $\phi_i$ 
integration contour to $e^{-i\theta} u_i$
with real $u_i$ and $\theta>0$ to move away from the singularities.
In contrast, if we had used other prescriptions containing product of terms in which
some have $i\eps$ and others 
have $-i\eps$, the contour could get pinched
between poles from opposite sides, since the number of denominator factors
given by the number of independent pairs of D-instantons grows faster than the
number of integration variables which is proportional to the number of D-instantons.
However other special choices that avoid singularities may be possible, and given any
such choice, there is another choice in which we change the sign of all the $i\eps$'s.
As long as we always take the average of such pair of choices, we shall get unitary
theory, provided the normalization constants $\NN$ are real.

Since \cite{1912.07170} found agreement between the matrix model results and
string theory results with the
Lorentzian prescription, one might ask if the agreement is spoiled when
we use the
unitary prescription. A conservative answer to this question is that since 
in two dimensional string theory we expect violation of unitarity in the closed string sector
anyway, one could continue to use the Lorentzian prescription for this theory and use
unitary prescription for the critical string theories. We have nevertheless
explored what happens to the agreement with the matrix model result if we use
the unitary prescription to define the instanton amplitudes in two dimensional string
theory. We find that for the leading two instanton contribution to the $n$
point amplitude of closed strings and the leading 3-instanton contribution to the
2-point amplitude of closed strings, the difference 
between the two prescriptions can be compensated for by a change in the normalization
constants $\NN$ for (2,1) and (3,1) ZZ-instantons\cite{0101152}. 
For example in the analysis of \cite{1912.07170}, the
normalization constants (in our notation) for the (2,1) and (3,1) instantons
would be given by $\NN_2 = -3i/(64\pi^2)$ and $\NN_3=5i/(192\pi^2)$ respectively.
In contrast in the unitary prescription the agreement with the matrix model results 
determines these constants to be $-i/(32\pi^2)$ and $i/(96\pi^2)$ 
respectively.\footnote{Similar results were observed independently
by Balthazar, Rodriguez and 
Yin\cite{private}.}  Eventually 
it may be possible
to explain these results as a result of contour deformation in the configuration space of
string fields, but this 
will require a better understanding of this configuration space.
Note that in the unitary prescription, the unitarity violation in the
two dimensional string theory can be attributed solely to the imaginary values of the
normalization constants $\NN$. 

Based on all the observations described above, we can summarize the status
of our understanding of D-instanton amplitudes in the critical superstring
theory as follows:
\begin{enumerate}
\item We have argued that for BPS instantons the normalization constant $\NN$  
for the instanton and
the anti-instanton are complex conjugates of each other. Even though this is
determined by examining a particular term in the effective action that is protected
by supersymmetry, since the normalization constant is the same for all amplitudes,
we conclude that the closed string effective action associated with the single instanton
and single anti-instanton amplitudes are real. Furthermore, since the normalization
constants for multi-instanton amplitudes are determined from that of single
instanton amplitudes, and since no tachyon appears on the multi-instanton or
multi-anti-instanton sectors anywhere in the moduli space of instantons, 
we can conclude that the
contribution to the closed string effective action from the multi-instanton and 
multi-anti-instanton sectors are also real. Therefore there is no violation of 
unitarity in the closed string sectors from these amplitudes.
\item The situation is different for amplitudes in the instanton anti-instanton sector and
for non-BPS D-instantons that carry tachyonic modes on their world-volume. In the
former case we can have real closed string effective action and therefore unitary 
amplitudes if we choose the unitary prescription, but there are other choices of
integration contour that leads to complex effective action.  
In the presence of non-BPS D-instantons,
the steepest descent contour runs along the imaginary axis for the tachyon field,
suggesting that the normalization constant $\NN$ should be imaginary. This would 
suggest imaginary contribution to the closed string effective action
unless $\NN$ vanishes.
\end{enumerate}
While we cannot fully resolve the ambiguities mentioned above, we have used
Picard-Lefschetz theory and general properties of the tachyon potential on non-BPS
D-instanton and instanton - anti-instanton system, to argue 
that for the former case $\NN$ can
indeed be taken to vanish, and in the latter case unitary prescription is the appropriate 
prescription for computing the effective action. 
These arguments depend on a particular natural 
choice of integration contour for the tachyonic
modes that keeps the tachyons real along the contour.
On the other hand, for
two dimensional bosonic string theory, this natural choice of contour does not exist,
forcing the theory to be non-unitary. It should be noted however that even in critical
superstring theories, 
other choices of integration contour that violate unitarity are also possible.

We shall now summarize the rest of the paper.
In \S\ref{scut} we describe why the
usual cutting rules, needed for the proof of unitarity, must be modified 
in the open closed string field theory\cite{9705241,1907.10632} on
D-instantons.  In \S\ref{sreal} we discuss some aspects of the reality properties
of string field theory action.
In \S\ref{seffective} we describe how we can prove the modified cutting rules,
required for unitarity,
for
amplitudes induced by a single D-instanton, provided we assume certain reality
property of the overall normalization constant that multiplies a D-instanton amplitude.
This normalization constant 
cannot be computed with the currently available techniques, but
there is indirect evidence that the required reality properties may fail
if the D-instanton system has open string tachyonic modes.
In \S\ref{smulti} we extend the result to multi-instanton amplitudes, where we
find a more direct relation between the existence of open string tachyonic modes and
the reality properties of the closed string effective action. We discuss the Lorentzian
and unitary prescriptions for integrating over the open string zero modes and show
how using the unitary prescription one can avoid the breakdown of reality properties
of the effective action, provided the overall normalization constants associated with
single instanton amplitudes are real.
In \S\ref{sexamples} we illustrate the
cutting rules for single D-instanton amplitudes with a specific example in 
two dimensional string 
theory\cite{dj,sw,gk,kr,9111035,0307083,0307195,1705.07151,1907.07688,
1912.07170}. We demonstrate that the cutting rules hold when the overall
normalization constant satisfies the required reality properties, and the breakdown
of unitarity found in  \cite{1907.07688} can be traced solely to the imaginary value
of the normalization constant.  
In \S\ref{sexamplemulti} we analyze the unitarity of 2-instanton amplitudes in two
dimensional bosonic string theory. 
We demonstrate that even if we choose the normalization constant appearing in the one
instanton sector in a way that gives unitary amplitude, the cutting rules still fail in the
two instanton sector
if we use the Lorentzian contour 
prescription of \cite{1912.07170} to integrate over the 
open string zero modes.
This failure can be traced directly to the appearance of a tachyonic mode of the
open string connecting the two D-instantons below a critical separation.
However we also show that unitarity is restored if we use the unitary prescription for
the integration contour, provided the overall normalization constants are chosen to be real.
In \S\ref{spicard} we use Picard-Lefschetz theory to argue that in type IIA and IIB
superstring theories, where the tachyon potential is bounded from below, we have natural
choice of integration contours for which
there is no violation of unitarity in the closed string sector due to
D-instantons.

We shall end this introduction with some final remarks.
Since the amplitudes in critical string
theory are not Borel summable, 
a critical reader may
want to take the point of view that 
one cannot trust the instanton contribution to these amplitudes
before we have understood the perturbative corrections fully. While this is
a valid criticism for general string amplitudes,  not every 
quantity that one can compute in perturbation theory has this problem. 
In particular, the matrix elements of $S(g_s^*)^\dagger S(g_s)$ 
between closed string states is
known to be given by those of the identity matrix 
to all orders in perturbation theory due to perturbative unitarity
of superstring theories. Therefore it makes sense to compute the D-instanton corrections
to these quantities. Any non-zero contribution to these matrix elements 
will signal violation of
unitarity within the closed string sector and possibly 
point to the existence of  new sectors that must be included in the theory. 
Another motivation comes from the study
of resurgence\cite{1206.6272,1511.05977,1601.03414,1802.10441} which shows
that formal power series expansions may be useful in
extracting non-perturbative information on the theory even when the perturbation series 
is not Borel summable. Understanding the systematic power series expansion of the
instanton contribution to the amplitude will be an additional input to such formal power
series expansions, besides the usual perturbation expansion.

\sectiono{Problems with ordinary cutting rules in open closed string field theory on the
D-instanton} \label{scut}

Let $S=1+iT$ be the S-matrix of a theory. Then the unitarity relation $S^\dagger S=1$ may
be expressed as
\be\label{e1}
\langle f|T|i\rangle - \langle f|T^\dagger|i\rangle   = i\, \langle f|T^\dagger \, T|i\rangle =
 i \sum_n \langle f|T^\dagger |n\rangle \langle n| T|i\rangle\, ,
\ee
where $|i\rangle$ and $\langle f|$ are the incoming and outgoing states respectively,
and the sum over $n$ on the right hand side runs over a complete set of states in the 
theory. As in \cite{1604.01783}, throughout this paper we shall refer to the matrix 
elements of
$T$ as amplitudes, without any extra normalization factor.
However, one small difference with \cite{1604.01783} is that there we defined $T$
through $S=1-iT$ whereas here we have defined it via $S=1+iT$. This is
responsible for the opposite sign on the left hand side of \refb{e1} compared to
\cite{1604.01783}. 

For perturbative amplitudes the right hand side of \refb{e1} 
is interpreted as sum over cut diagrams,
in which we draw a cut through the diagram dividing it into two parts, 
with all the incoming lines on the left side of the
cut and all the outgoing lines on the right side of the cut. The rules for evaluating the 
contribution from a cut diagram require
hermitian conjugating the part of the diagram to the right of the cut,\footnote{As in
\cite{1604.01783}, we define the hermitian conjugate 
of a diagram by reversing the sign of all
external momenta and charges and 
then taking the complex conjugate of the amplitude.}
and
replacing  the $i/(-\ell^2-m^2+i\eps)$ factor of each internal cut propagator by 
$2\pi \delta(\ell^2+m^2) \Theta(\ell^0)$ where $\ell$ is the momentum carried by 
the propagator from left to the right and $m$ is the mass of the state propagating along
the cut line. A cut on an external line has no effect and a cut does not pass through any
interaction vertex.
In that case the collection of the cut lines represent the state $|n\rangle$,
the left side of the diagram containing the original incoming states and outgoing
cut lines represent $\langle n|T|i\rangle$
and the right side of the diagram containing the original outgoing states and incoming
cut lines represent 
$\langle f|T^\dagger |n\rangle$. Therefore cutting rules, relating the difference between
an amplitude and its hermitian conjugate 
to the sum of cut diagrams, establish unitarity of the theory. 
There are some additional
subtleties involving sign factors for disconnected diagrams that can be found in 
\cite{1604.01783}.
Due to gauge invariance of string field theories,
there are additional complications due to possible contribution of unphysical states along the
cut propagator, and these have to be analyzed separately\cite{1607.08244}. 

For D-instanton induced amplitudes, the cutting rules required for proving unitarity are 
somewhat different. To see this, we need to recall a few facts about the D-instanton
induced amplitudes: 
\begin{enumerate}
\item The world
sheet description of the amplitudes in the presence of D-instantons 
involves integration over moduli spaces of punctured
Riemann surfaces as usual,
but now we also have Riemann surfaces with boundaries, with D-instanton boundary
condition at the boundary. This in particular involves imposing Dirichlet boundary condition
on the (Euclidean) time coordinate, and other non-compact
spatial coordinates. 
\item All single D-instanton induced amplitudes are accompanied
by a single 
multiplicative factor of $\NN e^{-C/g_s}$ where $\NN$ is a normalization constant
and $C/g_s$ is the instanton action, irrespective of how many boundaries or how many
disconnected components the world-sheet has.  Formally $\NN$ can be interpreted as
the exponential of the annulus diagram with ends lying on the D-instanton.
If there are $k$ D-instantons, then   
 the exponential factor is replaced by $e^{-k\, C/g_s}$ and the
normalization factor also changes to $\NN^k$.\footnote{If there are different types of 
D-instantons
with different actions, then in the exponent we shall get the sum of the actions of
different D-instantons and the normalization constant will be given by the product of
the normalization constants for different D-instantons.}
The $\NN^k$ factor accounts for exponential of sum of $k$ annulus diagrams, with the
$i$-th annulus having both its boundaries lying on the $i$-th D-instanton.
There are also annulus diagrams with the two ends lying on two different 
D-instantons, but these contributions are finite and depend on the locations of the
D-instantons, and must be included separately  in the
computation of the amplitude.
\item 
Typically these amplitudes suffer from various infrared problems, which can all be
systematically dealt with by representing them as sum of Feynman diagrams in open-closed
string field theory\cite{9705241,1907.10632}. 
In this formalism, 
the D-instanton induced amplitudes are built from Feynman diagrams whose interaction
vertices are given by 
correlation functions of open and closed string vertex operators and ghost fields on
appropriate Riemann surfaces, integrated over subspace of the moduli spaces of the
Riemann surfaces. These subspaces do not include any degenerate Riemann surface,
and therefore the interaction vertices are completely regular. 
\item In this open-closed string field theory, 
the open strings carry strictly
zero momentum due to Dirichlet boundary condition associated with D-instantons and therefore represent zero dimensional fields, but closed string states are regular fields
of the theory and the asymptotic states are built out of the closed strings.
Due to vanishing momenta, the path integral over the 
open string zero modes cannot be treated via Feynman diagrams, since the propagators
are infinite. One needs to remove their contribution from the internal propagators
of a Feynman diagram,
and carry out the path integral over these modes at the end 
of the calculation\cite{2002.04043}. The
relevant zero modes, after appropriate field redefinition, consist of a set of zero modes
in the ghost sector of string field theory,
one for each D-instanton, and the collective modes describing the location of each
D-instanton in space-time. The integration over the ghost zero modes may be interpreted
as division by the volume of the rigid $U(1)$ gauge group that exists on each D-instanton,
and may be absorbed into the overall normalization constant $\NN$
common to all amplitudes.
The integration over the collective modes need to be carried out explicitly at the end
of the calculation.

\item We shall divide the interaction vertices of this open closed string field theory
 into two types. If the 
Riemann surface associated with the interaction vertex does not have a boundary we shall
call it an ordinary vertex since it represents the usual vertex of a closed string field theory.
On the other hand if the Riemann surface has one or more boundaries on which we impose
D-instanton boundary condition, then we shall call this the D-instanton type vertex. 
\item The
ordinary interaction
vertices are accompanied by the usual momentum conserving delta
function, but the D-instanton type interaction vertices do not have such delta functions, since
Dirichlet boundary condition on the non-compact space-time directions break
translation invariance.
\item
The overall momentum conservation is restored at the end when we integrate over the 
collective modes of the open string describing the location of the D-instanton in space
and time. 
\end{enumerate}

Now we are in a position to explain why the ordinary cutting rules are not sufficient to
establish the unitarity relation given in \refb{e1}. There are several ways to see this.
We shall first consider the case of a single D-instanton.
\begin{enumerate}
\item Let $T$ and $T^\dagger$ on the left hand
side of \refb{e1} denote one instanton contributions to the amplitudes. Both are
proportional to $e^{-C/g_s}$. 
On the other hand, for a generic cut of the original diagram, the diagrams on
both sides
of the cut will have D-instanton type interaction vertices and will therefore carry
$e^{-C/g_s}$ factors. Therefore their product will be proportional to $e^{-2\, C/g_s}$.
This will be in clear disagreement with the left  hand side of \refb{e1}
\item Consistency with \refb{e1} requires that there should be
momentum conserving delta function accompanying the individual 
diagrams on either
side of the cut, since they are supposed to represent physical amplitudes
encoded in the matrix elements of $T$ and $T^\dagger$ appearing on the
right hand side of \refb{e1}. On the
other hand, 
if the cut had
D-instanton type interaction vertices on both sides, then
the original diagram had no such momentum conserving delta function for parts of the
diagrams on either side of the cut,
although there was an overall momentum 
conserving delta function for the
full diagram. 
Since the usual cutting rules do not generate an extra momentum conserving delta
function in the cut diagram, there is a clear conflict between the usual cutting rules
and \refb{e1}.
\item A generic cut of the original diagram will also cut the open string propagators.
If we try to represent such a cut diagram by a term on the right hand side of \refb{e1},
then
the sum over $|n\rangle$ in \refb{e1} will have to contain open string states as well
as closed string states. This is inconsistent with the fact that open string states on the
D-instanton are not asymptotic states. Therefore they should not appear in the
sum over intermediate
states in \refb{e1}.
\end{enumerate}
This suggests that we should somehow modify the cutting rules so that
instead of summing over all cuts of the original diagram, we sum over
only those cuts for which all the D-instanton induced interaction vertices are
on one side of the
cut. In this case the amplitude on only one side of the cut will have $e^{-C/g_s}$
factor and therefore the product of the amplitudes on the two sides of the cut
will be proportional to $e^{-C/g_s}$ as required. 
Also, the amplitude on the side of the cut 
that involves ordinary interaction vertices, will satisfy 
momentum conservation since each 
vertex
conserves momentum. Therefore the conservation of total momentum in the original
diagram will also lead to conservation of momentum on the side of the cut that
carries D-instanton type vertices. 
Finally since open strings can only begin and end on D-instanton type vertices, and
since all the D-instanton type vertices are on one side of the cut, there are no cut
open string propagators, in accordance with the expectation that open strings are
not asymptotic states.

Required cutting rules for multi-instanton contribution to the amplitude may be analyzed
in a similar way. If there are $k$ instantons then a given boundary of the world-sheet may
have  $k$ possible boundary conditions -- one for each D-instanton. We shall consider
the general case where they are different D-instantons, but the discussion also applies
to the case of identical D-instantons since they
can still be
distinguished by their locations in space-time. Let us call the boundary condition for 
the $i$-th D-instanton the $i$-type boundary. 
An interaction vertex of the open closed string field theory
will now carry a set of labels $(i_1,\cdots, i_m)$ if boundaries
of type $i_1,\cdots, i_m$ are present on the associated Riemann surface. 
Note that a given boundary may have many type of segments since the vertex 
operator of an open string
starting on the $i$-th D-instanton and ending on the $j$-th D-instanton, 
inserted on a boundary,
connects segments of type $i$ and $j$. All such segments have to be included
in the labels $(i_1,\cdots, i_m)$ of an interaction vertex.
Following
the same argument as before, one can argue that the right hand side of \refb{e1} now
should represent a sum over cut diagrams such that all boundaries of a given type are present
on one side of the cut. In other words, the left side of the cut will contain a subset
$S\subset \{1,\cdots, k\}$ of the boundary types and the right side of the cut
will involve the
complementary subset $S^c$ of the boundary types. Therefore all
the interaction vertices
on the left of the cut must be of type $(i_1,\cdots , i_m)$, with $i_1,\cdots, i_m\in S$ and
all the interaction vertices on the right of the cut must be of type $(j_1,\cdots, j_p)$ with
$j_1,\cdots, j_p\in S^c$. If 
$C_i/g_s$ denotes the action of
the $i$-th D-instanton then
the  weight factors associated with the left and
the right sides of the cut, regarded as separate amplitudes as on the right hand
side of \refb{e1},
will have the factors $e^{-\sum_{i\in S}\, C_i/g_s}$ and $e^{-\sum_{i\in S^c}\, C_i/g_s}$
respectively, and 
their product gives $e^{-\sum_{i=1}^k\, C_i/g_s}$.
On the other hand if each side had vertices of all types, then each will
have a weight factor $e^{-\sum_{i=1}^k\, C_i/g_s}$ and their product would have given 
$e^{-2\, \sum_{i=1}^k\, C_i/g_s}$, 
in disagreement with the weight factor $e^{-\sum_{i=1}^k\, C_i/g_s}$ of the
original amplitude. 

This shows that the desired cutting rules in open closed string field theory on the
D-instanton are different from those in ordinary quantum field theories.
Let us now review what goes wrong in the proof of the conventional cutting rules in 
open closed string field theory on the D-instanton.
For this we recall the main assumptions of
\cite{1604.01783} that went behind the proof of the cutting rules for perturbative 
string amplitudes:
\begin{enumerate}
\item The amplitudes are given by the usual sum of Feynman diagrams expressed 
as integrals over loop momenta.
\item The action from which the Feynman rules are derived is real.
\item The propagators have the usual pole structure, but we do not make any assumption of
how many particles the theory has or what their masses are.
\item The interaction vertices fall off
exponentially in the direction of imaginary energy and real spatial momenta. This
assumption is not essential for the proof of the cutting rules, but is a property of string
field theory that ensures ultra-violet finiteness of the amplitudes.
\item The interaction
vertices have the usual momentum conservation laws. 
\end{enumerate}

For D-instanton induced amplitudes,
one of the assumptions that fails in the proof of cutting rules 
is momentum conservation
at the interaction vertices of open-closed string field theory relevant for 
computing the amplitudes. As already discussed, this
failure is due to the breaking of translational invariance by the
D-instanton boundary condition. The overall momentum conservation is restored
at the end after integration over a particular collective mode of the
open string field,  describing the average location
of all the D-instantons in space-time. 
However the individual interaction vertices of open closed
string field theory do not satisfy the momentum conservation rule. This in turn 
prevents us from proving the cutting rules in the open closed string field theory in the 
usual form. 
For example, in the analysis of \cite{1604.01783}, the constraints on
internal momenta due to momentum conservation at the interaction vertices were
used to determine what kind of contour deformations are allowed in the complex
plane of the internal energies. This in turn was used to determine the regions of
integration that contribute to the anti-hermitian part of an amplitude, leading to
the cutting rules in the
form described earlier.

The first assumption, that the  amplitudes are given by usual sum over Feynman
diagrams,  also fails for D-instanton amplitudes. The overall factor of $\NN\, e^{-C/g_s}$
accompanying a D-instanton amplitude cannot be assigned to either an interaction vertex
or a propagator of the Feynman diagram, since this factor does not depend on the number
of propagators and interaction vertices in the diagram. 
For this reason any non-linear relation between
the amplitudes derived from the analysis of Feynman diagrams, like the cutting rules, will
fail to hold in its usual form.

There is another issue in open-closed string field theory on D-instantons
that was not present in the
analysis of the cutting rules in closed string field theory. The reality of the action needed
for the proof of the cutting rules is formulated naturally in the Lorentzian theory. However
since D-instantons are solutions in the Euclidean theory, the open string fields are defined
naturally in the Euclidean space. 

We shall see in \S\ref{seffective} and \S\ref{smulti} that
these problems can be addressed by integrating out the open string fields to define an
effective action of the closed string fields in the Euclidean space and then analytically
continuing this action to the Lorentzian space. The interaction vertices of this 
effective field theory will satisfy the usual
momentum conservation laws. Therefore we can now try to apply the cutting rules directly
in this closed string effective field theory. The reality of the effective action is not
guaranteed however, and we shall see that this can fail when the D-instanton system
has open string tachyons. This in turn leads to failure of unitarity of the closed string
sector by itself. However when the effective action is real, the usual cutting rules hold
in the closed string effective field theory, leading to unitary amplitudes. Furthermore,
we can also translate these cutting rules to the original open closed string field
theory, and show that these cutting rules precisely correspond to the ones that are
needed for the unitarity of the theory.

\sectiono{Reality properties of
Euclidean and Lorentzian actions}  \label{sreal}

Since D-instantons are defined as solutions in Euclidean string theory, they are used
to first compute corrections to the Euclidean action of string field theory. This then has
to be Wick rotated to Lorentzian signature, to check if it satisfies the properties 
required for the proof of the cutting rules. The main properties that will be relevant 
are momentum conservation at the interaction vertices and the reality of the action.
For this reason we shall begin in 
\S\ref{seuclid} by reviewing some standard relations between Euclidean and 
Lorentzian action. 
\S\ref{sfield} contains a discussion of string field theory actions and their
reality properties.

\subsection{Relation between the Euclidean and the Lorentzian actions} \label{seuclid}

In this section we shall set up some notations on Wick rotation. Even though this is
standard textbook material, we have included this discussion, since various factors of
$i$ and their signs will play a crucial role in our analysis.

Let us denote by $x=(x^0, \cdots x^{D-1})$ the space-time coordinates in the Lorentzian
space and by $x_E=(x_E^0,\cdots, x_E^{D-1})$ the space-time coordinates in the 
Euclidean space. 
The conjugate momentum variables will be denoted by 
$p=(p_0,\cdots, p_{D-1})$ and
$(p^E_0,\cdots, p^E_{D-1})$ respectively.\footnote{We shall use both subscript 
$E$ and
superscript $E$ to denote Euclidean variables.}
Then the usual analytic continuation rule sets the spatial coordinates / momenta
in the two formalisms to be equal and the time coordinates are related as,
\be\label{eELreln}
x^0 = - i x_E^0, \qquad p_0=i\, p^E_0\, ,
\ee
so that the time evolution operator $e^{- i H x^0}$ in the Lorentzian theory is mapped
to $e^{- H x_E^0}$ in the Euclidean theory. 
Note that in the Lorentzian theory we have used $p_0$ as the independent variable
since this is conjugate to $x^0$.
When we refer to $p^0$ it should be regarded as $-p_0$. 

It is instructive to recall how the analytic continuation is done. For this we introduce the
interpolating variables:
\be \label{einter}
x^0_\theta = e^{-i\theta} u, \qquad p_0^\theta = e^{i\theta} v\, ,
\ee
with real variables $u$ and $v$. For $\theta=0$ we interpret $u$ as $x^0$ and
$v$ as $p_0$, while for $\theta=\pi/2$, we interpret $u$ as $x_E^0$ and
$v$ as $p_E^0$. 
This can be summarized by saying that in position space we rotate clockwise
as we go from the Lorentzian to the Euclidean theory and rotate anti-clockwise
as we go from the Euclidean to the Lorentzian theory. On the other hand,
in momentum space we rotate anti-clockwise
as we go from the Lorentzian to the Euclidean theory and rotate clockwise
as we go from the Euclidean to the Lorentzian theory.

If we consider an integral over $x^0_\theta$ or $p_0^\theta$ 
that appear {\it e.g.} in a
Fourier transformation, then
if the integration 
contour is rotated by changing $\theta$, 
the combination $x^0_\theta
p_0^\theta$ that appears in the exponent remains real.  
Since we have:
\be
dx^0_\theta = e^{-i\theta} du, \qquad dp_0^\theta = e^{i\theta} dv\, ,
\ee
and since under contour rotation the integration measure remains $dx^0_\theta$ or
$dp_0^\theta$, 
it follows that under the Wick rotation we shall have,
\be \label{emesrot}
d x^0=-i d x_E^0, \qquad dp_0 = i dp^E_0\, .
\ee
On the other hand the relation $p^0=-p_0$ is simply a change of variable and not a
contour rotation. Therefore the integrals do not pick up any sign when we  
replace $p_0$ by $p^0$
since the change in measure is compensated for by a change in the integration limits.
Since $p^0=-p_0$ corresponds to energy $\omega$ in the Lorentzian theory, 
we shall refer to 
$-p^E_0$ as  the Euclidean energy $\omega_E$.

We shall denote by
$S_E$ and $S$ the Euclidean and the Lorentzian actions, expressed as momentum
space integrals with sum of products of fields and momentum factors in the integrand.
We shall assume that $S_E$ and $S$ have been
normalized such that the
Euclidean path integral is weighted by $e^{S_E}$ and the Lorentzian path integral is
weighted by $e^{i S}$. We shall show that with this definition $S$ is obtained from 
$S_E$, with both represented as momentum space integrals, by
making the following replacements:
\begin{enumerate}
\item
We replace in the integrand
the Euclidean momentum conserving delta function by the Lorentzian 
momentum conserving delta function, and integral over the Euclidean momenta
$p_E$ by
integral over Lorentzian momenta $p$.
\item We replace, in the argument of the scalar fields, the Euclidean momenta by 
Lorentzian
momenta i.e.\ we have 
\be
\phi_E(p_E) \to \phi(p)\, .
\ee
Note that this is a replacement rule and not an equality. The latter is discussed in
\refb{eequality}.
\item Inside the argument of any 
explicit function $f$ of Euclidean momenta multiplying the integrand, we should
express the Euclidean momenta in terms of Lorentzian momenta using
\refb{eELreln}. For example a factor of $(p^E_0)^2$ in  $f$ will be
replaced by $-(p_0)^2$ in the Lorentzian theory.
\item For tensor fields, the replacement rule will have extra factors of $i$ for every 
0 component of the fields, {\it e.g.}
\be \label{e36flo}
A_E^0(p_E) \to i A^0(p) = -iA_0(p)\, .
\ee
\item For fermion fields the relationship   between the Euclidean and the Lorentzian fields
is more complicated. In general this will
require us to work with complex fields in both, the Euclidean and the Lorentzian theory,
since the reality properties of fermions differ in the two cases. Later we shall describe a way
to avoid this problem.
\end{enumerate}
We shall illustrate this by taking the example of a scalar field $\phi$ in the momentum
representation.
Let us consider
a term in $S_E$, expressed as a momentum space integral of the form
\ben \label{eec1}
&& \int d^Dp^E_{(1)}\cdots d^D p^E_{(n)} \, \delta^{(D)}(p^E_{(1)}+\cdots +p^E_{(n)}) \, 
\prod_{i=1}^n \phi_E(p^E_{(i)}) \nonumber \\
&=& \int d^Dp^E_{(1)}\cdots d^D p^E_{(n-1)} \, \left\{\prod_{i=1}^{n-1} \phi_E(p^E_{(i)}) \right\}
\phi_E(-p^E_{(1)}-\cdots-p^E_{(n-1)})\, .
\een
Let $\phi$ be the field in the momentum representation in the Lorentzian space-time.
The relation between $\phi(p)$ and $\phi_E(p_E)$ can be found 
from the corresponding relations between the coordinate space
fields $\wt\phi(x)$ and $\wt\phi_E(x_E)$,
\be\label{e3939}
\wt\phi(x)=\wt\phi_E(x_E), \qquad \phi(p) = \int d^D x\, e^{-ip.x}\, \wt\phi(x), 
\qquad \phi_E(p_E) = \int d^D x_E\, e^{-ip_E.x_E}\, \wt\phi_E(x_E)\, .
\ee
\refb{emesrot} and \refb{e3939} now give,
\be \label{eequality}
\phi(p) = -i\, \phi_E(p_E)\, .
\ee
Using \refb{eequality} and \refb{emesrot} we can express 
the right hand side of 
\refb{eec1} as:
\ben \label{eec2pre}
&& i^n \, (-i)^{n-1}\, \int d^Dp_{(1)}\cdots d^D p_{(n-1)} \, 
\left\{\prod_{i=1}^{n-1} \phi(p_{(i)}) \right\} \, \phi(-p_{(1)}-\cdots - p_{(n-1)})
\nonumber \\
&=&
i \int d^Dp_{(1)}\cdots d^D p_{(n)} \, \delta^{(D)}(p_{(1)}+\cdots +p_{(n)}) \, 
\prod_{i=1}^n \phi(p_{(i)}) \, .
\een
Identifying this as the contribution to $i\, S$, we see that the contribution to $S$
is given by:
\be\label{eec2}
\int d^Dp_{(1)}\cdots d^D p_{(n)} \, \delta^{(D)}(p_{(1)}+\cdots +p_{(n)}) \, 
\prod_{i=1}^n \phi(p_{(i)}) \, .
\ee
As claimed, this is obtained from the Euclidean action
by replacing the Euclidean momentum conserving delta function by the Lorentzian 
momentum conserving delta function, the euclidean momenta by
Lorentzian momenta and the fields defined in the euclidean space by fields
defined in the Lorentzian space. Generalization of these rules to tensor fields, and
to cases where we have explicit momentum dependent functions multiplying the
integrand, is straightforward. 

With these rules, a real scalar function $f(p_{(i)}^E)$
of Euclidean momenta, multiplying the integrand in \refb{eec1},
will transform to a real scalar function of the Lorentzian
momenta multiplying the integrand in \refb{eec2},
since the zeroth components of momenta occur in pairs in a scalar function. Note that
in momentum space a real function means that the function is invariant under 
the simultaneous operation of complex conjugation and change of sign of all the
momenta in its argument.
On the other
hand if the momenta are contracted to the index of a tensor field, then the extra
$i$ from \refb{eELreln} will be compensated by the $i$ coming from the relation between
the Lorentzian and the Euclidean version of the tensor field. Therefore these
replacement rules do not generate any extra factor of $i$, and a real 
Euclidean action gets transformed to a real Lorentzian action. 
Conversely an imaginary term in the Euclidean action will generate an imaginary term
in the Lorentzian action.
The exception to this 
rule comes from terms involving the $\ve$ tensor. We shall not discuss these terms in
detail, but in physical examples such terms are accompanied by a factor of $i$
in the Euclidean action which goes away in the Lorentzian version. 

Note that for this argument to work, it is essential that the function $f(p_{(i)}^E)$ is
analytic in the complex $p_{(i)}^E$ plane. Otherwise one can find counterexamples.
Let us for example suppose that the Euclidean theory has a term like $e^{-|p_0^E|}$
where $p^E$ denotes some combination of the momenta $p_{(i)}^E$. 
Since this remains invariant under simultaneous operation of complex conjugation and
change of sign of the momenta, it is a real function. Now it follows from
the discussion below \refb{einter} that positive $p_0^E$ is analytically continued to positive 
$p_0$ and negative $p_0^E$ is analytically continued to negative $p_0$.
Therefore we see using \refb{eELreln} that $e^{-|p_0^E|}$
will analytically continue to the function $e^{ip_0}$ for positive $p_0$ and $e^{-ip_0}$
for negative $p_0$. This can be summarized by saying that the analytically
continued function is $e^{-i|p_0|}$. This is not a real function, indicating that a real term in the
Euclidean action  leads to a complex term in the Lorentzian action. This can be traced 
to the non-analyticity of the function $f$ at $p^E_0=0$ that allows us to have terms
of the form $e^{-|p_0^E|}$.
In string field theory action such terms are absent since the possible
non-analyticities of this kind can come from integration near the boundaries of the moduli
space representing degenerate Riemann surfaces and these regions are excluded
from the definition of the interaction terms of string field theory. 
Of course in the full amplitude we must also
include contributions from the boundaries of the moduli space, and these contributions
do lead to non-analyticity of the amplitude in the complex momentum plane. As a
result, they produce imaginary parts of the amplitude for Lorentzian signature, consistent
with cutting rules\cite{1604.01783}. We shall illustrate this in the context of matrix model
in \S\ref{sexamplemulti}. 
There, \refb{e6.7}, representing a contribution to the two point closed
string amplitude, is a non-analytic function of the external energy
$\omega_2$ (after we strip off the momentum conserving delta function). 
In contrast, \refb{e6.14}, representing a contribution to the
action, is an analytic function of the energy.

In the presence of fermions, the relation between the reality properties of Euclidean and
Lorentzian action is more complicated since the reality properties of the fermions themselves
are different in the Euclidean and the Lorentzian theory. A given term 
in the Euclidean action involving
fermions will be called real if after analytic continuation to Lorentzian 
signature\refb{e36flo}, and 
imposing reality conditions on the fermions as given {\it e.g.} in \cite{1606.03455},  
it gives a real term in the Lorentzian action.

\begin{figure}
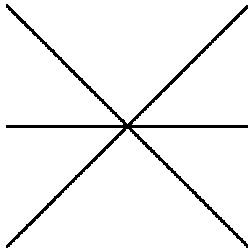

\begin{center}
\figstar
\end{center}

\vskip -.8in

\caption{A Feynman diagram without any internal
propagator, contributing to $n$-point amplitude.
\label{figstar}
}
\end{figure}

Since the weight factor in the euclidean path integral measure is $e^{S_E}$, 
the contribution to an amputated $n$-point Green's function from a Feynman
diagram with an $n$-point vertex and no internal propagator, as shown in Fig.~\ref{figstar},
 will be given 
directly by the integrand of the 
term in the action responsible for the interaction vertex, without
any extra normalization factor (except for the symmetry factors
$k!$ for $k$ identical external particles). 
Similarly in the Lorentzian theory, where the
weight factor is $e^{iS}$, the amputated $n$-point Green's function from a
similar Feynman diagram will be
given by $i$ times the integrand of the term in the action that produces the $n$-point vertex.
This also gives the contribution to the S-matrix for on-shell external
states. Therefore the contribution
to the T-matrix from such a Feynman diagram will be given by the integrand of the
term in
the action without any extra normalization. Of course in all these cases there
are also more complicated Feynman diagrams with internal propagators that also
contribute to the amplitudes. In the Euclidean theory we shall call the Green's
functions as amplitudes whereas in the Lorentzian theory we shall call the 
T-matrix elements as amplitudes. Therefore for the contribution from diagrams
without internal propagators, the amplitudes are related to the integrand of the
terms in the action
without extra normalization
in both the Euclidean and the Lorentzian theories. From this it follows that the
T-matrix elements in the Lorentzian theory are given by direct analytic continuation
of the amplitudes in the Euclidean theory without any extra factor of $i$ and/or minus
sign.

These observations will be useful while constructing the action of the closed string
effective field theory from the amplitudes in the open closed string field theory.

\subsection{Reality properties of string field theory action} \label{sfield}

In this section we shall discuss some results on the reality properties of string
field theory actions.

We shall begin by reviewing the reality property of the ordinary interaction vertices
of open closed string field theory, as defined in \S\ref{scut}.
Since these interaction vertices are identical to the 
interaction vertices of the closed string field theory, they can be expressed
as correlation functions of closed string vertex operators and other ghost
insertions on Riemann surfaces without 
boundaries, integrated over a subspace of the moduli
space of Riemann surfaces that avoids all degeneration points. Such integrals are 
manifestly finite and manifestly real with appropriate choice of reality condition
on the string fields\cite{sonoda,9206084,9705038,1606.03455}. 
Indeed a general feature of (super-)string
perturbation theory is that all amplitudes are formally 
real, 
in the sense that they can be 
represented as integrals  over the moduli space of Riemann surfaces 
with real integrands,
and the imaginary parts, required for unitarity, come from having to 
regulate the divergences
from the boundaries of the moduli spaces\cite{sundborg,amano,marcus,sundborg1,
9302003,9404128,9410152,berera2,1307.5124}. 
Since the region of integration describing 
the string field theory interaction vertices
do not include the boundaries of the moduli spaces, they correspond to real terms in the
action.

Some qualifications are necessary, however, on what one means by real integrand.
As discussed before, in momentum space, reality of the integrand will mean that 
the integrand is invariant under simultaneous operation of complex conjugation and
change of sign of all the external momenta. However, as discussed in 
\cite{9206084,1606.03455}, even in the
perturbative sector of closed string field theory, this does not hold -- 
together with complex conjugation and
change of sign of the external momenta, one also needs to include an 
involution operation on the
moduli space of Riemann surfaces that changes the sign of the complex structure
on the Riemann surface. After integration over part of the moduli space
that is relevant for defining the interaction vertex, this gives us back the usual notion
of reality of the interaction terms in the action.

The relevant string field theory in the presence of D-instantons is
open closed string field theory which has both open and closed string fields. 
The interaction vertices  of this theory are
given by integrals of correlation functions of open and closed string
vertex operators and world-sheet ghost fields on moduli spaces of Riemann surfaces
with boundaries, and the integration runs over parts of the moduli space that do
not include any degenerate Riemann surfaces. 
However,
as alluded to at the end of 
\S\ref{scut}, and will be discussed in detail in \S\ref{seffective} and \S\ref{smulti}, 
we shall be working with the closed string effective action obtained by integrating out the
open strings living on the D-instanton. The 
interaction vertices of this effective field theory will also be given by similar 
integrals, but now the external states will only have closed strings and the
integration region will include degenerate Riemann
surfaces associated with open string degeneration. Furthermore, the interaction vertices
will carry overall multiplicative 
factors given by products of $\NN\, e^{-C/g_s}$ factors described in 
\S\ref{scut}. This leads to the following possible sources of complex contribution to the
closed string effective action:
\begin{enumerate}
\item The integrands that need to be integrated over the moduli spaces of Riemann 
surfaces may be complex.
\item $\NN$ could be complex.
\item The divergences in the integrand from near the boundaries of the moduli spaces
associated with open string degeneration may
force us to define the integral via analytic continuation, and this
may lead to imaginary terms in the effective action. 
\end{enumerate}
We can eliminate the first possibility by the following argument.
The main ingredient in the proof of reality of the
closed string field theory action is to establish a choice of basis states for the expansion
of the off-shell closed
string field in which all three point functions are real. Since the presence of
boundaries on the Riemann surface can be represented by insertion of an off-shell
closed string given by the boundary state associated with the D-brane, one would
expect that as long as the boundary state is real in the same basis, the 
integrand that appears in the definition of the interaction vertex of closed string effective
field theory is also  real.\footnote{This argument shows the reality of the integrand in a
region of the moduli space in which the effect of the boundaries can be regarded as 
disk one point functions connected to the rest of the diagram via  closed string 
propagators. Since the integrand is non-singular in the interior of the moduli
space, we would expect that this property can be made to hold in the entire moduli space
by choosing the data on local coordinates at the punctures and the locations of the
picture changing operators appropriately.}
Even if the boundary state is complex, as long as there is
another D-instanton with complex conjugate boundary state, which has the same action
and normalization constant $\NN^*$, the net contribution to the closed string effective action
will be real, since for every Riemann surface with boundaries ending on some combinations
of D-instantons, 
there is a another Riemann surface obtained by replacing each boundary state
by its complex conjugate boundary state, and the sum of these two contributions to the
integrand is still real. An example of this is provided by type IIB string theory, where
D-instanton and anti-D-instanton produce complex conjugate contribution to the closed
string effective action\cite{9701093}.  

We shall
formally summarize our results on reality, described above, as:\\
\noindent{\bf The reality postulate}: {\it The closed string effective action, 
obtained by integrating out all the open string modes on the
D-instanton, can be formally expressed as the product of normalization constants
$\NN\, e^{-C/g_s}$ 
and integrals over the moduli spaces of
Riemann surfaces with boundaries with real integrand. Therefore the
imaginary contribution to the effective action comes either from the imaginary
part of $\NN$ or from boundaries of the moduli space
associated with open strings degeneration, where the integrand may diverge, forcing us
to define the integral in a different way.
}

As in the case of perturbative closed string field theory, 
some qualifications are necessary on what one means by real integrand. 
Since the world-sheet theory refers to some fixed locations of 
D-instantons in Euclidean space-time, the notion of reality, as stated, needs to be
defined for fixed locations of the D-instantons,
even though we'll eventually have to integrate over these
locations.
If we are working in Euclidean signature
space-time, then 
the notion of reality is the same as that in the case of perturbative closed string
field theory, i.e. 
the integrand is invariant under simultaneous operation of complex conjugation,
change of sign of all the external momenta and an involution on the
moduli space of Riemann surfaces that changes the sign of the complex structure
on the Riemann surface. 
However, when external fermions are present, there is additional
complication since the reality properties of fermions are different for Euclidean and
Lorentzian signature. For this reason it is useful to define
the notion of reality directly in the Lorentzian space by analytically continuing 
the integrand to Lorentzian signature, but still keeping the D-instanton positions at
Euclidean time. This will ensure 
that eventually when we need to integrate over these locations,
no extra factor of $i$ arises from having to rotate the integration
contour of the D-instanton locations, and, more importantly, we do not need to worry 
about whether we might encounter poles while rotating the integration contour. 
Now, since from the Lorentzian perspective
the D-instantons are located
at imaginary time, the complex conjugation must also be accompanied by the change in
sign of the instanton locations. Therefore after analytic continuation to Lorentzian
signature, the reality of the integrand means that the integrand is invariant under
the simultaneous operation of complex conjugation, reversal
of sign of the external momenta, the involution on the moduli space involving change
of sign of the complex structure on the Riemann surface 
and the reversal of sign of the D-instanton
locations along the Euclidean time.\footnote{The coordinate describing the
average time coordinate of all the D-instantons can be left out of this, since integration over
this mode produces the overall Euclidean energy conserving delta function, which,
following the algorithm described in \S\ref{seuclid}, gets mapped to Lorentzian energy
conserving delta function. However the differences between the Euclidean time 
coordinates of different D-instantons will have to change sign.}

By working with the closed string effective action we avoid having to determine the reality 
condition on the open string fields on the D-instanton, since these fields are
integrated out. The only exceptions are the
zero modes which have to be `integrated in' for dealing with the divergences coming
from the boundaries of the moduli spaces, and then
integrated separately at the end. The effect of
integration over these modes will be discussed
in subsequent sections.

\sectiono{One instanton effective action and its reality properties} \label{seffective}

In this section we shall discuss the construction of the one instanton effective action
of closed strings and analyze its reality properties and the cutting rules.

\subsection{Effective action} \label{ssingleeff}

\begin{figure}
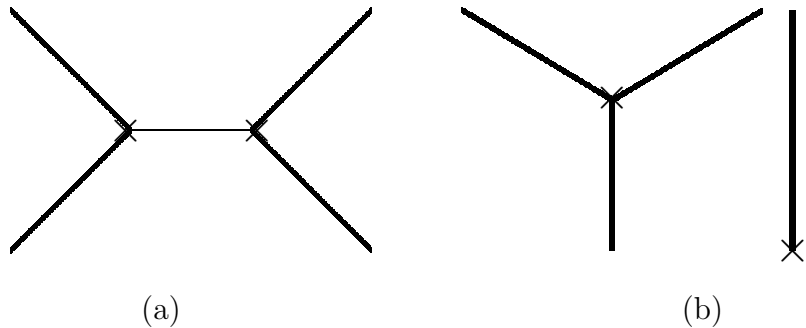

\begin{center}
\hbox{\hskip .5in \figfive}
\end{center}

\vskip -.8in

\caption{Examples of amplitudes in the original open closed string field theory on a
single D-instanton that become part of the interaction vertex of the closed
string effective field theory. The thick lines represent closed string propagators, 
the thin lines represent
open string propagators and the $\times$'s denote D-instanton type interaction vertex.
\label{fig5}
}
\end{figure}

In an ordinary quantum field theory, if we want to integrate out a set of fields $U^c$ and
construct the effective action of the complementary set of fields $U$, we define 
the $n$-point interaction vertex of the effective field theory to be the sum of all connected
diagrams for $n$-point amputated Green's function that do not have any propagator
of a field in the set $U$. In the current problem $U$ stands for the set of closed
string fields and $U^c$ stands for the set of open string fields on the D-instanton.
Formally, 
the procedure for constructing the interaction vertex of the closed string effective field
theory is similar, except that we also sum over disconnected contribution to the Green's 
function as long as each connected component is made of D-instanton type interaction
vertices of the original open closed string field theory. 
The inclusion of disconnected diagrams in the definition of the interaction vertex
of the effective field theory may be unfamiliar, but 
is necessary due to the fact that 
(1) the individual connected components do not satisfy momentum conservation and
(2) irrespective of how many connected components a diagram has, we have a single
factor of $\NN\, e^{-C/g_s}$ multiplying the amplitude. This is also reflected in the fact that
the amplitude  is
to be evaluated in the open string zero mode background which needs to be
integrated at the end after summing over disconnected diagrams. In fact we
shall also include 0-point D-instanton type vertices with no external open and closed
strings but just Riemann surfaces with boundaries. As usual, these zero point
interaction vertices will involve integration over subspaces of the moduli spaces of the
corresponding 
Riemann surfaces  that are not covered by Feynman
diagrams with internal open and  / or closed string propagators. The only exception
will be the annulus diagram that is included in the overall normalization constant 
$\NN$.
Fig.~\ref{fig5} 
shows
examples of some closed string four point 
amplitudes in the original open closed string field
theory that become part of the four point interaction vertex in the closed string
effective field theory. We shall call the ordinary $n$-point interaction vertex of
the original open closed string field theory the ordinary $n$-point 
interaction vertex of the effective field
theory, and the sum of all other contributions to the $n$-point 
interaction vertex of the effective field
theory the D-instanton type interaction vertex. It then follows from the Feynman rules
of open closed string field theory that the D-instanton
type interaction vertex of the effective field theory will be accompanied by a
factor of $\NN\, e^{-1/g_s}$ irrespective of how many D-instanton type interaction
vertices of the original theory it has. We shall show in \S\ref{sgauge} that 
the action of the closed string effective 
field theory constructed this way 
satisfies the quantum BV master equation\cite{1609.00459} 
and therefore has the gauge
invariance required for the consistency of the theory.

\begin{figure}
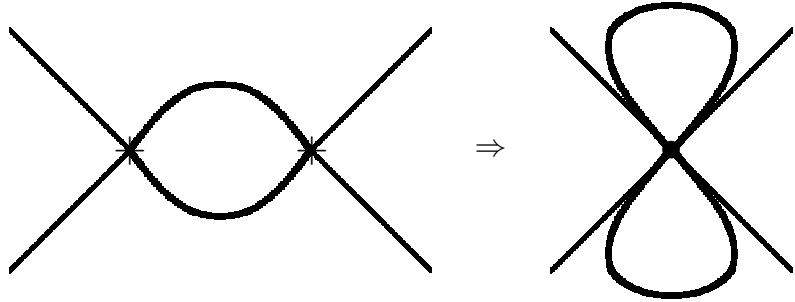

\begin{center}
\hbox{\hskip .5in \figfour}
\end{center}

\vskip -.8in

\caption{Translating a Feynman diagram of the original open closed string
field theory (left) into a Feynman 
diagram of the closed string effective field theory (right). 
The interaction vertices in the left diagram 
are D-instanton type interaction vertices of the open closed
string field theory and the vertex labelled by $\bullet$ on the right diagram 
is a D-instanton type interaction vertex of the closed string effective
field theory. The propagators displayed in this diagram are closed string
propagators. If the original diagram had any open string propagator, it would have
been fused into the interaction vertex of the effective field theory.
\label{fig4}
}
\end{figure}

We can also give the 
relation between the Feynman diagrams built from this effective field theory
and those built from the original open closed string field theory. Given any Feynman
diagram of the
original open closed string field theory, 
we draw an auxiliary Feynman diagram in which we fuse all the
D-instanton type interaction vertices into a single composite 
vertex, and also collapse all the open string
propagators (which must begin and end at the 
D-instanton type vertices) into this composite vertex.  
We keep all the closed string propagators displayed as usual. 
The composite vertex constructed this way is part of the D-instanton type
interaction vertex of the
closed string effective field theory. This has been illustrated in Fig.~\ref{fig4}. Note that 
there is only one composite vertex, i.e.\ only one D-instanton type vertex of the
closed string effective field theory in the entire diagram. This reflects the fact that
since we are considering single D-instanton amplitude, we allow only one power of
$e^{-C/g_s}$. Therefore the D-instanton type interaction vertex of the
closed string effective field theory, 
that we get by integrating out the open string fields, is used only once.
As will be discussed in \S\ref{smulti}, 
diagrams where the vertex is used multiple times are part of
the multi-instanton contribution to the amplitude.
It is also useful to note that if in the original Feynman diagram of the open closed
string field theory none of the D-instanton type vertices  were connected to an external
closed string, either directly or via internal closed string propagators, 
then after collapsing them into the composite vertex, we would get a
disconnected bubble diagram of the closed string effective field theory. However, such
diagrams are excluded from the beginning in the open closed string field theory 
Feynman rules due to
cancellation between the numerator and the denominator factors in the 
path integral\cite{2002.04043}.

\subsection{Divergences}

This construction of the closed string effective field theory described above
suffers from various divergences that need to be
treated carefully\cite{2002.04043}. 
The ordinary interaction vertices of the effective field theory, being identical to the 
ordinary interaction vertices of the original open closed string field theory, have the
interpretation of correlation functions of closed string vertex operators and other ghost
insertions on Riemann surfaces with boundaries, integrated over a subspace of the moduli
space of Riemann surfaces that avoids all degeneration points. Such integrals are 
manifestly finite. 
Let us now consider the D-instanton type interaction
vertex in the closed string effective field theory.
This is also given by integrals of correlation functions of closed string
vertex operators and world-sheet ghost fields on moduli spaces of Riemann surfaces,
but the Riemann surfaces have one or more boundaries with D-instanton boundary
condition. The part of the moduli space of these Riemann surfaces over which we need
to integrate for computing the effective action now includes one or more boundaries
associated with open string degeneration. 
Therefore the integrals could diverge from these regions, requiring special treatment.
This has been discussed in detail in \cite{2002.04043}, but we shall 
summarize the main points below.
\begin{enumerate}
\item We may get divergences of the form $\int_0^1 dq\, q^{-\alpha-1}$ where $q$ is
related to the Schwinger parametrization of the open string propagators, and is usually
referred to as the sewing parameter due to its role in the world-sheet description
of the amplitude. The divergences
arising from the terms with $\alpha>0$ 
can be traced to the propagation of open string tachyons. 
Open closed string field theory instructs us to use the replacement rule,
\be \label{erp1}
\int_0^1 dq\, q^{-\alpha-1} \quad \to \quad -\alpha^{-1}\, ,
\ee
for dealing with these divergences.
\item We may also get logarithmic divergences of the form $\int_0^1 dq \, q^{-1}$. These can
be traced to the result of integration over the open string
zero modes. The prescription described
in \cite{2002.04043} was to 
remove these contributions and then integrate over the zero modes
separately. The procedure for removing these contributions may be formally stated as:
\be \label{erp2}
\int_0^1 dq\, q^{-1} \quad \to \quad 0\, .
\ee
Since we leave the zero modes unintegrated, we need to treat them as external states.
Following this algorithm we arrive at the Wilsonian effective action, expressed as
function of closed string fields and open string zero mode fields.
\end{enumerate}

We shall now list the zero modes over which we have to carry out the 
integrals, and the strategy we follow to compute these integrals:
\begin{enumerate}
\item There are open string zero modes, which, after field redefinition, can be
related to the collective coordinates of the
D-instanton describing the location of the instanton in Euclidean space-time. 
The effect of integration over the collective coordinates will be described shortly.
The field redefinition that relates the
open string modes to the collective coordinates gives rise to Jacobian in the integration
measure, which can be regarded as additional  multiplicative
contribution to the closed string effective
action. This could include, in particular,
the closed string moduli fields including the dilaton, leading to
a moduli dependent multiplicative factor in the effective action\cite{private}.
\item There is an open string zero mode from the ghost sector associated with the rigid
U(1) gauge transformation on the D-instanton. After an appropriate field redefinition,
the path integral over this zero mode may be regarded as division by the volume of the
U(1) gauge group. Since this is a finite constant that can be absorbed into the definition
of the normalization constant $\NN$, we shall drop this integral from our analysis.
However the Jacobian that arises from the required
change of variable gives rise to additional multiplicative
contribution to the closed string effective action. 
\item Finally, due to the breakdown of the Siegel gauge, there is an extra auxiliary open string
field from the ghost sector and we have to explicitly evaluate its contribution to the Feynman
diagrams. This gives extra terms that are finite and computable. 
\item
The
integration over the collective coordinates is done at the end, and 
produces the momentum 
conserving delta function that involves total momentum of the external closed strings
in all the D-instanton type vertices. Therefore this produces the momentum conserving delta
function at the D-instanton type vertex of the closed string effective field theory.
The original integration is done in Euclidean space, producing the delta function with 
argument given by total Euclidean momentum.  According to \refb{eec1}, 
\refb{eec2} this translates to a 
momentum conserving delta function in the Lorentzian space.
\item In superstring theories there will also be fermionic collective 
coordinates that need to
be integrated over. 
This means that when we construct the D-instanton type 
interaction vertex of the effective field theory, we need to sum over Feynman diagrams
of the original open closed string field theory that have not only external closed string
states, but also certain number of open string states representing the zero mode
fields. There can be additional contribution to the effective action from the Jacobian
associated with the change of variables from the open string fields to the fermionic
collective coordinates.
\end{enumerate}

From the analysis above, we see that the D-instanton type interaction
vertex of the 
closed string effective 
field theory has the usual momentum conserving delta function. Furthermore
a single vertex of this type is accompanied by a single factor of $\NN\, e^{-C/g_s}$. 
This removes the first hurdle in the proof of cutting rules in this effective
field theory. 
We shall analyze the reality of the
effective action in \S\ref{ssinglereal}.

\subsection{Gauge invariance} \label{sgauge}

In string field theory, the kinetic term in the action is given by the matrix
element of the world-sheet BRST
operator between a pair of string fields. If $P$ is any projection operator that
commutes with the BRST operator, then there is a way to integrate out all $P$
non-invariant fields and write down an effective action for the $P$ invariant 
fields, both in the classical theory\cite{0112228,0306332,2006.16270} and 
in the quantum theory\cite{1609.00459}.
The resulting string field theory preserves all the algebraic structures
of the original theory and satisfies the quantum Batalin-Vilkovisky (BV)
master equation.
Since the BRST operator in the combined Hilbert space of open snd closed strings is
block diagonal, we can use this procedure to integrate out all the open string modes on
the D-instanton
and write down an effective action for closed string fields. 

There are however a few differences between the process of integrating out a subset of
fields described in \cite{1609.00459} and the one 
described here, due to the inclusion of disconnected diagrams, and also due to the
presence of the normalization factor $\NN\, e^{-C/g_s}$. 
To understand this difference, let us denote by $S_0$ the action of
purely closed string field theory, and by $S_1$ the additional term in the action that
we get by integrating out the open string modes in the sense 
of  \cite{1609.00459}. $S_1$ will be given by
appropriate correlation functions of closed string vertex
operators on the {\it connected} 
world-sheet with boundaries,  
integrated over appropriate regions of the moduli spaces of Riemann
surfaces that include open string degeneration but no closed string degeneration. 
We shall proceed by ignoring possible divergences in $S_1$ from the boundaries associated
with open string degeneration, but will comment on this at the end of this subsection.
Then
the effective action that would appear in the analysis of \cite{1609.00459} would
be $S_0+S_1$, and it would satisfy the BV master equation\cite{bv1,bv}
\be \label{emas0}
{1\over 2} \{ S_0+S_1, S_0+S_1\} + \Delta (S_0+S_1) =0\, ,
\ee
where $\{A,B\}$ denotes the anti-bracket of $A$ and $B$ and $\Delta$ is the BV Laplacian:
\be \label{edefanti}
\{A,B\} = {\p_r A\over \p \psi^s} \, {\p_l B\over \p \psi_s^*} - {\p_r A\over \p \psi_s^*}
\, {\p_l B\over \p \psi^s}, \qquad \Delta A = {\p_r\over \p\psi^s} {\p_l\over \p\psi_s^*} A\, .
\ee
Here $\psi^s$ are the field variables, $\psi_s^*$ are the conjugate anti-fields carrying
opposite grassmann parity, and $\p_l$ and $\p_r$ denote left and right derivatives
respectively.
Since $S_0$ itself satisfies the BV master equation, we actually have
two equations:
\be \label{emast1}
{1\over 2} \{ S_0, S_0\} + \Delta S_0 =0\, ,
\ee
and
\be \label{emast2}
\{ S_0, S_1\} +{1\over 2} \{S_1, S_1\} + \Delta S_1 =0\, .
\ee
We have used the fact that $\{A,B\}$ is invariant under the exchange of $A$ and $B$
if both $A$ and $B$ are grassmann even.

In contrast, the effective action that we have here includes contribution from
disconnected world-sheets, 
as long as
every disconnected component has a boundary ending on the D-instanton. 
Furthermore, all diagrams with boundaries
will be multiplied by a factor of
\be
K\equiv \NN\, e^{-C/g_s} \, .
\ee
Since the sum of the disconnected diagrams may be represented as $e^{S_1}-1$, the
total effective action may be written as
\be
S_{\rm total} = S_0 + K\, \left(e^{S_1}-1\right) + \OO(K^2)\, ,
\ee
where the $\OO(K^2)$ terms denote multi-instanton effective action.
We now need to check if $S_{\rm total}$ satisfies the BV master equation. For this
we compute:
\be\label{emaster1}
{1\over 2} \{S_{\rm total} , S_{\rm total} \} = {1\over 2} \{ S_0, S_0\} + 
K\, \{S_0, S_1\} \, e^{S_1} +\OO(K^2) \, .
\ee
We also have
\be \label{emaster2}
\Delta S_{\rm total} = \Delta S_0 + K\, \Delta \, e^{S_1} = \Delta S_0 +
K \,  \left(  {1\over 2}\,
\{S_1,S_1\} +\Delta S_1\right) \, e^{S_1}\, ,
\ee
where we have used the standard identity involving $\Delta$ and the anti-brackets
that follows from \refb{edefanti}.
Adding \refb{emaster1} and \refb{emaster2} and using \refb{emast1}, \refb{emast2},
we get,
\be \label{emaster3}
{1\over 2} \{S_{\rm total} , S_{\rm total} \} + \Delta S_{\rm total} =\OO(K^2)\, .
\ee
As will be demonstrated in \S\ref{sgauge2}, 
the order $K^2$ terms will cancel against the contribution from the 2-instanton
effective action.
This shows that the one instanton effective action satisfies the desired master equation.

Note that in the above analysis we have ignored the issues of divergences and have
pretended that the effective action can be computed as if we can represent its
interaction terms as
appropriate integrals over the moduli spaces of Riemann surfaces with no 
modification. Therefore the
proof of gauge invariance is formal. However, since the treatment of divergences
merely reorganizes the way we carry out the integration over the string fields, and does
not add or subtract any ad hoc gauge non-invariant terms, 
we do not expect this procedure to cause
breakdown of gauge invariance.

\subsection{Reality of the effective action} \label{ssinglereal}

We shall now turn to the study of reality properties of the effective action. 
The situation here is different from the case of effective field theory obtained by integrating
out heavy physical fields, where we do expect the effective action to acquire imaginary
part above a certain momentum 
threshold, describing the possibility of production of heavy states and
violation of unitarity in the light sector alone. Here the degrees of freedom that have
been integrated out are open strings that are not allowed asymptotic states, and we
need the reality condition to hold exactly for the proof of unitarity.

We shall begin with the ordinary interaction vertices of the theory.
Since these interaction vertices are identical to the 
ordinary interaction vertices of the closed string field theory, they generate real terms
in the closed string effective action\cite{9206084,9705038,1606.03455}.

The D-instanton type interaction vertices in the open closed
string field theory are
given by integrals of correlation functions of open and closed string
vertex operators and world-sheet ghost fields over the 
moduli spaces of Riemann surfaces
with boundaries, and the integration runs over parts of the moduli space that do
not include any degenerate Riemann surfaces. 
Formally, after integrating out the open string modes, the
interaction vertices of closed string effective field theory will also be given by similar 
integrals, but now the vertex operators will only be of closed strings and
the integration region will include degenerate Riemann
surfaces associated with open string degeneration. 
According to the reality postulate of \S\ref{sfield} whose validity we shall assume, 
the integrand itself will be real after factoring out the overall normalization constant
$\NN$, and the possible imaginary contributions can only arise
from having to carefully treat the divergences that arise from the boundaries of the
moduli spaces associated with open string degeneration.
This, and the normalization constant $\NN$,
will be the only source of violation of reality of the Lorentzian
effective action for closed strings, obtained by integrating out the open strings.

Let us therefore consider the effect of divergences from the boundaries of the moduli spaces.
\refb{erp1} gives the procedure for dealing with power law divergences associated with
tachyonic internal states. We see that this replaces a formally real but divergent
integral by a real
number. Therefore this replacement does not affect the reality property of the amplitude
given as integrals over the moduli spaces of Riemann surfaces. On the other hand, 
the first step for
dealing with logarithmically divergent integrals is to use the replacement rule 
\refb{erp2} that replaces a formally real but divergent integral by 0. This also does not
affect the reality property of the amplitude. The second step is to actually carry out the
integration over the zero modes following the procedure described in \S\ref{ssingleeff}. 
We shall now examine if these could lead to
violation of the reality condition. 

First let us consider the terms in the effective action 
that arise from the two Jacobians
associated with field redefinition of the collective mode and the ghost zero
mode, and the contribution from the auxiliary ghost field propagators.
It can be shown that these computations 
can all be reduced to the computation of certain amplitudes
in the original open closed string field theory, and are free from all 
divergences\cite{appear}. 
Therefore by the general result
that string theory amplitudes are real unless they
are divergent, 
these contributions do not affect the reality
property of the interaction vertices. The only exception is the possibility of an
overall complex constant factor-- since we have not determined the reality condition
on the open string fields, we can redefine them by multiplicative complex constants
leading to a multiplicative constant in the Jacobians.
This will only affect the overall normalization constant $\NN$, which in any case
cannot be computed with the present technology and will have to be determined
by other considerations. 

We now turn to the effect of the final integration over the collective coordinates.
As already discussed, this produces the momentum conserving
delta function involving the total euclidean momentum carried by all the external states
of the interaction vertex of the effective field theory. 
According to \refb{eec1}, \refb{eec2} this translates to a 
momentum conserving delta function in the Lorentzian space. Therefore this integral
also does not affect the reality property of the interaction vertex.

In superstring theories there will also be fermionic collective coordinates that need to
be integrated over. 
This means that in the amplitude needed to construct the interaction vertex, we need to
insert the vertex operators of open string fermion zero modes besides the usual closed
string vertex operators. 
Furthermore, since open strings live in the Euclidean theory, this computation needs
to be performed in the Euclidean theory. This could be potentially problematic since
the reality property of spinors in the Euclidean and the Lorentzian signature spaces
are different.
We can avoid this problem as follows. If there are
$2N$ real fermion zero modes, we can combine them into $N$ complex conjugate
pairs. Now the result of insertion of a given complex conjugate pair of zero modes
in a given
Feynman 
diagram can be regarded as part of the coefficient of the $q^{-1}$ term in an auxiliary
Feynman diagram in which we replace the insertion of the complex conjugate pair of
zero modes by an open string propagator with sewing parameter $q$. This exercise can
be repeated for each complex conjugate pair of zero modes leading to $N$
open string propagators with sewing parameters $q_1,\cdots, q_N$. 
At the end we have an
amplitude with only external closed strings, and
the amplitude with open string zero modes is part of the coefficient of the
$\prod_{i=1}^N q_i^{-1}$  term in the amplitude. 
Since the integrand of the amplitude with external closed strings is real by our reality
postulate, the coefficient
of $\prod_{i=1}^N q_i^{-1}$ in the integrand is also real, establishing the reality of the
amplitude with $2N$ open string zero mode field insertions.
This argument breaks down if the D-instanton has complex boundary state, but in that
case the contribution from the D-instanton with complex conjugate boundary state
will provide the complex conjugate contribution, restoring the reality of the effective
action.

Of course,
due to the various issues involving differences in the reality properties
of fermions in the Euclidean and Lorentzian spaces, there may be extra factors of $i$
coming from integration measure 
over the fermionic zero modes. However such factors 
are common to all amplitudes and may be absorbed into
the overall normalization constant $\NN$. 

Therefore we conclude that 
the action of the closed string effective field theory, obtained after integrating out the open
string fields on the D-instanton, is real provided the normalization constant $\NN$ is
real.
Since at present we do not have a systematic procedure for determining $\NN$
from first principles, we have to rely on indirect means for determining $\NN$ and
checking its reality. These methods include comparison with matrix model results
for two dimensional string theory and use of S-duality for type IIB string theory.
In the former case $\NN$ will turn out to be imaginary whereas for type IIB 
superstring $\NN$ for D-instanton and anti-D-instanton 
can be shown to be complex conjugates of each other, leading to real effective action.
This is evident from the reality of the instanton
induced term in the effective action needed for consistency with 
S-duality\cite{9701093}. 
The limited data available to us, as well as the fact that $\NN$ is formally given by the
one loop determinant of open string modes on the D-instanton, 
suggests that the violation of the reality
of $\NN$ is correlated with the presence of the tachyonic mode on the D-instanton.

\subsection{Cutting rules and unitarity} \label{ssinglecut}

Since for real $\NN$ 
the interaction vertices of the closed string effective field theory satisfy the usual
momentum conservation and reality conditions,
we can apply the
results of \cite{1604.01783} to conclude that the anti-hermitian 
part of any amplitude is given by
the sum over all the cut diagrams of the Feynman diagrams of the closed
string effective field theory.
We can now  translate the cutting rules of this effective  
field theory into the cutting rules of the original
open closed string field theory as follows. 
In the effective field theory with the usual cutting rules, the cut does not
pass through any vertex. Therefore the single D-instanton type vertex present
in the entire Feynman diagram must lie on one side of the cut. When we
interpret this in terms of the Feynman diagrams in the original
open closed string field theory, 
this implies that all the
D-instanton type vertices 
must lie on one side of the cut. Furthermore, since all internal open string
propagators must have both ends attached to a D-instanton type interaction 
vertex, no open string propagator is
cut. As discussed in \S\ref{scut}, this is exactly the kind of cutting rules we need for
the proof of unitarity of D-instanton induced amplitudes.

The cutting rules are one of the ingredients in the proof of unitarity.  In a gauge theory
like string field theory, we also need to show that in the cut propagator the contribution
from the unphysical states cancel so that the sum over cut diagrams can be interpreted 
as the
right hand side of \refb{e1} with the sum over $n$ running over physical states only. 
For closed string field theory this was proved in \cite{1607.08244} using the 
BV formalism. Since the closed string effective field theory also satisfies
the BV master equation, the same proof will hold for this theory as well.

\sectiono{Multi-instanton contribution to the effective action and its reality
properties} \label{smulti}

We shall now analyze the $k$ D-instanton contribution to the closed string effective
action and study its reality properties. We shall first
consider the case of $k$ different D-instantons, carrying normalization factors
$\NN_1 e^{-C_1/g_s},\cdots, \NN_k e^{-C_k/g_s}$, and then discuss some additional
subtleties that may arise for identical D-instantons. 

\subsection{Construction of the effective action ignoring the zero modes} \label{seffmulti}

First we shall ignore the effect of the zero modes and divergences from the boundaries
of the moduli spaces of Riemann surfaces and study how the 
Feynman diagrams of open-closed string field theory can be organized into Feynman
diagrams of closed string effective field theory. This will help us read out the
interaction vertices of the closed string effective field theory.  Let us first consider
the case $k=2$. 
\begin{enumerate}
\item In this case the Riemann surfaces corresponding to the amplitudes
have two types of D-instanton boundary condition, corresponding to the first or the
second D-instanton.
 In the open closed 
string field theory there are now three different types of D-instanton interaction vertex,
type 1, type 2 and type 12, reflecting the fact that associated Riemann surfaces have 
boundaries of type 1, type 2 and types 1 and 2. 
As in \S\ref{seffective}, we must also include in this discussion the zero point
vertices, associated with Riemann surfaces with boundaries but no
punctures, and disconnected diagrams.
\item Since for two instanton amplitudes
the underlying Riemann surfaces must have at least one boundary of each type,
we can classify the Feynman diagrams in the open closed string
field theory into two possible types -- 
(a) those carrying interaction vertices of type 1 and type 2, but not of 
type 12 and (b) those carrying at least one interaction vertex of type 12 and 
possibly other
interaction vertices of any type. 
\item Let us now
compare this with the diagrams generated by the closed string effective field
theory with two insertions of  single D-instanton type vertex constructed in
\S\ref{seffective}, one of type 1 and the other of type 2. It is straightforward to see
that when we reinterpret this as Feynman diagrams in the original open closed string
field theory, then these generate all the diagrams of type (a) but none of the diagrams
of type (b). Therefore we must interpret the Feynman diagrams of type (b) as being
generated by a new interaction vertex of closed string effective field theory, obtained by
fusing all the D-instanton type interaction vertex and the open string propagators 
into a single composite vertex. 
\item Put another way, the new $n$-point 
interaction vertex of the
closed string effective field theory will contain all $n$-point connected and
disconnected amplitudes
of the original open closed string field theory that use at least one 12 type interaction
vertex, and arbitrary number of possibly other D-instanton type vertices and internal
open string
propagators, but no internal closed string propagator or ordinary interaction vertex.
\end{enumerate}

The annulus diagrams require special care. 
These diagrams do not carry any factor of $g_s$ and therefore must be
resummed as an exponential multiplying the integrand. 
There are different types of annulus diagrams depending on the boundary
conditions at the two boundaries. The diagrams where both boundaries carry the
same boundary condition can be exponentiated and their contribution can be absorbed
into a factor of $\NN_1\NN_2$ in the overall normalization. The annulus
diagrams with the two boundaries 
having two different boundary conditions 
must be regarded as a contribution to the 12 type zero point function, and therefore
Feynman diagrams containing one or more factors of 
such annuli and possibly 
other D-instanton type vertices and open string propagators, but no internal closed string propagator or ordinary interaction
vertex, must be included in the 
contribution to the two instanton interaction vertex in the effective field theory. 
There are however some subtleties that we discuss below:
\begin{enumerate}
\item In open closed string field theory, part of this annulus diagram comes from a 
Feynman diagram where a pair of closed string one point functions on the disk, with
1 and 2 type boundary conditions, are connected by a closed string propagator.
This has been shown in Fig.~\ref{fig7}. 
So only part of the annulus contribution, that is not included in Fig.~\ref{fig7}, should
be counted as a contribution to the 12 type zero point interaction vertex 
in open closed string field
theory, and this is the part that needs to be included in the construction of the
2-instanton interaction vertex of closed string effective
field theory following the procedure described above.
If we regard the annulus as a cylinder and
denote by $2 t$ the ratio of the circumference
to the height of the cylinder, then the Feynman diagram of Fig.~\ref{fig7} covers
the integration range $0\le t\le a$ where $a$ is some real number. The precise value
of $a$ depends on the choice of the off-shell closed string one point interaction 
vertex. Therefore the part of the annulus
contribution that should be included in the definition of the two instanton vertex of the
closed string effective field theory is given by:
\be\label{ea12}
A_{12} = \int_a^\infty dt \, t^{-1}\, Z_{12}(t)\, ,
\ee
where $Z_{12}(t)$ is the partition function of the 12 open strings:
\be\label{ez12t}
Z_{12}(t) = Tr_{12} e^{-2\pi t L_0}\, .
\ee
Usually we would have a factor of $1/2$ in \refb{ea12}, but we have included in 
\refb{ea12} also the contribution from 21 open strings, assuming that $Z_{21}=Z_{12}$.

\begin{figure}
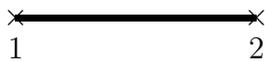

\begin{center}
\hbox{\hskip 1.5in \figseven}
\end{center}

\vskip -.8in

\caption{A closed string exchange diagram that gives part of the annulus diagram with
different boundary conditions on the two boundaries.
\label{fig7}
}
\end{figure}

\item Now consider a general Feynman diagram of the open closed string field
theory with external closed strings but no internal closed string or ordinary interaction 
vertices and no factor of the annulus diagram.
If the
diagram under consideration has components with 12 type
interaction vertex, then it is part of the contribution to the 2-instanton
interaction vertex
of the closed string effective field theory, and 
we must multiply this by $e^{A_{12}}$ to 
take into account the effect of all possible insertions of
annulus diagrams with two boundaries
lying on two different instantons. On the other hand 
if the original diagram does not have any 12 type 
interaction vertex, then it is not a part of the 2-instanton interaction vertex of the 
closed string effective field theory.
Now we should multiply this by $e^{A_{12}}-1$ to get a
contribution to the interaction vertex of the closed string effective field theory. 
Diagrammatically
this may be interpreted by saying that a Feynman diagram in the open closed string field
theory containing only external closed strings and no internal closed strings or
ordinary interaction vertices, can be regarded as a contribution to the 2-instanton
interaction vertex of the closed string effective field theory if after fusing the interaction
vertices and / or annuli
sharing common boundary type,
we are left with a single composite vertex.
\end{enumerate}
With this prescription the two instanton interaction vertex in the closed string
effective field theory will have an overall multiplicative factor of 
$\NN_1\NN_2\, e^{- (C_1+C_2)/g_s}$, and the rest of the contribution to the 
interaction vertex
will have to be computed explicitly following the algorithm described above.
We shall call this the 12 type interaction vertex of the closed string effective field theory.

Next  consider the case $k=3$. In this case the Riemann surfaces associated with the
amplitudes have three different types of boundaries, and the corresponding
open closed string field
theory has many different types of D-instanton type vertices -- types 1, 2, 3, 
12, 23, 13 and
123. Now given a general Feynman diagram in this theory, the question that we need
to ask is if this diagram can be
associated with a Feynman diagram of the closed string effective field theory using
the interaction vertices that have already appeared in the analysis of one and 
two instanton
amplitudes. If the answer
is in the negative then we need a new interaction vertex of the closed string effective field
theory to generate this diagram in the open closed string field theory. We shall
first illustrate this with some examples and then give the general result. 

Let us suppose that the Feynman diagram of the open closed string field theory has
D-instanton type interaction vertices of type 1, 2, 12 and 3. Then we can get this from a
Feynman diagram of the closed string effective field theory with one 2-instanton 
interaction vertex and one 1-instanton interaction vertex, obtained as
follows. We fuse all the 1, 2 and 12 type interaction vertices, as well as any
open string propagator connected to these vertices, into a single composite vertex
and interpret this as part of a 2-instanton interaction vertex
of the closed string effective field 
theory. We also fuse all the 3 type interaction vertices, as well as any
open string propagator connected to these vertices, into a single composite vertex
and interpret this as part of a 1-instanton vertex of the closed string effective field 
theory. Therefore for this class of diagrams we do not need to introduce a new
interaction vertex in the closed string effective field theory. 
On the other hand if the Feynman diagram of the original open closed string
field theory contained 12 and 23 type vertices, or 123 type vertex, then there will be 
no way to get this from the 1 and 2-instanton interaction 
vertices of closed string effective field
theory. We need to fuse all the interaction vertices of the Feynman diagram,
as well as any
open string propagator connected to these vertices, into a single composite vertex
and interpret this as part of a 3-instanton interaction vertex of the closed string effective field 
theory. 

Based on these examples, we see that a
3-instanton interaction vertex of the closed string effective field theory
with $n$ external closed strings is 
obtained by summing over all Feynman diagrams of open closed string field
theory with $n$ external closed strings, satisfying the following properties:
\begin{enumerate}
\item
The diagram can have internal open string propagators and
D-instanton type vertices of the open closed string field theory,
but cannot have any ordinary interaction vertex or internal closed string
propagator.
\item
If we fuse each pair of interaction vertices sharing a common boundary type, and
repeat this for all pairs, then
at the end we should be left with a  single composite vertex.
\end{enumerate}

We now turn to the discussion of annulus diagrams which we have ignored so far.
Annuli with the same
boundary condition on the two boundaries
are included in the overall normalization factor $\NN_1\NN_2\NN_3
e^{-(C_1+C_2+C_3)/g_s}$.
Now consider a diagram in the open closed string field theory with no ordinary interaction
vertex and no internal closed string propagator. Its contribution to the 3-instanton vertex
of closed string effective field theory is determined as follows.
\begin{enumerate}
\item If after following the rules described above for fusing the interaction vertices
we are left with a single composite vertex, then the contribution from the diagram
is multiplied by a factor of $\exp[A_{12} + A_{23}
+A_{13}]$, with $A_{ij}$ defined as in \refb{ea12}, to determine its contribution to the
3-instanton vertex.
\item If after following the rules for fusion
the diagram has two composite
vertices, one containing all world-sheet boundaries of type $i$ and $j$ and the other
containing all world-sheet boundaries of type $k$,
then it should be multiplied by $e^{A_{ij}} (e^{A_{ik} + A_{jk}} -1)$, so that
we are forced to include at least one factor of $A_{ik}$ or $A_{jk}$ from the
annulus diagram. Diagrams with only $A_{ij}$ insertions are already present in the
Feynman diagram of the closed string effective field theory with one insertion
of two instanton vertex of type $ij$ and one insertion of two instanton vertex of
type $k$. 
\item Finally if a diagram in open closed string field theory has only 
interaction vertices of type 1, 2 and 3 so that it has three composite vertices after we use
the rules for fusion, 
then it needs to be multiplied by
\be
(e^{A_{12}}-1)(e^{A_{23}}-1) (e^{A_{13}}-1) + (e^{A_{12}}-1)(e^{A_{23}}-1) 
+ (e^{A_{23}}-1) (e^{A_{13}}-1) + (e^{A_{12}}-1)(e^{A_{13}}-1) \, .
\ee
The first term represents the contribution where there is at least one annulus of each type,
12, 23 and 13. The second term represents the contribution where there is 
at least one annulus of type 12 and one annulus of type 23, but no annulus of type 13. 
The other two terms are obtained as permutations of the second term. 
This includes all diagrams containing one of the following configurations: 1) at least
one 12 and one 23 type annulus, 2) at least one 13 and one 23 type annulus and
3) at least one 12 and one 13 type annulus.
\end{enumerate}

We are now ready to generalize this to $k$-instanton amplitudes. 
In \S\ref{scut} we have already introduced the notion of a label $(i_1,\cdots,i_\ell)$ for 
each
interaction vertex of open closed string field theory specifying the type of boundary 
segments
that the Riemann surface associated with the vertex has. For a given Feynman diagram
in this open-closed string field theory, we now introduce an auxiliary diagram, in which
we fuse each pair of interaction vertices of the original diagram, 
that share a label, into a
composite vertex, and also fuse all the open string propagators connected to the
original vertices into the composite vertex. For example, if the original Feynman
 diagram had interaction vertices of types 12, 23, 36 and 45, then in the auxiliary diagram
 the vertices of type 12, 23 and 36 will be fused into a single composite vertex, and the
 vertices of type 45 will be fused together (if there are more than one such vertices) 
 to form a second composite vertex. If the auxiliary Feynman
 diagram, constructed this way, has more than one composite vertex, then it can
 be regarded as coming from the Feynman diagram of the closed string effective
 field theory based of $\ell$ instanton vertices with $\ell<k$. If on the other hand the
 auxiliary diagram has only one composite vertex, i.e.\ if all the D-instanton type
 vertices of the original Feynman diagram are fused into a single composite
 vertex, then it needs
 to be interpreted as coming from a Feynman diagram of closed string effective field
 theory that uses a new vertex -- the $k$-instanton interaction vertex.
 Conversely, a $k$-instanton interaction vertex of the closed string effective field theory,
 with $n$ external closed strings, is 
obtained by summing over all Feynman diagrams of open closed string field
theory with $n$ external closed strings, 
satisfying the following properties:
\begin{enumerate}
\item
The diagram can have internal open string propagators and
D-instanton type vertices of the open closed string field theory,
but cannot have any ordinary interaction vertex or internal closed string
propagator.
\item
If we fuse all pairs of interaction vertices sharing a common boundary type, then
at the end we should be left with a single composite vertex.
\end{enumerate}
This is exactly analogous to the situation for $k=2$ and $k=3$.

We need to augment this rule by giving the algorithm for including annulus diagrams 
with different boundary conditions. Let us call an annulus with $i$ type boundary
condition on one boundary and $j$-type boundary condition on the other boundary
an $ij$ type annulus. 
Given a Feynman diagram in the open closed string field theory
with only external closed strings and no internal closed strings and ordinary
interaction vertices,  we must sum over
those annulus insertions for which, after fusing all the interaction 
vertices and annuli sharing a common
label, we are left with a single composite vertex. The required insertions
may be expressed as combinations of $e^{A_{ij}}$ with $1\le i<j\le k$. The
overall multiplicative factor for this vertex will be given by $\prod_{i=1}^k \left\{\NN_i
e^{-C_i/g_s}\right\}$, with the $\NN_i$'s including contributions from the
annuli whose both boundaries have
the same boundary condition.

\subsection{Effect of divergences and open string zero modes} 

Let us now turn to the analysis of possible divergences. We begin with the
configuration where the D-instantons are widely separated so that the open
strings connecting a pair of D-instantons have positive mass$^2$ and do not
cause any divergences. In this case the divergences will arise from tachyonic modes
and zero modes living on individual D-instantons. Tachyons can be handled
using the replacement rule \refb{erp1} and will not require any further discussion. 
However the zero modes
require special care. For this let us again turn to the case of two instantons. 
At a generic point in the D-instanton moduli space, where they
are at different locations, we have zero modes from the 11 open strings and
22 open strings, and we need to learn how to integrate over them. As usual we
begin by leaving them unintegrated and treat them as external states of the diagram
to construct the Wilsonian effective action.
First consider the diagrams in the open closed string field theory
where we have only type 1 and type 2 D-instanton vertices
but no type 12 vertex or annulus. In this case we fuse all the 1 type vertices together 
and all the 2 type vertices together to construct a pair of composite vertices.
Let us call them composite vertex 1 and 2 respectively.
All the external 11 string zero modes will attach themselves to the composite vertex
1, since in the original open closed string field theory they can only attach to type 1
or type 12 type vertices. Similarly all the external 22 string zero modes will attach
themselves to the composite vertex 2. As a result the integration over these zero modes
become completely independent of each other
and reduce to two copies of the integration procedure
described in \S\ref{ssingleeff}. Therefore integration over these zero modes produce exactly
the factors needed to make each composite vertex into a full 1-instanton vertex of the
closed string effective field theory, including the momentum conserving delta functions
associated with each composite vertex. 

Next consider the case where the original Feynman diagram in the open closed string
field theory has a 12 type vertex and / or a 12 type annulus. 
In this case there is only one composite vertex after we
fuse all the D-instanton type vertices of the original diagram into a single 
composite vertex. Therefore
both the 11 and 22 zero mode fields will attach to the same composite vertex and the
integral over the zero modes no longer factors into two independent integrals.
We can proceed as before. We now expect two ghost zero modes that,
after field redefinition, represents the effect of division by the rigid $U(1)\times U(1)$
gauge transformation on the two D-instantons. The Jacobian associated with this
field redefinition is no longer given by the product of two independent Jacobians 
associated with individual D-instantons and need to be recomputed. There are also
a pair of auxiliary open string fields from the ghost sector whose contribution to the
Feynman diagrams must be explicitly computed, producing additional contribution to
the two instanton interaction term. Finally there are now two sets of zero modes,
associated with the space-time
coordinates of the D-instantons. After appropriate field redefinition,
one of them, which we shall denote by $\phi$, 
will represent the average of the positions of the two D-instantons. The dependence on
this mode will be proportional to $e^{ip.\phi}$ where $p$ is the total
momentum carried by the closed strings connected to the composite vertex, and integration
over $\phi$ gives the momentum conserving delta function at the 2-instanton 
interaction vertex
of the closed string effective field theory. The other set of zero modes $\chi$ represent 
the difference between the 
locations of the two D-instantons after appropriate field redefinition. 
The amplitude will have complicated but
computable dependence on $\chi$ and the integration over $\chi$ needs to be 
computed explicitly to find the 2-instanton vertex of the effective field theory.

Since for large separation between the D-instantons the 12 open strings become
very massive, the 
contribution to the interaction vertex of the
closed string effective field theory from any Riemann surface that includes
both a 1-type boundary and a 2-type boundary,
is exponentially suppressed for large
$\chi^2$. 
Therefore in these contributions,
the $\chi$ integral does not have any divergence from the large
$|\chi|$ region. 
This will be seen explicitly in an example in \S\ref{sexamplemulti}, where  the
function $f(\chi)$ in \refb{e6.14} has such exponential suppression factor.
The full amplitude of course does receive contribution from the large $|\chi|$ region, but
these come from regions of the moduli space associated with closed string degeneration,
and are represented, in the open closed string field theory, by Feynman diagrams
involving closed string exchanges. These are not part of the definition of the
2-instanton interaction
vertex of the closed string effective field theory.

As we reduce $\chi^2$, the mass$^2$ of some of the states of the 12 open string
may change from being positive  to being
negative  below a critical separation between the two D-instantons. 
These may lead to additional divergences that were not present in one instanton
amplitudes. As we shall see
in \S\ref{smultireal}, such divergences may be regulated by appropriate $i\eps$
prescription, but could lead to complex contribution to the closed string effective
action.

The generalization of this analysis to the $k$ instanton case is straightforward. For
any Feynman diagram of the original open closed string theory, once we fuse the
interaction vertices into composite vertices following the rules mentioned earlier,
for every composite vertex there will be a momentum conserving delta function from
integration over the collective coordinates, 
consistent with the interpretation that every composite
vertex is part of an interaction vertex of the closed string effective field theory. The new
vertex that arises from these Feynman diagrams -- the $k$ D-instanton vertex -- will
have a single momentum conserving delta function from integration over the collective
coordinates that describes the average location of all the $k$ D-instantons. The dependence
of the Wilsonian effective action on the other collective
coordinates, describing the differences between the locations of the D-instantons, will be
complicated but computable, and the $k$-instanton interaction vertex of the
closed string effective field theory will include integration over these collective 
coordinates. As before, the integrand will be exponentially suppressed for large
separation between the D-instantons, and there is no divergence in the integral from this
region. However at finite separation, when some of the open strings connecting
different D-instantons turn tachyonic, the integrand may have additional divergences that
need to be dealt with by $i\eps$ prescription. This will be discussed
in \S\ref{smultireal}.

\subsection{Identical D-instantons}

The case of $k$ identical D-instantons can be analyzed in more or less the
same way since different D-instantons can still be distinguished by their locations in
space-time. We shall of course get extra factor of $1/k!$ in the expression for the effective
action since exchanging the locations of the instantons does not generate new
configurations. The extra complication arises from the fact that in the limit when the
instantons coincide, additional massless states appear. These could lead to new
divergences from two sources. First, the $ij$ type annulus diagram receives a 
contribution proportional to $1/m^2$ from a pair of bosonic open string states of 
 mass $m$ and
a contribution proportional to $m^2$ from a pair of fermionic open string 
states of mass 
$m$.\footnote{Since for every state of the open string stretched from the
$i$-th instanton to the
$j$-th instanton, there is a degenerate state of the open string stretched from 
the $j$-th 
instanton to the
$i$-th instanton, the states come in pairs.} These could diverge in the $m\to 0$ limit,
but if there are more fermionic / ghost modes of this type than bosonic modes then these
divergences could be avoided. The other possible source of divergences is from
internal propagators of $ij$ type open strings which could give $1/m^2$ factors. 
These divergences become more severe at higher orders when there can be
more $ij$ type open string propagators. As long as
the integrals over the original zero modes give finite contribution from the limit of
coincident D-instantons, we shall not worry about the appearance of these new
zero modes, but for high order diagrams when the integrals diverge we need to
treat these zero modes also via Wilsonian effective action. This means that when the
separation between the D-instantons go below some chosen cut-off, we explicitly
remove the contribution of the new light modes from the internal propagators and
regard them as background, and at the end 
explicitly perform the integrals over these new light modes. 

As discussed in \S\ref{seuclid}, and elaborated in \S\ref{s6.1}, \ref{stwo}, the  
connection between the
reality properties of the Euclidean and the Lorentzian action, that we are using in our 
analysis, relies on the effecting action being an analytic function of the external
energies in the complex plane. 
One could wonder if the extra massless modes on coincident D-instantons 
could introduce non-analyticity of the
effective action in the complex energy plane, thereby spoiling this connection. 
However, since derivatives
with respect to external energy bring down powers of the separation $\chi$ between the
instantons in the {\em numerator} of the integrand,  
and since the extra massless modes arise when the 
separations
vanish, these derivatives do not make the integrand more singular.
Therefore we do not expect the presence of these massless modes to introduce 
non-analyticities in the complex energy plane.

Finally, we would like to note that in special cases, massless open string modes may
also appear on a pair of non-identical D-instantons when they coincide. The effect of these
modes may be analyzed in the same way as described above.

\subsection{Gauge invariance} \label{sgauge2}

We shall now verify the gauge invariance of the effective action.
We shall first analyze the two
instanton contribution to the
closed string effective action, and then give a general argument for the
gauge invariance of the multi-instanton contribution to the action.

As in \S\ref{sgauge}, we shall begin with the result of \cite{1609.00459}, 
demonstrating the gauge
invariance of the closed string effective action from connected world-sheet. For 
two identical instantons, these contributions may be of three types. 
First we have the contribution from world-sheet
without boundaries. This was called $S_0$ in \S\ref{sgauge}. 
Then we have the contribution
from the world-sheet with at least one boundary, with all the boundaries ending on the
same D-instanton. This contribution is given by $2S_1$, with $S_1$ being the same as
what appears in \refb{emas0}. The factor of 2 is related to the fact that the boundaries may
lie on either the first instanton or the second instanton. Finally we have the contribution
from world-sheet with two or more
boundaries, with at least one boundary lying on the first instanton
and at least one boundary lying on the second instanton. We shall denote this
contribution by $S_2$. Then $S_0+2S_1+S_2$ will satisfy the BV master equations:
\be\label{emast5}
{1\over 2}\, \{S_0+2S_1+S_2, S_0+2S_1+S_2\} + \Delta (S_0+2S_1+S_2)=0\, .
\ee
Using \refb{emast1} and \refb{emast2}, this can be reduced to
\be\label{emast6}
\{S_0,S_2\} + 2\{S_1,S_2\} +\{S_1,S_1\} + {1\over 2} \{S_2, S_2\} +\Delta S_2=0\, .
\ee

In the current problem, we include in the definition of the 2-instanton effective action the
disconnected diagrams and also include a multiplicative factors of $K^2$
where $K=\NN e^{-C/g_s}$ as in \S\ref{sgauge}. Including the 0-instanton and 1-instanton
contributions to the effective action introduced in \S\ref{sgauge}, the total effective action up
to 2-instantons is given by,
\be \label{emaster5}
S_{\rm total}=S_0 + K\, (e^{S_1}-1) + {1\over 2} \, K^2 \, (e^{S_2}-1) \, e^{2\, S_1}\, .
\ee
In the last term the factor of $e^{S_2}-1$ implies that the diagram must contain at least 
one connected
component with both types of boundaries. As discussed in 
\S\ref{seffmulti}, if such a component
is absent, then the contribution to the amplitude may be regarded as the one 
coming from a pair of one instanton interaction vertices of the effective field theory, and
does not represent a genuine 2-instanton contribution to the effective action. $e^{2S_1}$
factor reflects that we can have arbitrary number of connected components whose
boundaries all end on the first instanton and also an arbitrary number of connected 
components whose boundaries all end on the second D-instanton. The factor of $1/2$
reflects that the exchange of the two D-instantons does not generate a new configuration.

We now need to verify that $S_{\rm total}$ satisfies the BV master equation to order
$K^2$. We have already seen that it satisfies the master equation up to order $K$. Therefore
we only have to examine the terms of order $K^2$. We have, to this order,
\ben \label{emaster6}
{1\over 2} \, \{ S_{\rm total}, S_{\rm total}\} + \Delta\, S_{\rm total}&=&
K^2 \Bigg[ {1\over 2}\,
e^{S_2+2\, S_1}\, \{S_0, S_2\}  + e^{2\, S_1} (e^{S_2}-1) \{S_0, S_1\} +{1\over 2} 
\, e^{2 \, S_1} \, \{S_1, S_1\} \nonumber \\ &&
\hskip -.2in +\, {1\over 2} \, e^{S_2+2\, S_1} \left( \Delta S_2  +  {1\over 2}
\{S_2, S_2\}\right)  + (e^{S_2}-1)\, e^{2 \, S_1}  (\Delta S_1+ \{S_1, S_1\})
\nonumber \\ && 
\hskip -.2in +\,  e^{S_2+2\, S_1}  \{S_2, S_1\}
\Bigg]\, .
\een
Using  \refb{emast2} and \refb{emast6} one can show that the right hand
side vanishes. Therefore we have:
\be \label{emaster7}
{1\over 2} \, \{ S_{\rm total}, S_{\rm total}\} + \Delta\, S_{\rm total} =0\, .
\ee

This proves gauge invariance of the closed string effective field theory up to two instanton
order when the instantons are identical. The case where the instantons are not identical
can be analyzed with minor modifications of the above argument. Let us 
denote by $S_1$
and $S_1'$ the  single instanton 
effective actions associated with connected world-sheet for the first
and the second
instanton respectively, by $K$ and $K'$ the corresponding normalization factors, 
and by $S_2'$ the analog of $S_2$ appearing in \refb{emast5}, except that the two
instantons now refer to different instantons.
Then in
\refb{emast5}, $2\, S_1$ will be replaced by $S_1+S_1'$, $S_2$
will be replaced by $S_2'$ and besides \refb{emast2},
we also have a similar equation involving $S_1'$. Using this we can get the analog of
\refb{emast6}:
 \be\label{emast6alt}
\{S_0,S_2'\} + \{S_1,S_2'\} + \{S_1',S_2'\} +\{S_1,S_1'\} 
+ {1\over 2} \{S_2', S_2'\} +\Delta S_2'=0\, .
\ee
The analog of \refb{emaster5} now takes the form
\be \label{emaster5alt}
S_{\rm total}=S_0 + K\, (e^{S_1}-1) +K'\, (e^{S_1'}-1)  
+  K K' \, (e^{S_2'}-1) \, e^{S_1+S_1'}\, ,
\ee
and our goal will be to show that the order $KK'$ term in
\be 
{1\over 2} \, \{ S_{\rm total}, S_{\rm total}\} + \Delta\, S_{\rm total}
\ee
vanishes. This is given by:
\ben
&& K\, K' \Bigg[ e^{S_1+S_1'} \{S_1, S_1'\} +e^{S_1+S_1' + S_2'} \, \{S_0, S_2'\}
+ e^{S_1+S_1'} \left(e^{S_2'}-1\right) \{S_0, S_1+S_1'\} \nonumber \\ &&
+e^{S_1+S_1' + S_2'} \left( \Delta S_2' + {1\over 2} \{S_2', S_2'\} + 
\{S_2', S_1+S_1'\}  \right) \nonumber \\ &&
+ e^{S_1+S_1'} \left(e^{S_2'}-1\right) \left(\Delta S_1
+ \Delta S_1' + {1\over 2} \{S_1, S_1\} +  {1\over 2} \{S_1', S_1'\} +
\{ S_1, S_1'\}
\right)
\Bigg]  \nonumber \\ &&
=0\, ,
\een
where in the last step we have used \refb{emast2}, its analog with $S_1$ replaced
by $S_1'$ and \refb{emast6alt}.

We expect the gauge invariance of the
effective action 
to hold also at higher instanton order, since the original amplitudes
were gauge invariant, and our procedure for constructing the effective action simply
organizes these amplitudes into Feynman diagrams of closed string effective field theory.
The method described above becomes
complicated due to complicated combinatorial factors. However, one can proceed following
an alternate route in which one directly proves the main identities
of \cite{9206084}, or 
equivalently
the quantum $L_\infty$ algebra\cite{9209099}, 
for the interaction vertices of the closed string effective action constructed here.
The key ingredient used in the proof in \cite{9206084} 
is that the elementary interaction vertices,
together with all other Feynman diagrams with propagators, cover the integration over
the moduli spaces of Riemann surfaces in a one to one fashion. This property is also
true by construction for the effective action described here, except that we include both
connected and disconnected Riemann surfaces.
This in turn would 
prove the gauge invariance of
closed string field theory constructed here.

\subsection{Reality of the effective action} \label{smultireal}

The discussion of reality of the effective action proceeds as in the case of single
instanton effective action. The main new ingredient is the integration over the
collective coordinates labelling the differences between the D-instanton locations. 
The integrand is constructed from the Wilsonian effective action obtained by
integrating out the massive and the tachyonic modes of the open string. 
The reality postulate of \S\ref{sfield}  ensures that after factoring out the overall 
normalization constant $\NN_1\cdots \NN_k$, the integrand is invariant under 
simultaneous operation of complex conjugation, reversal of sign of external momenta,
change of sign of the complex structure of the Riemann surface and change of sign of the
collective coordinates. This would suggest that after integration over the moduli spaces of
Riemann surfaces and the collective coordinates, we should obtain real contribution to
the effective action, except for possible complex contribution from the 
$\NN_1\cdots \NN_k$ factor multiplying the $k$-instanton effective action.

We should not however forget the possibility that the apparent 
divergences from the
boundaries of the moduli spaces associated with open string degeneration may lead to
additional imaginary contribution. These can be analyzed more or less in the same way
as in the single instanton case and can be shown not to introduce complex contribution, with
one class of exceptions. 
As already discussed, when the separation between a pair of D-instantons passes
through a
critical value, some of the open string states 
connecting the pair may make transition from being massive to tachyonic.
In such cases we need an $i\eps$ prescription to  integrate 
through these singular points. 
These may
lead to additional factors of $i$ that need to be accounted for. 

We shall illustrate this by an example. Let us consider a two instanton system and
let $\chi$ be the collective coordinate labelling the difference between the 
space-time locations of the
pair of D-instantons. Let us further suppose
that the spectrum of open strings stretched between the two D-instantons has a
complex conjugate pair of states with mass$^2$ given by
\be \label{echimass}
{\chi^2\over 4\pi^2} - M^2\, ,
\ee
for some real constant $M$ so that for $|\chi|<2\pi M$ the pair becomes tachyonic. 
The first term in \refb{echimass} is the contribution due to the tension of the open string
stretched over a distance $|\chi|$ in $\alpha'=1$ unit, and the $-M^2$ term is the mass$^2$
of the state for coincident D-instantons. 
Integrating out a pair of such open string modes will generate a factor in the amplitude
proportional to
\be \label{eres}
{4\pi^2\over \chi^2 - 4\pi^2 M^2}\, ,
\ee
from one loop determinant, i.e.\ the annulus diagram with the two boundaries lying
on the two D-instantons. There may also be
other singular factors, {\it e.g.} factors of
$1/ (\chi^2 - 4\pi^2 M^2)$ from internal 12 string propagators.
Now when we carry out the integration over $\chi$,
the integral may no longer be real, since we need appropriate contour prescription
around the point $\chi=\pm 2\pi M$. 

The divergence
in \refb{eres} at $\chi=2\pi M$
may appear to be unrelated to the
divergences from the boundary of the moduli space, but this 
is not so. Indeed, the expression for the part of the amplitude
given in \refb{eres}, when expressed as integral over the moduli space of Riemann
surfaces, has  no divergence in the integrand at 
$\chi=2\pi M$. Instead, what we have is an analog of \refb{erp1} with $\alpha$ replaced by
a term proportional to $(\chi^2 - 4\pi^2M^2)$. As $\chi$ passes through $2\pi M$, 
$\alpha$ passes through 0, and although the integrand remain finite and real at fixed $q$,
new divergences appear at $q=0$ for $\chi\le
2\pi M$, forcing us to adopt some contour prescription.

One possible choice of contour that was used in 
\cite{1912.07170} is  as follows. 
Since according to the general guideline discussed in \S\ref{seuclid} the 
position space 
integration contour is rotated anti-clockwise as we go from the Euclidean to the 
Lorentzian theory, if we require that the $\chi$ integration contour does not cross
a pole during this rotation,
we need to
choose the original poles to be below the real axis for $\chi^0>0$ and above the
real axis for $\chi^0<0$. 
This can be summarized by saying that \refb{eres} should be replaced by\cite{talk}
\be \label{eres1}
{4\pi^2\over \chi^2 - 4\pi^2 M^2 +i\eps}\, .
\ee
The $i\eps$ generates an imaginary contribution to the Euclidean effective action
after integration over $\chi$,
which translates to an imaginary term in the Lorentzian effective action.
We shall see an explicit example of this in \S\ref{sexamplemulti}. We shall call this
procedure the Lorentzian contour prescription. In the multi-instanton amplitude there
may be several singular factors of this type, and in the Lorentzian prescription we 
shall replace each such singular factor by a similar $i\eps$ prescription. This has been
given in \refb{esingone}.

It is however not necessary to choose the contour using the prescription given in
\refb{eres1}. An alternate prescription that
does not generate an imaginary part of the interaction vertex of the closed string effective
field theory and preserves unitarity is
where we take the average of the Lorentzian prescription \refb{eres1} and its complex
conjugate, replacing all the $i\eps$'s by $-i\eps$.
We shall call this the unitary prescription. 
If we are to choose a uniform prescription that works for all string theories, then the
unitary prescription seems more physical since a  Lorentzian prescription will lead
to unitarity violation also in critical superstring theory, {\it e.g.} from the D-instanton -
anti-D-instanton contribution to type IIB string theory amplitudes.
The actual computation is somewhat easier in the Lorentzian prescription since
we can evaluate the integral over $\chi$ by rotating the integration contour to the
Lorentzian space\cite{1912.07170}. 
However once the result in the Lorentzian prescription has been
obtained, 
the difference is easy to compute using the relation:
\be \label{eprin}
{1\over \chi^2 - 4\pi^2 M^2-i\eps}= {1\over \chi^2 - 4\pi^2 M^2 +i\eps}
+ 2\, i\, \pi\, \delta(\chi^2 - 4\pi^2 M^2)\, .
\ee
We shall describe a simpler algorithm to compute the result in the unitary
prescription at the end of \S\ref{s6.1}. The generalization of the unitary prescription
at higher orders and / or for multiple instantons has been described
below \refb{esingone}.

Ref.~\cite{1912.07170} used the Lorentzian prescription to evaluate the 
D-instanton contribution to the
amplitudes of two dimensional string theory and found agreement with the matrix model
results. We shall see  
in \S\ref{sexamplemulti} that at least in the 2-instanton and 3-instanton sectors,
the unitary prescription is also consistent with the matrix
model results, with a different choice of the coefficients of the contribution from type
$(2,1)$ and type  (3,1)
ZZ-instantons.

In terms of the original integral over the tachyonic modes, the dependence on the
contour prescription can be understood as follows. For $\chi>2\pi M$ the tachyons
have positive mass$^2$ and the steepest descent contours run along the real axes,
while for $\chi<2\pi M$ the tachyons
have negative mass$^2$ and the steepest descent contours run along the imaginary axes.
This abrupt jump at $\chi=2\pi M$ makes the choice of integration cycle ill defined. 
By deforming the $\chi$ integration contour around $2\pi M$ 
into the complex plane, we ensure that the
steepest descent contours for the tachyons rotate continuously from the real plane
to the imaginary plane. However, now the combined integration cycle over $\chi$ and
the tachyons depends on whether the $\chi$ contour passes below or above $2\pi M$
in the complex plane, leading to different results for the integral.

In principle, however, 
there should be no ambiguity in carrying out the integral over the open string
modes, at least in the critical superstring theory where the tachyon potential is bounded 
from below, with the minimum describing  the perturbative vacuum\cite{0410103}. 
Since a particular D-instanton configuration represents a saddle point of the Euclidean
path integral, the integration over the open string modes need to be carried out
along the steepest descent contours (Lefschetz thimbles\cite{1001.2933,1009.6032,1206.6272,1511.05977,1601.03414,1802.10441}). 
The ambiguities we encounter here should be resolved once we
express the original integration contour over the string fields as a union of the
thimbles. 
At present this is not known since our
understanding of the configuration space of string fields is limited. However, we shall
make some guesses in \S\ref{spicard}.

In type IIB string theory, certain contribution to the effective action from the instanton
- anti-instanton sector can be predicted\cite{0510027,1404.2192,2008.02713}. 
It will be interesting to explore whether this can 
be reproduced using the techniques described above for constructing the multi-instanton
contribution to the closed string 
effective action. I wish to thank Michael Green for
raising this interesting question.

\subsection{Cutting rules}

We can now use the approach
 of \cite{1604.01783} to prove the 
 cutting rules in the closed string effective field theory, provided the effective
 action satisfies
 the reality condition. We can also translate these rules to the original Feynman diagrams
 of open closed string field theory as follows. Since the closed string effective field
 theory satisfies ordinary cutting rules, the cut does not pass through any
 interaction vertex. Therefore when we represent this as a cut diagram in the open closed
 string field theory, 
 all the D-instanton labels 
 of a given type must lie on one side of the cut. 
 Let $U$ denote the set of labels carried by the interaction vertices 
 on the left side of the cut and
 the complementary set $U^c$ denote the set of labels carried by the interaction
 vertices on the
 right side of the cut. Then the interaction vertices on the left of the cut 
 are of type $(i_1,\cdots,i_m)$ with 
 $i_1,\cdots, i_m\in U$ and the interaction vertices on the right of the cut 
 are of type $(j_1,\cdots, j_p)$ with
$j_1,\cdots, j_p\in U^c$. This is exactly the desired cutting rule discussed in
\S\ref{scut}. Furthermore, since 
every internal open string propagator of the original Feynman diagrams is
hidden inside the interaction vertices of the closed string effective field theory, an open
string propagator is never cut, in agreement with the fact that open string states are
not asymptotic states and should not appear as part of the states $|n\rangle$ in
\refb{e1}.
The momentum conservation for diagrams on either side of the cut follows from the
momentum conservation rules at the interaction vertices of closed string effective field
theory.
This allows us  to identify
the contributions from two sides of the cut 
as matrix elements of $T^\dagger$ and $T$ between closed string states.

\sectiono{One instanton amplitudes in two dimensional string theory} \label{sexamples}

In this section we shall verify the results on one instanton amplitude, described
in \S\ref{seffective}, 
in the context of two dimensional bosonic string
theory.

\subsection{Leading one instanton contribution}

The world-sheet theory of the two dimensional string
consists of a free scalar describing the time
coordinate and a $c=25$ Liouville theory describing the space coordinate, besides the
usual $b,c$ ghost system.  The physical closed string spectrum of this theory has a
single massless scalar field in two dimensions -- called the tachyon.  
In this theory the leading D-instanton contribution to the scattering amplitude for one 
incoming closed
string tachyon of energy $\tilde\omega_1=-\omega_1>0$ and $(n-1)$ 
outgoing closed string tachyons of energies
$\omega_2,\cdots, \omega_n>0$ was computed in \cite{1907.07688} as the product
of $n$ disk one point functions, with the result:
\be\label{e2d1}
T_{\rm 1-inst}=
\NN\, e^{-1/g_s} \, \prod_{i=1}^n \{2 \, \sinh(\pi|\omega_i|)\}\, 2\pi \delta\left(\sum_{i=1}^n
\omega_i\right)
\, .
\ee
The result \refb{e2d1} is valid for arbitrary signs of $\omega_i$, with the understanding that
negative $\omega_i$ denotes an incoming particle of energy $-\omega_i$. The value of the
constant $\NN$ will be described shortly.

We shall now write down an interaction term in
the closed string effective action that produces the term in the amplitude given in
\refb{e2d1}. 
Clearly there is no unique 
off-shell
continuation, but we shall write down the term that follows naturally from the open-closed
string field theory. It takes the form:
\be\label{eaction}
{1\over n!} \, \NN \, e^{-1/g_s}\, \int \prod_{i=1}^n
\left\{ {d\omega_i} \,dP_i \, 2\sinh(2\pi P_i)
\, \Phi_C(\omega_i, P_i) e^{-b (P_i^2-\omega_i^2/4)}
\right\} \, 2\, \pi \, \delta\left(\sum_{j=1}^n \omega_j\right)\, ,
\ee
where $\Phi_C(\omega,P)$ denotes the closed string tachyon field of energy
$\omega$ and Liouville
momentum $P$, in the convention in which the on-shell condition takes the form 
$P=|\omega|/2$. $b$ is a positive constant that is inherited from string field theory
definition of disk one point function of off-shell external states, 
\be \label{estub}
2\, \sinh(2\pi P) \, \exp[-b (P^2 - \omega^2/4)]\, ,
\ee
and is known as the
stub length. Note that $b$ has no effect on on-shell amplitudes, but acts as a
damping term for off-shell Euclidean momenta, thereby making integration over
internal momenta ultra-violet finite as long as the ends of the integration contour
for internal energies are pinned at $\pm i\infty$\cite{1604.01783}. 
The overall normalization of \refb{eaction}
follows from \refb{e2d1} and our discussion in \S\ref{seuclid}, In particular, since
\refb{e2d1} gives the expression for the T-matrix, there is no extra factor of $i$ or $-1$
in relating the normalization constants in \refb{e2d1} and \refb{eaction}.

The normalization constant $\NN$ was computed in \cite{1907.07688} from 
comparison of the
result with the matrix model, yielding the result:
\be\label{e2d2}
\NN = {i\over 8\pi^2}\, .
\ee
Note that since \cite{1907.07688} stated the result for the S-matrix, they 
had a normalization constant $-1/ (8\pi^2)$. 
Since in our convention the amplitude refers to the T-matrix, 
we have the factor
of $i$ in the expression for $\NN$. 
The presence of $i$ in \refb{e2d2} shows that the single D-instanton terms in the
effective action are imaginary.
It follows from our results in \S\ref{seffective} that
this will be the sole source of unitarity violation
for all single D-instanton induced amplitudes in two dimensional bosonic string theory. 

One could ask if it is possible to see the factor of $i$ directly in the string theory analysis.
A tentative answer is that this is the result of integration over the open string tachyon on
the D-instanton.\footnote{This is a somewhat
different point of view from the one given in 
footnote 2 of \cite{1907.07688}.} 
If we denote the open string tachyon field by $T$ and it mass$^2$ by $-M^2$,
then the integration over the tachyon mode in the Euclidean
path integral has to be performed by taking the integration contour to lie along
the imaginary axis, and generates a factor,
\be
\int_{\mp i\infty}^{\pm i\infty} dT \, e^{M^2 T^2} 
=\pm i \, {\sqrt \pi\over M}\, .
\ee
This factor of $i$ agrees with the one appearing in \refb{e2d1}.
However since we do not yet have a full derivation of $\NN$ in
string theory, this is not a definitive conclusion. 
It will be interesting to find a definitive
criteria for determining the (lack of) reality of $\NN$ for some given D-instanton in a
given string theory, since a complex $\NN$ would signal transition of closed strings
to other (D-brane) states in the theory.\footnote{Since $\NN$ reflects the ratio of
integration measures around the D-instanton and the perturbative vacuum, one could
in principle compute $\NN$ by regarding the D-instanton and the perturbative vacuum as
different classical solutions of the same underlying open closed string field theory.}

Since for real $\NN$ the term \refb{eaction} is real and there is no violation
of unitarity, we are led
to the conclusion that just by studying single D-instanton amplitude in
two dimensional string theory with the current level of understanding, we cannot 
come to a definitive conclusion that the theory is non-unitary -- we need 
comparison with the matrix model to determine that $\NN$ is imaginary. 
We shall test this in
\S\ref{sunit} by showing that the amplitudes satisfy cutting rules for real $\NN$.
In \S\ref{suniv} we shall discuss the effect of having complex $\NN$.

\subsection{Unitarity of one instanton amplitude for real $\NN$} \label{sunit}

Within the context of two dimensional string field theory,
assuming $\NN$ to be real does not violate any of the basic principles -- at least at
the level we are analyzing the theory.  
In order to illustrate the point that the violation of unitarity in two dimensional string
theory arises from the factor of $i$ in $\NN$, we shall now 
show how the one instanton contribution would have satisfied the correct cutting
rules if $\NN$ had been real. 
This will be based largely on the analysis of \cite{private,2003.12076}, 
but  for
the sake of completeness
we have included it here 
with the new viewpoint and appropriate factors of $i$.

We shall consider the case of a two point closed string
tachyon amplitude, with the incoming tachyon of energy 
$\tilde\omega_1=-\omega_1>0$ and an
outgoing tachyon of energy $\omega_2>0$. 
We shall focus
on a particular world-sheet contribution that involves the product of a disk one point 
function for the outgoing closed string tachyon of energy $\omega_2$ and annulus
one point function for the incoming closed string tachyon of energy $\tilde\omega_1
$.
The divergent part of this amplitude associated with closed string degeneration is
given by\cite{1907.07688}:
\ben\label{egim1}
&& \hskip -.3in  2\, \pi \, \delta(\tilde\omega_1-\omega_2)\, 
\NN\, e^{-1/g_s}\, g_s\, 
\sinh(\pi\,\omega_2)\, 2^{11/2}\, \pi\,  \int_0^\infty dP_1 \int_0^\infty dP_2 \ 
\CC(\tilde\omega_1/2, P_1, P_2)  \nonumber 
\\ && \hskip -.3in
\sinh(2\pi P_1)\, \sinh(2\pi P_2)\, \int_c^\infty ds \, \int_0^{1\over 4} dx\, 
 s^{1/2} \, \exp\left[-2\pi s \left\{ (1-2x)P_1^2 + 2x P_2^2 - x
\left({1\over 2}-x\right)\tilde\omega_1^2\right\}
\right]\, ,\nonumber\\
\een
where the lower limit $c$ of $s$ integration is some positive number and
$\CC(P_1,P_2,P_3)$ is the genus zero three point function of three Liouville
primaries carrying momenta $P_1$, $P_2$ and $P_3$\cite{9403141,9506136}. 
The factor of $1/2$ in the
first argument of $\CC$ arises due to the fact that in the normalization convention of 
\cite{1705.07151}, 
which we are using, the on-shell condition for a closed string tachyon of energy
$\omega$ and momentum $P$ is $\omega=2\, P$. 
As discussed in \S\ref{seffective}, like all string theory amplitudes this amplitude is
formally real\footnote{The relevant test is not reality but the hermiticity of an
amplitude, which requires us to 
test invariance under complex conjugation, together with
exchange of the incoming and the outgoing
states, or equivalently change of the signs of all external momenta in the 
convention that an incoming particle of momentum $p$ is regarded as an
outgoing particle of momentum $-p$. 
In \refb{egim1} the exchange of the incoming and the outgoing states 
simply exchanges $\tilde\omega_1$ and $\omega_2$. This has
no effect since $\tilde\omega_1=\omega_2$ after using energy conservation.}  
for real $\NN$ and the possible imaginary parts would have to arise from
the process of making sense of the divergences. In this case the relevant divergences 
arise from the $s\to\infty$ limit.
By making change of variables
\be 
t_1 = 2\, \pi\, s\, (1-2\, x), \quad t_2 = 4\pi \, s\,  x\, .
\ee
and symmetrizing the integral under $t_1\leftrightarrow t_2$, $P_1\leftrightarrow P_2$ exchange, we can
rewrite this as\cite{2003.12076}
\ben \label{egim2}
&&  2\, \pi \, \delta(\tilde\omega_1-\omega_2)\, 4\, \NN\, e^{-1/g_s}\, g_s\, 
\sinh(\pi\omega_2)\, \pi^{-1/2}\,  \int_0^\infty dP_1 \int_0^\infty dP_2 \
\CC(\tilde\omega_1/2, P_1, P_2) \nonumber 
\\ && 
\sinh(2\pi P_1)\, \sinh(2\pi P_2)\, \int_{2\pi c \le t_1+t_2<\infty} dt_1 \, dt_2\, 
 (t_1+t_2)^{-1/2} \, \exp\left[-t_1 \, P_1^2 -t_2\,  P_2^2 +{t_1 t_2\over t_1+t_2} \,
 {\tilde\omega_1^2\over 4}
\right]\, .\nonumber\\
\een
Now it is easy to see that for $\tilde\omega_1\ge 2(P_1+P_2)$, 
the integration over $t_1,t_2$
diverges for $t_1,t_2\to\infty$, but there is no divergence when any one of them 
remains finite even if the other one approaches $\infty$. Using this, we can
express the divergent part of \refb{egim2} as:
\ben \label{egim25}
&&  2\, \pi \, \delta(\tilde\omega_1-\omega_2)\, 4\, \NN\, e^{-1/g_s}\, g_s\, 
\sinh(\pi\omega_2)\, \pi^{-1/2}\,  \int_0^\infty dP_1 \int_0^\infty dP_2 \
\CC(\tilde\omega_1/2, P_1, P_2) \nonumber 
\\ && \sinh(2\pi P_1)\, \sinh(2\pi P_2)\, 
\int_b^\infty dt_1 \, \int_b^\infty dt_2\, 
 (t_1+t_2)^{-1/2} \, \exp\left[-t_1 \, P_1^2 -t_2\,  P_2^2 +{t_1 t_2\over t_1+t_2} \,
 {\tilde\omega_1^2\over 4}
\right]\, ,\nonumber \\
\een
for some positive constant $b$. The difference between \refb{egim25} and
\refb{egim2} is real and finite for real $\NN$ and will not be discussed further.
Using the results of \cite{1607.06500}, \refb{egim25} can be rewritten as:
\ben\label{egim3}
&&
-{1\over 2}\, i\,  \NN\, e^{-1/g_s}\, g_s\, 
2\, \sinh(\pi\omega_2)
\,   \int_0^{\infty} dP_1 \int_0^{\infty} dP_2 \, 
C(\tilde\omega_1/2, 
P_1, P_2) \, 2\, \sinh(2\pi P_1)\, 2\, \sinh(2\pi P_2) \nonumber \\ && \hskip .2in
\, \int {d\wt\omega\over 2\pi} \, {1\over -{1\over 4} (\tilde\omega_1-\wt\omega)^2 + P_1^2-i\eps}
\, {1\over -{1\over 4} \wt\omega^2 + P_2^2-i\eps} \exp\left[-b (P_1^2+P_2^2)
+{b\over 4}\wt\omega^2 + {b\over 4} (\tilde\omega_1-\wt\omega)^2\right]\nonumber \\\cr
&&\hskip 1in \times \ 2\, \pi \, \delta(\tilde\omega_1-\omega_2)\, . \een

\begin{figure}
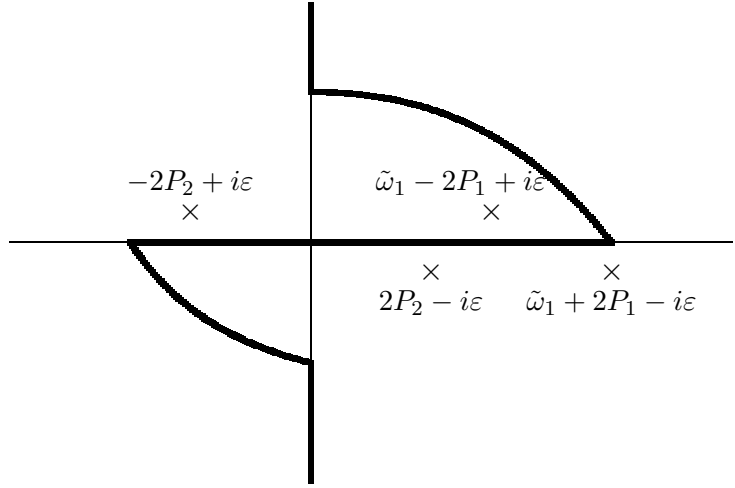

\begin{center}
\figcontour

\end{center}

\caption{The thick contour in this
figure shows the choice of $\tilde\omega$ integration contour in
\refb{egim3}. The $\times$'s mark the locations of the poles in the complex $\tilde\omega$
plane.
\label{figcontour}
}
\end{figure}

\noindent As shown in Fig.~\ref{figcontour}, 
the $\wt\omega$ integration contour runs along the real axis in the region where the
poles are situated, but then returns to $\pm i\infty$ by remaining at large distances
in the first and the third quadrant. This makes the integral finite due to the exponential
suppression factor.
Note that the equality of \refb{egim25} and \refb{egim3} 
is an identity for $\tilde\omega_1^2<0$, which can be proved by using the Schwinger
parameter representation for the two propagators in \refb{egim3} and then carrying out
the integration over $\wt\omega$ by deforming the integration contour to lie along the
imaginary axis.
 For $\tilde\omega_1^2>0$, \refb{egim25} diverges due to the
divergence in the $t_i$ integral for large $t_1,t_2$ but \refb{egim3} is finite.
Therefore \refb{egim3} is the analytic continuation of \refb{egim25} for $\tilde\omega_1^2>0$.

We can now compute the imaginary part of \refb{egim3} by deforming the $\wt\omega$
integration contour to lie along the imaginary axis, and picking up the residue from the
pole at $\tilde\omega_1-2P_1+i\ve$ 
that it crosses on the way. The integral along the imaginary axis gives real
result, since the effect of complex conjugation can be compensated by a change
in the sign of the integration variable. The residue at the pole however gives a complex
contribution to the final result. The final result for $2\, \times$ 
the imaginary part, assuming that $\NN$ is
real, is given by\cite{2003.12076}:
\ben \label{egim22pre}
&& 2\, i\, \NN\, e^{-1/g_s}\, g_s \, 
2\, \sinh(\pi\omega_2)  \int_0^\infty d\omega_3 \int_0^\infty d\omega_4 \
\CC(\tilde\omega_1/2, \omega_3/2, \omega_4/2) \, 
{2\, \pi}\,  \delta\left(\omega_3+\omega_4-{\tilde\omega_1}\right)
\nonumber \\ && \hskip 1in \, (2\omega_3)^{-1} \, (2\omega_4)^{-1}\, 
\{2\, \sinh(\pi \omega_3)\}\, \{2\, \sinh(\pi \omega_4)\} \, 2\, \pi \, \delta(\tilde\omega_1-\omega_2) 
\, . 
 \een

\begin{figure}
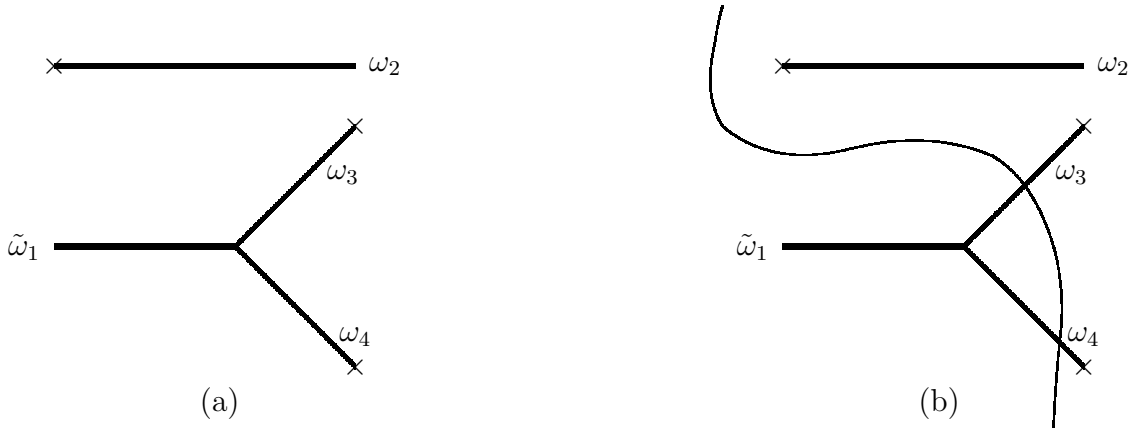

\begin{center}
\hbox{\figthreenew \hskip 1.5in \figthree}
\vskip -.6in

\hbox{\hskip 1.5in (a) \hskip 3.5in (b)}
\end{center}

\caption{The figure on the left shows a
Feynman diagram of open closed string field theory,
describing part of the contribution to 
the product of an annulus one point
function and a disk one point function. A $\times$ denotes the one point closed string
interaction vertex associated with one point function on the disk and a free end
marks an external state. All energies flow from the left to the right.
The figure on the right is a cut of this diagram, with the
thin curve denoting the
cut, 
drawn such that all the D-instanton type interaction vertices are to the right of the cut.
The corresponding Feynman diagrams in the closed string effective field theory are obtained
by collapsing all the $\times$'s into a single composite vertex. These have been shown in
Fig.~\ref{fig12}.
\label{fig3}
}
\end{figure}

\begin{figure}
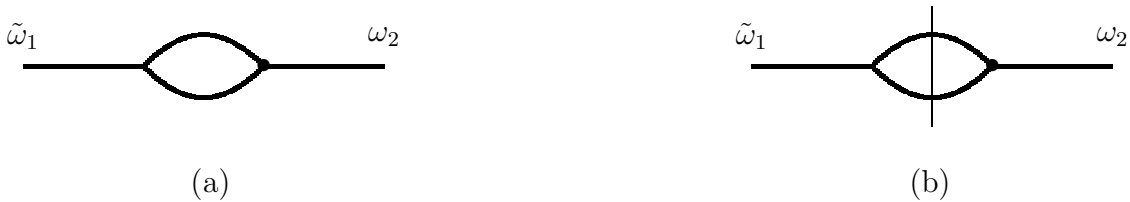

\begin{center}
\hbox{\figtwelvea \hskip 1.5in \figtwelve}
\vskip -.8in

\hbox{\hskip 1.3in (a) \hskip 3.5in (b)}
\end{center}

\vskip -.2in

\caption{The figure shows the representations of Fig.~\ref{fig3}(a) and (b) 
as Feynman diagrams of
closed string effective field theory. $\bullet$ denotes a D-instanton type interaction
vertex of the effective field theory.
\label{fig12}
}
\end{figure}

We shall now give the Feynman diagram interpretation of \refb{egim3} and
\refb{egim22pre} in the open closed string field theory.
\refb{egim3} can be interpreted as the contribution from the Feynman diagram
shown in Fig.~\ref{fig3}(a). The $\times$ denotes a disk one
point function and produces the factors of $2\sinh(2\pi P) \exp[-b(P^2-\omega^2/4)]$,
given in \refb{estub}, 
from the one
point function of the Liouville primary of momentum $P$ and energy $\omega$
on the disk. 
$\omega_2$ and $\tilde\omega_1$ label the outgoing and incoming energies of
the original Feynman diagram and $\omega_3=\tilde\omega_1-\wt\omega$ and 
$\omega_4=\wt\omega$  label
internal energies,
all flowing from left to right. 
$C(\tilde\omega_1/2, 
P_1, P_2)$ represents the three point interaction vertex of three closed strings.
For off-shell external states, 
we could include additional damping factors in this vertex by replacing $b$ by $b'>b$ in
\refb{egim25} and \refb{egim3}, but have chosen not to do so.
The two factors in the second line of \refb{egim3} represent the
propagators of the internal lines in Fig~.\ref{fig3}(a), carrying energies
$\omega_3=\tilde\omega_1-\wt\omega$ and 
$\omega_4=\wt\omega$. 
Integration over the
zero mode $\phi$, labelling the position of the D-instanton along the Euclidean
time,   produces
the energy conserving delta function $2\pi \delta(\tilde\omega_1-
\omega_2)$
after  Wick rotation. The factors of
$i$ can be understood as follows. If we were computing the S-matrix element, then
we would get factor of $-i$ from each of the two propagators, a factor of $i$ from the
closed string three point vertex and a factor of $i$ from the D-instanton induced vertex
of the effective field theory, obtained by merging the three $\times$ in the diagram.  
The product of these factors give 1.
Note that we do not have separate $i$ from each $\times$ -- this is part of the Feynman
rule of the theory since we integrate out the open string zero modes in the Euclidean
space where there are no $i$'s in the interaction vertices and then rotate the effective
field theory action to the Lorentzian space. To compute the T-matrix element we need
to multiply the result by a factor of $-i$. This gives the overall multiplicative 
factor of $-i$ in \refb{egim3}. 
The factor of $1/2$ in the front is a combinatorial factor due to having two
identical internal lines carrying energies $\omega_3$ and $\omega_4$. This may be
clearer from Fig.~\ref{fig12}(a) that gives the representation of the same term as
a Feynman diagram in the closed string effective field theory.

Let us now turn to \refb{egim22pre}  and
show that for real $\NN$, it
can be interpreted as the contribution to
$T-T^\dagger$ from the
cut diagram shown in Fig.~\ref{fig3}(b) in accordance with \refb{e1}. 
According to the
cutting rule, to compute the difference between \refb{egim3} and its hermitian
conjugate, we simply have to replace the  propagators carrying energies $\omega_3$
and $\omega_4$ by cut propagators,
\be\label{ereplacecut}
 -{i\over -{1\over 4} \omega_i^2 + P^2-i\eps} \quad \Rightarrow  \quad
 2\, \pi\, \delta\left(-{1\over 4} \omega_i^2 + P^2\right)\, , \qquad i=3,4,
 \ee
and take hermitian conjugate of the diagram on the right of the cut.
Since for real $\NN$ the diagram on the right of the cut is hermitian, taking its
hermitian conjugate has no effect. This gives the contribution to $T-T^\dagger$ according
to cutting rules to be:
\ben\label{egim3new}
&&
-{i\over 2}\,  \NN\, e^{-1/g_s}\, g_s\, 
2\, \sinh(\pi\omega_2)
\,   \int_0^{\infty} dP_1 \int_0^{\infty} dP_2 \, 
C(\omega/2, P_1, P_2) \, 2\, \sinh(2\pi P_1)\, 2\, \sinh(2\pi P_2) \nonumber \\ && \hskip .2in
\, \int {d\wt\omega\over 2\pi} \, 2\pi i \, \delta\left(-{1\over 4} (\tilde\omega_1-\wt\omega)^2 + P_1^2\right)
\, 2\pi i \delta\left(-{1\over 4} \wt\omega^2 + P_2^2\right)  2\, \pi \, \delta(\tilde\omega_1-\omega_2)\, . \een
It is straightforward to verify that this
precisely reproduces \refb{egim22pre}.
Physically the cut diagram is given by the product of a genus zero
contribution to closed string $1\to 2$ amplitude and the hermitian
conjugate of the leading D-instanton contribution
to the closed string $1\to 2$ amplitude, given by the product of three disk one point
functions. Fig.~\ref{fig12}(b) shows the representation of the same cut diagram in the
closed string effective field theory.

\subsection{Unitarity violation in the matrix model} \label{suniv}

We now turn to the case when $\NN$ is imaginary as given in \refb{e2d2}, and explore
whether the analog of \refb{e1} can still give us some useful information. For this let us first
rewrite \refb{e1} is a way that will be convenient to generalize to the case when $\NN$
is not real. Let $T_k$ denote the contribution to $T$ from the $k$ instanton sector, 
containing terms proportional to $e^{- k/g_s}$. Then for hermitian $T$,
comparing  terms proportional to $e^{-1/g_s}$ on two sides of \refb{e1}, we get:
\be\label{e1ag}
\langle f|T_1|i\rangle - \langle f|T_1^\dagger|i\rangle    =
 i \sum_n \langle f|T_1^\dagger |n\rangle \langle n| T_0|i\rangle
 +  i \sum_n \langle f|T_0^\dagger |n\rangle \langle n| T_1|i\rangle\, .
\ee
Now when $\NN$ is imaginary, then we should expect an analog of
\refb{e1ag} with $T_1$ replaced by $iT_1$ to hold, since up to order $e^{-1/g_s}$,
$T_0+iT_1$ will represent a T-matrix constructed from real effective
action in which $\NN$ has been replaced by $i\NN$. Therefore we now expect a
relation of the form:
\be \label{eright}
\langle f|iT_1|i\rangle - \langle f|(iT_1)^\dagger|i\rangle    =
 i \sum_n \langle f|(iT_1)^\dagger |n\rangle \langle n| T_0|i\rangle
 +  i \sum_n \langle f|T_0^\dagger |n\rangle \langle n| (iT_1)|i\rangle\, .
\ee
Due to the extra factors of $i$, this relation is no longer useful for proving unitarity
of the closed string amplitudes. Nevertheless, as we shall discuss now, this identity can
still be used to extract useful information on T-matrix elements.

The first term on the right hand side of \refb{eright}
represents the contribution from the cut diagram Fig.~\ref{fig3}(a) with
$\NN$ replaced by $i\NN$. The second term on the right hand side
represents the contribution from
another diagram, obtained by mirror reflecting Fig.~\ref{fig3} about the
vertical axis and  exchanging $\omega_2$ and $\tilde\omega_1$. This gives the same
contribution as the first term. Therefore the right hand side of \refb{eright} is given
by twice the term given in \refb{egim22pre}, with $\NN$ replaced by $i\,\NN$.
After cancelling the $i$ from two sides of \refb{eright}, we get the matrix element of
$T_1+T_1^\dagger$ for $1\to 1$ scattering:
\ben\label{eright1}
(T_1+T_1^\dagger)  
&= & 2\, \pi \, \delta(\tilde\omega_1-\omega_2) \, 4\, i\, \NN\, e^{-1/g_s} \, g_s\,  2\, \sinh(\pi\omega_2)  \int_0^\infty d\omega_3 \int_0^\infty d\omega_4 \
\CC(\tilde\omega_1/2, \omega_3/2, \omega_4/2)
\nonumber \\ && \hskip 1in {2\, \pi}\,  \delta\left(\omega_3+\omega_4-{\tilde\omega_1}\right)
\, (2\omega_3)^{-1} \, (2\omega_4)^{-1}\, 
\{2\, \sinh(\pi \omega_3)\}\, \{2\, \sinh(\pi \omega_4)\} \nonumber \\ 
&=&2\, \pi \, \delta(\tilde\omega_1-\omega_2) \,
 i\, \NN\, e^{-1/g_s} \, g_s\,  2\, \sinh(\pi\omega_2)  \int_0^\infty d\omega_3 \int_0^\infty d\omega_4 \
\tilde\omega_1 \, 
{2\, \pi}\,  \delta\left(\omega_3+\omega_4-{\tilde\omega_1}\right)
\nonumber \\ && \hskip 2in \, \, 
\{2\, \sinh(\pi \omega_3)\}\, \{2\, \sinh(\pi \omega_4)\}\, ,
 \een
where we used $\CC(\tilde\omega_1/2, \omega_3/2, \omega_4/2)=
\tilde\omega_1\omega_3\omega_4$ when $\tilde\omega_1
=\omega_3+\omega_4$\cite{1705.07151}.
Using $S=iT$, we now see that the matrix element of $S_1-S_1^\dagger$ for
$1\to 1$ scattering is given by:
 \ben
 (S_1-S_1^\dagger)  
&= & -2\, \pi \, \delta(\tilde\omega_1-\omega_2) \, \NN\, e^{-1/g_s} \, g_s
\,  2\, \sinh(\pi\omega_2)  \int_0^\infty d\omega_3 \int_0^\infty d\omega_4 \
\tilde\omega_1 \, 
{2\, \pi}\,  \delta\left(\omega_3+\omega_4-{\tilde\omega_1}\right)
\nonumber \\ && \hskip 2in \, 
\{2\, \sinh(\pi \omega_3)\}\, \{2\, \sinh(\pi \omega_4)\} \nonumber \\ 
&=& -2\, {i\over \pi} \, \delta(\tilde\omega_1-\omega_2) \, e^{-1/g_s} \, g_s\, 
\tilde\omega_1\, \sinh^2(\pi\tilde\omega_1)\, 
\left\{ {\pi \, \tilde\omega_1\over \tanh(\pi\tilde\omega_1)} -1\right\}\, , 
 \een
where in the last step we have used \refb{e2d2} and explicitly carried out the
integration over $\omega_3$ and $\omega_4$.
This agrees with the dual matrix model result given in \cite{1907.07688}. 
Similar agreement
has been found for all $1\to n$ scattering amplitudes\cite{2003.12076}.

\sectiono{Two instanton amplitudes in two dimensional string theory} \label{sexamplemulti}

In this section we shall examine the source of
unitarity violation in 2-instanton amplitudes in 
two dimensional string theory and show that the results are consistent
with the general conclusions of \S\ref{smulti}. In \S\ref{s6.1} we shall follow \cite{1912.07170} 
and evaluate the
leading 2-instanton contribution to the two tachyon amplitude in two dimensional
string theory using the Lorentzian prescription for the integration contour. However we
shall not assume the value of $\NN$, allowing it to be an arbitrary complex number.
The amplitude will turn out to have an imaginary part even for real $\NN$.
In \S\ref{stwo} we shall analyze whether this imaginary part is consistent with the
cutting rules. For this we analyze 
the same amplitude as a sum of Feynman
diagrams in open closed string field theory, as well as in the closed string effective
field theory. We shall find that while a part of the imaginary contribution 
can be traced to the
contribution from a cut diagram, the rest of the contribution to the imaginary
part cannot be regarded as coming from a cut diagram. 
However the difference can be accounted for by the contribution from the
imaginary part of the closed string effective action originating in the 2-instanton sector.
This shows violation of unitarity in the theory even when $\NN$ is real. Furthermore, the
imaginary part of the 2-instanton 
effective action can be traced directly to the $i\eps$ in the Lorentzian prescription, 
needed for dealing with the  extra
open string tachyon modes that appear on the D-instanton  system when the 
two D-instantons come close. In \S\ref{sunires} we show that if we had used the
unitary prescription for evaluating the amplitude then for real $\NN$, the imaginary
part of the amplitude would have been consistent with the cutting rules, thereby
showing that we get a unitary amplitude. We also show that in this case the
two instanton contribution to the closed string effective action is real. 
In \S\ref{smattwo} we compare our
conventions with those in 
\cite{1912.07170}, and 
verify that what we call the Lorentzian prescription indeed agrees with that given in
\cite{1912.07170} when we put all the factors of $i$ in the proper places and choose
$\NN$ to be $i/(8\pi^2)$.
We then show that if we had used the unitary prescription for computing the two
instanton contribution in this theory, the agreement between the string theory results
and the matrix model results found in \cite{1912.07170} continues to hold, provided we
choose a different normalization constant for the (2,1) instanton amplitude.
In \S\ref{sthree} we extend this analysis to the three instanton sector.

\subsection{Leading two instanton contribution to the closed string two point function}
\label{s6.1}

In this subsection we shall compute the leading two instanton contribution to the two
point function of closed string tachyon following \cite{1912.07170}, 
with an incoming closed string of energy $\tilde
\omega_1=-\omega_1>0$ and an outgoing closed string of energy $\omega_2>0$.
In order to understand the analytic properties of various results as function of
$\omega_1,\omega_2$, we shall express our result in a way that will be valid
for both signs of the $\omega_i$'s.
There will be an overall normalization factor of $\NN^2  e^{-2/g_s}$, with the $\NN^2$
factor containing the effect of annulus amplitude with both boundaries lying on the same
D-instanton. However we need to also include the effect of the annulus with the two
boundaries lying on different D-instantons. This is given by\cite{1912.07170}:
\be \label{efulla12}
\AAA_{12}=
\int_0^\infty dt \, t^{-1} \left[ e^{-t\{(\phi_{(1)}-\phi_{(2)})^2 -4\pi^2\}/2\pi} - e^{-t\{\phi_{(1)}-\phi_{(2)}\}^2 /2\pi}
\right] = \ln \, {\{\phi_{(1)}-\phi_{(2)}\}^2 \over \{\phi_{(1)}-\phi_{(2)}\}^2 -4\pi^2}\, .
\ee
This includes the contribution to the partition function from open strings stretching from
instanton 1 to instanton 2 and  from open strings stretching from
instanton 2 to instanton 1 -- otherwise there would have been a factor of $1/2$
multiplying this expression. Note the subtle difference betwwen $\AAA_{12}$ and 
$A_{12}$ defined in \refb{ea12}. In the latter, the lower limit of integration is a constant
$a>0$ since we subtract the contribution given by exchange of closed strings 
between a pair of disks in the underlying open closed string field theory. In contrast
$\AAA_{12}$ represents the full annulus contribution. $\AAA_{12}$ is the relevant 
contribution while computing the full amplitude, whereas $A_{12}$ is the relevant
contribution for computing the interaction vertex of the closed string effective field
theory.

\begin{figure}
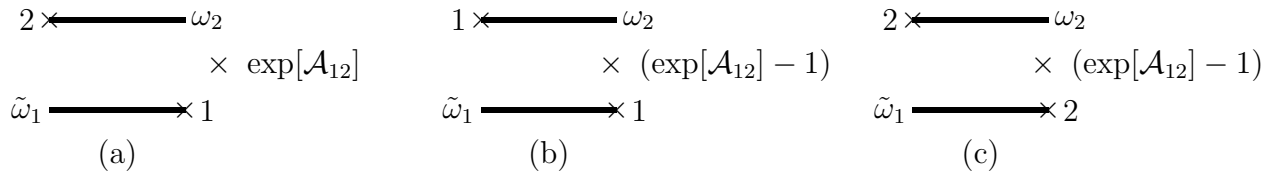

\begin{center}
\hbox{\hskip -.5in \figeight\qquad \fignine\qquad \figten}
\end{center}

\vskip -.8in

\caption{The leading order two instanton contribution to the two point function of closed
strings. As usual $\times$ denotes the disk one point function.
$\AAA_{12}$ denotes the full annulus contribution with the two boundaries
lying on two different instantons. We also have a diagram
in which 1 and 2 are exchanged in Fig.(a).
The $-1$ in the last two terms ensures that the
underlying world-sheet has at least one boundary of each type. 
\label{fig8}
}
\end{figure}

The leading 2-instanton 
contribution to the closed  string two point function will come from the product of 
$\NN^2 e^{-2/g_s} \exp[\AAA_{12}]$ times the product of two disk one point functions. 
However, if the boundaries of both the disks lie on the same D-instanton, then the
multiplicative factor is $\NN^2 e^{-2/g_s} (\exp[\AAA_{12}]-1)$, so that after expanding
the exponential, each term represents a world-sheet diagram that has each type of
boundary appearing at least once.
These contributions have been shown in Fig.~\ref{fig8}. Even though they look like 
Feynman diagrams of open closed string field theory, they should be interpreted as
the full world-sheet contribution to the amplitude since we use $\AAA_{12}$ instead
of $A_{12}$ in the exponent. For example,
Fig.~\ref{fig8}(a) represents an amplitude where the incoming closed string vertex
operator
is inserted at the center of the disk whose boundary lies on instanton 1 and
the outgoing closed string vertex
operator
is inserted at the center of the disk whose boundary lies on instanton 2. This
amplitude is multiplied by $\exp[\AAA_{12}]$. There is a similar diagram (not shown
in the figure) where the
boundary conditions of the two disks are interchanged.  They give identical contributions and
their total contribution to the amplitude in the \underline{Euclidean theory} is
given by:
\ben\label{e6.1}
&&\hskip -.15in \NN^2 \, e^{-2/g_s}\, 
\int \, d\phi_{(1)} \, d\phi_{(2)} \, e^{-i \omega^E_1 \phi_{(1)} - i \omega^E_2 \phi_{(2)}}\, 
2\, \sinh(2\pi P_1)\, 2\, \sinh(2\pi P_2) 
\exp\left[ \AAA_{12} \right] \nonumber \\
&& \hskip-.3in =  \NN^2 \, e^{-2/g_s}\, 4\, \sinh(2\pi P_1)\, 
\sinh(2\pi P_2)\, \int \, d\phi_{(1)} \, d\phi_{(2)} \, e^{-i \omega^E_1 \phi_{(1)} 
- i \omega^E_2 \phi_{(2)}}\, 
\left[ 1 + { 4\pi^2\over 
(\phi_{(1)}-\phi_{(2)})^2 -4\pi^2}\right]\, . \nonumber \\
\een
Note that even though we have a sum of two diagrams, we do not have a factor
of 2. This is
due to the fact that the two D-instantons are
identical and therefore exchanging them does not give a new contribution as long
as we integrate $\phi_{(1)}$ and $\phi_{(2)}$ over the entire real line. Since we
are working in the Euclidean theory, there are no factors of $i$ in the overall 
normalization. On the right hand side of \refb{e6.1}, 
the contribution of the first term inside the square
bracket has separate factors of $\delta(\omega_1^E)$ and $\delta(\omega_2^E)$, 
and
corresponds to 
a disconnected diagram in the closed string effective field theory with a pair of
D-instanton type interaction vertices. We shall not consider this contribution any further.

Fig.~\ref{fig8}(b) represents an amplitude where the incoming closed string vertex
operator
is inserted at the center of the disk whose boundary lies on instanton 1 and
the outgoing closed string vertex
operator
is inserted at the center of the disk whose boundary also lies on instanton 1.
Fig.~\ref{fig8}(c) represents a similar contribution where the boundaries of both disks
lie on instanton 2. This is related to Fig.~\ref{fig8}(b) by the exchange of the positions
of the two D-instantons and give the same result after integration over the positions.
Therefore we shall combine the two contributions. The result takes the form:
\ben\label{e6.2}
&&\NN^2 \, e^{-2/g_s}\, 
\int \, d\phi_{(1)} \, d\phi_{(2)} \, e^{-i \omega^E_1 \phi_{(1)} - i \omega^E_2 \phi_{(1)}}\, 
2\, \sinh(2\pi P_1)\, 2\, \sinh(2\pi P_2)
\left\{\exp\left[\AAA_{12}
\right]-1\right\} \nonumber \\
&=& \NN^2 \, e^{-2/g_s}\, 4\, \sinh(2\pi P_1)\, 
\sinh(2\pi P_2)\, \int \, d\phi_{(1)} \, d\phi_{(2)} \, e^{-i \omega^E_1 \phi_{(1)} 
- i \omega^E_2 \phi_{(1)}}\, 
 { 4\pi^2\over 
(\phi_{(1)}-\phi_{(2)})^2 -4\pi^2}\, . \nonumber \\
\een
Again we see that despite having two contributions there is no factor of two, since 
exchanging two identical D-instantons does not give a new contribution as long as
we integrate $\phi_{(1)}$ and $\phi_{(2)}$ individually over the entire range.

Introducing new variables $\phi$ and $\chi$ via,
\be 
\phi_{(1)} = \phi + {1\over 2}\chi, \qquad \phi_{(2)} = \phi -{1\over 2} \chi\, ,
\ee
and using the Lorentzian prescription \refb{eres1}
for dealing with the singularities at $\chi=2\pi$,
we can express the sum of \refb{e6.1} and \refb{e6.2}, leaving out the first term
inside the square bracket of \refb{e6.1}, as
\ben\label{e6.5}
&& \hskip -.1in
\NN^2 \, e^{-2/g_s}\, 4\, \sinh(2\pi P_1)\, 
\sinh(2\pi P_2)\, \int \,d\phi \, d\chi\, e^{-i\phi(\omega^E_1+\omega^E_2)}\, 
{4\pi^2\over \chi^2 -4\pi^2 +i\eps}\, \nonumber \\ &&
\hskip 3in \times \, \left[e^{-i\chi(\omega^E_1-\omega^E_2)/2}
+ e^{-i\chi(\omega^E_1+\omega^E_2)/2}\right]
\nonumber \\
&& \hskip -.25in = \NN^2 \, e^{-2/g_s}\, 16\, \pi^2 \, \sinh(2\pi P_1)\, 
\sinh(2\pi P_2)\, 2\pi \delta(\omega^E_1+\omega^E_2) \int \, {d\chi\over \chi^2-4\pi^2+i\eps}\, \left[e^{-i\chi(\omega^E_1-\omega^E_2)/2}
+1\right] \, . \nonumber \\
\een
We shall now analytically continue $\omega^E_i$ to the Lorentzian energies $\omega_i
=i\omega^E_i$ by replacing $\omega^E_i$ by $e^{-i\theta}\omega_i$ with real
$\omega_i$ and deforming
$\theta$ from 0 to $\pi/2$. This can be done by simultaneously deforming the $\chi$
integration contour as $\chi = e^{i\theta} u$ with $u$ real, so that 
$e^{\pm i\omega_i^E\chi}$
remain bounded along the integration contour and the integral converges for all
$\theta$. At $\theta=\pi/2$ we have,
\be\label{e6.6repl}
\omega_i^E = - i\, \omega_i, \qquad \chi = i \, u\, .
\ee
Now as discussed in \S\ref{seuclid} under this deformation the amplitudes in the
Euclidean theory directly get mapped to the T-matrix elements without any extra factor
of $i$. Therefore we have:
\ben \label{e6.7}
T_{\rm 2-inst} &=&i \, \NN^2 \, e^{-2/g_s}\, 16\, \pi^2
\, \sinh(\pi|\omega_1|)\, \sinh(\pi 
|\omega_2|)\, 2\pi\delta(\omega_1+\omega_2) \nonumber \\ 
&& \int\,  du \,
{1\over -u^2 -4\pi^2} \left[ e^{i\omega_2 u} +1\right]\nonumber \\
&=&  -i\, \NN^2 \, e^{-2/g_s}\, 8\, \pi^2
\,  \sinh^2(\pi \omega_2)\, 
2\pi\delta(\omega_1+\omega_2) \, \left(1 + e^{-2\pi|\omega_2|}\right)\, ,
\een
where we have used the relation 
$\omega_1=-\omega_2$ and the
on-shell condition $|\omega_i|=2P_i$ to simplify the coefficient of 
$\delta(\omega_1+\omega_2)$. 
As will be discussed in \S\ref{smattwo}, 
this
agrees with the result of \cite{1912.07170} after appropriate change of notation.

\begin{figure}
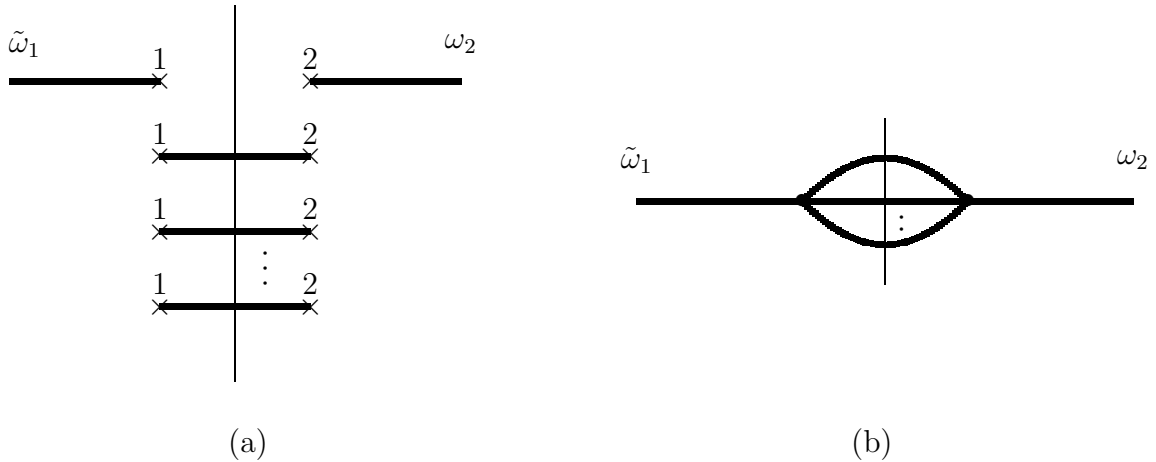

\begin{center}
\hbox{ \figeleven \qquad \figthirteen}
\end{center}

\vskip -1in

\hbox{\hskip 1.7in (a) \hskip 3in (b)}

\caption{(a) A cut diagram in open closed string field theory 
that can contribute to the imaginary part of the leading two 
instanton contribution to $1\to 1$ closed string amplitude. The thin vertical line
represents the cut. (b) The same cut diagram represented in the closed string
effective field theory, with the $\bullet$'s representing the interaction vertices of
this effective field theory.
\label{fig89}
}
\end{figure}

Before we proceed, we make a few remarks on the result given in \refb{e6.7}:
\begin{enumerate}
\item 
We can also obtain \refb{e6.7} by first evaluating the $\chi$ integration in 
\refb{e6.5} and then doing analytic
continuation, keeping in mind that the analytic continuation needs to be done via 
the second and the fourth quadrants of the $\omega_i^E$ plane, or equivalently, the first
and the third quadrants of the $\omega_i$ plane. Therefore positive (negative) $\omega_i^E$
analytically continues to positive (negative) $\omega_i$.
\item
Appearance of $|\omega_2|$ in \refb{e6.7} shows that the
amplitude is non-analytic at $\omega_2=0$. 
This singularity
can be traced to possible divergences from the large $|\chi|$ region in \refb{e6.5} 
-- although the integral is
finite, sufficient number of derivatives with respect to $\omega_2^E$ will generate
powers of $\chi$ in the numerator and make the integral diverge at $\omega_i^E=0$.
If on the left hand sides of \refb{e6.1} and \refb{e6.2} 
we had performed the $\chi$ integration before doing the $t$ integration in the
expression for 
$\AAA_{12}$ given in \refb{efulla12}, then the divergence would come from the $t\simeq 0$
region, associated with closed string degeneration.
In string field theory this non-analyticity can be traced to the singularities
associated with the cut diagrams shown in Fig.~\ref{fig89}. 
\item 
In contrast, we shall see in \S\ref{stwo}
that the contribution to the closed string effective action, given in
\refb{e6.14}, is an analytic function of $\omega_2$ in the complex $\omega_2$
plane since the integrand in the $\chi$ integration
will have exponential suppression from the large $|\chi|$
region. This can be traced to the fact that in the construction of the effective action
we shall use $A_{12}$ defined in \refb{ea12} where 
the $t$ integral has a lower cut-off. 
\item
From \refb{e6.7} we see that even if $\NN$ is taken to be real, as is needed for
unitarity of the one instanton amplitude, 
the 2-instanton amplitude acquires an imaginary part. This by itself does not signal
violation of unitarity -- we need to check if the imaginary part arises from the cutting
rules. 
We shall see in \S\ref{stwo} that part of \refb{e6.7} does come from sum over
cuts of the Feynman diagrams of string field theory shown in Fig.~\ref{fig89}.
The rest can be traced to the $i\eps$ factor in \refb{e6.5} that produces
an imaginary part of the two instanton interaction vertex in
closed string effective field theory.
This is turn will show that the amplitude violates unitarity.
\end{enumerate}

Let us now compute the same amplitude using the unitary prescription. This is done by
using the identify \refb{eprin} to express the amplitude as a sum of \refb{e6.5} and an
additional contribution proportional to $\delta(\chi^2 - 4\pi^2)$ in the integrand. For these
extra terms, we can 
evaluate the $\chi$ integral using this $\delta$-function and then replace $\omega_i^E$
by $-i\omega_i$ as given in \refb{e6.6repl}. This gives, for $\omega_2>0$:
\be
i\, \NN^2 \, e^{-2/g_s}\, 4\, \pi^2
\,  \sinh^2(\pi \omega_2)\, 
2\pi\delta(\omega_1+\omega_2) \, \left(2+e^{-2\pi\omega_2}+e^{2\pi\omega_2}
\right)\, .
\ee
Adding this to \refb{e6.7} we get the amplitude in the unitary prescription:
\be \label{e6.7prime}
T'_{\rm 2-inst} =  i\, \NN^2 \, e^{-2/g_s}\, 4\, \pi^2
\,  \sinh^2(\pi \omega_2)\, 
2\pi\delta(\omega_1+\omega_2) \, \left(e^{2\pi|\omega_2|}- e^{-2\pi|\omega_2|}\right)\, .
\ee
We shall check in \S\ref{sunires} that this imaginary 
contribution can be traced fully to the
cut diagrams of Fig.~\ref{fig89}, and is therefore consistent with unitarity.

We shall now discuss an alternate procedure for arriving at \refb{e6.7prime} that will be
easy to generalize to multi-instanton amplitudes. For this, note that under complex 
conjugation, an expression like \refb{e6.5} will have the sign of its $i\eps$ reversed.
The sign of the exponent also changes, but this can be undone by a $\chi\to -\chi$
change of integration variable. Therefore to get the
result for the unitary prescription, we can take 
the result for the Lorentzian prescription, analytically continue it back to the Euclidean
external momenta, complex conjugate the result, continue the result back to the
Lorentzian momenta, and finally average over the initial and the final results. This can
be achieved directly in the Lorentzian prescription by complex conjugation and
change of the signs of the exponents\footnote{Note that this is a formal
substitution and does not entail evaluating the result for opposite sign of the energy.
For example, although in \refb{e6.7} the final factor inside the parentheses is
$e^{-2\pi|\omega_2|}$ and therefore vanishes under antisymmetrization under
$\omega_2\to -\omega_2$,  
this is not relevant for our discussion. This is because the euclidean continuation
of $e^{-2\pi|\omega_2|}$ is $e^{-2\pi i |\omega_2^E|}$, and the exponent
changes sign under complex conjugation. Therefore when we analytically continue it
back to the Lorentzian signature, we get $e^{2\pi|\omega_2|}$ and the sign of the
exponent still changes.}  in the terms involving exponentials of the
energies
and then averaging over the initial
and the final results. For example, 
since complex conjugation of \refb{e6.7} produces an
overall sign due to the presence of the factor of $i$, this procedure simply
anti-symmetrizes the final factor inside the parenthesis under the
replacement $|\omega_2|\to -|\omega_2|$,
producing the factor $(e^{-2\pi|\omega_2|}-e^{2\pi|\omega_2|})/2$. This reproduces
\refb{e6.7prime}.

\subsection{Feynman diagram representation of the amplitude}
\label{stwo}

\begin{figure}
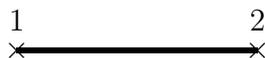

\begin{center}
\figsixteen
\end{center}

\vskip -1.3in

\caption{The closed string contribution to the annulus partition function with different
boundary conditions on the two boundaries.
\label{fig16}
}
\end{figure}

We shall now describe how the results of \S\ref{s6.1} can be organized into sum of
Feynman diagrams in the open closed string field theory. For this we need to first 
split the annulus contribution $\AAA_{12}$ given in \refb{efulla12} into the contribution
from closed string exchange diagram shown in Fig.~\ref{fig16}, where a pair of
disk one point functions are connected by the closed string propagator, and the
contribution $A_{12}$ from the open string loops. 
Contribution from Fig.~\ref{fig16} is given by:
\ben
&& \int_{-\infty}^\infty {d\omega_E\over 2\pi} \int_{0}^\infty dP \, e^{-i\omega_E (\phi_{(1)}
-\phi_{(2)})} \, {1\over P^2 + {\omega_E^2\over 4}} e^{-2\,b \, (P^2 +\omega_E^2/4)}
4\, \sinh^2 (2\pi P) \nonumber \\ &=&
\int_{2b}^\infty ds\,  \int_{-\infty}^\infty {d\omega_E\over 2\pi} \int_{0}^\infty dP \, e^{-i\omega_E (\phi_{(1)}
-\phi_{(2)})} \, e^{-s (P^2 + \omega_E^2/4)} \left( e^{4\pi P} + e^{-4\pi P} -2
\right)\, ,
\een
where in the first step we have used the off-shell closed string one point function
on the disk as given in \refb{estub} and the closed string propagator given on the
left hand side of \refb{ereplacecut}
after
euclidean rotation and in the
second step we have used the Schwinger parameter representation of the
propagator.  Since the integrand is even function of $P$,
we can extend the range of integration over $P$ from $-\infty$ to $\infty$ at the cost
of including a factor of 1/2, and then carry out integration over $\omega_E$ and $P$ by
standard rules of Gaussian integration. After changing variable to
\be
t\equiv 2\pi / s\, ,
\ee
the final result may be written as,
\be \label{e6.12new}
\int_0^a dt \, t^{-1} \left\{ e^{-t(\chi^2 -4\pi^2)/2\pi} - e^{-t\chi^2/2\pi}
\right\} , \quad \chi \equiv \phi_{(1)}-\phi_{(2)}\, ,
\ee
where,
\be \label{erelnab}
a\equiv {\pi\over b}\, .
\ee
Comparing this with \refb{efulla12} we see that the remaining contribution, that was called
$A_{12}$ in \refb{ea12} and that enters the definition of the two instanton
interaction vertex, is given by:
\be \label{eactuala12}
A_{12} = \int_a^\infty dt \, t^{-1} \left\{ e^{-t(\chi^2 -4\pi^2)/2\pi} - e^{-t\chi^2/2\pi}
\right\} \, .
\ee
Note that $A_{12}$ differs from $\AAA_{12}$ given in \refb{efulla12} in that in
\refb{eactuala12} the lower limit on the $t$ integral is $a$ while in \refb{efulla12} the
lower limit is 0. We shall simplify the analysis by working in the small $a$ limit -- but
this will still be different from the $a=0$ case. In this
case for any finite $\chi\equiv \phi_{(1)}-\phi_{(2)}$ the difference between $A_{12}$ and
$\AAA_{12}$ becomes small for small $a$, but for any small but fixed $a$, the 
contribution to \refb{eactuala12} for large $\chi$ gets suppressed as 
$e^{-a\chi^2/2\pi}$, while
there is no such suppression for $\AAA_{12}$.

\begin{figure}
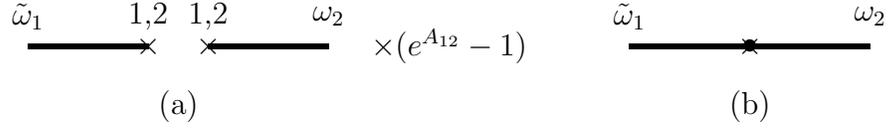

\begin{center}
\figseventeen
\end{center}

\vskip -1.1in

\caption{(a) A Feynman diagram in the open closed string field theory giving two 
instanton contribution to the closed string two point function. 1,2  label at the 
vertex means that it can refer to either the first or the second D-instanton.
(b) Representation of
the same Feynman diagram in closed string effective field theory. 
$\bullet \hskip -.1in {\Large\times}$ represents a
two instanton interaction vertex in the closed string effective field theory. 
\label{fig17}
}
\end{figure}

With this result, we are now ready to compute the contribution to the closed string two
point function from various Feynman diagrams of open closed string field theory.
We first consider the Feynman diagram of Fig.~\ref{fig17}.
Fig.~\ref{fig17}(a) shows the Feynman diagram representing this contribution in open
closed string field theory, while Fig.\ref{fig17}(b) shows the same contribution as
a Feynman diagram in closed string effective field theory. 
This contribution is related to the full 
result for the same amplitude to this order, given by \refb{e6.5}, by a
multiplication function $f(\chi)$ in the integrand, where
$f(\chi)$ is some even, real function of $\chi$ that
takes value 1 for $a\chi^2<<1$, but has a suppression
factor of
$e^{-a\chi^2/2\pi}$ for $a\chi^2>>1$. Therefore this may be written as:
\be \label{e6.13a}
\NN^2 \, e^{-2/g_s}\, 16\, \pi^2 \, \sinh(2\pi P_1)\, 
\sinh(2\pi P_2)\, 2\pi \delta(\omega^E_1+\omega^E_2) 
\int \, {d\chi\over \chi^2-4\pi^2+i\eps}\, f(\chi) \left[e^{-i\chi(\omega^E_1-\omega^E_2)/2}
+1\right] \, .
\ee
Due to the exponential suppression of the integrand for large $\chi$, we can
keep the $\chi$ integration contour along the real axis even when $\omega_i^E$ are
complex. Finiteness of the integral (and its derivatives with respect to $\omega_i^E$) 
shows that the integral is an analytic function of
$\omega_i^E$ in the whole  complex plane. 
 Therefore its value for $\omega_i^E= -i\omega_i$ for real $\omega_i$ can be obtained
simply by substitution of $\omega_i^E$ by $-i\omega_i$:
\be \label{e6.14}
\NN^2 \, e^{-2/g_s}\, 16\, \pi^2 \, \sinh(2\pi P_1)\, 
\sinh(2\pi P_2)\, 2\pi \delta(\omega_1+\omega_2) 
\int \, {d\chi\over \chi^2-4\pi^2+i\eps}\, f(\chi) \left[e^{-\chi(\omega_1-\omega_2)/2}
+1\right] \, .
\ee
Here we have used the result of \S\ref{seuclid} that in relating the Euclidean action to the
Lorentzian action, or the Euclidean Green's function to the T-matrix elements, the
momentum conserving $\delta$-functions have their arguments replaced by
Lorentzian momenta, and there are no extra factors of $i$ in the overall
normalization. 
The reality of the amplitude requires it to be invariant under the simultaneous operation
of complex conjugation and change of sign of the $\omega_i$'s. Since the latter
operation can be undone by a $\chi\to -\chi$ change of variable in the integration,
the only source of the imaginary part of this amplitude is the
$i\eps$ in the denominator. This gives a factor
proportional to $\delta(\chi^2-4\pi^2)$. Since in the small $a$ limit, 
$f(\chi)=1$ for finite $\chi$, we can replace  $f(\chi)$ by 1
at $\chi=\pm 2\pi$. 
This gives the imaginary part of the amplitude to be:
\be \label{eim1}
-{i} \, \NN^2 \, e^{-2/g_s}\, 4\, \pi^2 \, \sinh^2(\pi\omega_2)\, 
2\pi \delta(\omega_1+\omega_2) 
\left[e^{2\pi  \omega_2}
+ e^{-2\pi  \omega_2} +2\right] \, ,
\ee
where we have used the on-shell condition $P_i=|\omega_i|/2$.

We now make a few comments on the result \refb{eim1}.
\begin{enumerate}
\item
\refb{eim1} differs from \refb{e6.7}. This difference can be traced to the difference
between $\AAA_{12}$ and $A_{12}$ provided by the damping factor $f(\chi)$. 
The reason that the argument given above cannot be applied to the full
amplitude \refb{e6.5} is that
\refb{e6.14} does not make sense without the damping factor $f(\chi)$ since the
integral diverges for $\chi\to\infty$ or $\chi\to -\infty$ depending in the sign of
$\omega_1-\omega_2$. When we regulate this divergence we may get additional
imaginary terms. Indeed, these divergences are related to on-shell closed string
propagation in the intermediate state shown in Fig.~\ref{fig18}, 
and the imaginary part associated with this
will be computed in \refb{eim2}.
\item
Since Fig.~\ref{fig17}(b) represents a 2-instanton
interaction vertex of the closed string effective field theory with two external
closed strings, the $i$ in \refb{eim1} shows that the 2-instanton action of the closed
string effective field theory has imaginary contribution. This will lead to violation of
unitarity.
We could explicitly write down the term
in the effective 
action corresponding to \refb{eim1} following \refb{eaction}, but will not do so.
\item
As already discussed,
the expression \refb{e6.14} is analytic in the full complex $\omega_2$
plane, since due to the exponential suppression factor proportional to $e^{-a\chi^2/2\pi}$
hidden inside $f(\chi)$, the integration over $\chi$ converges for any complex $\omega_2$.
This is what is expected of any term in the closed string effective action. This is to be
contrasted with the amplitude given in \refb{e6.7} that has non-analyticity at $\omega_2=0$
due to the cuts associated with Fig.~\ref{fig89}. 
\end{enumerate}

\begin{figure}
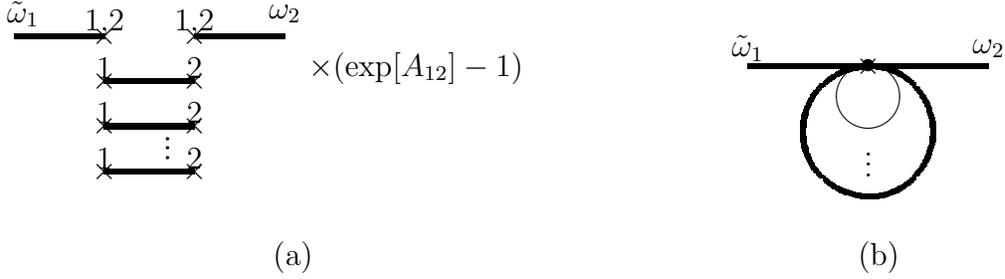

\begin{center}
\hbox{\figfourteen \hskip 2in \figfifteen}
\end{center}

\vskip -.8in

\hbox{\hskip 1.7in (a) \hskip 2.8in (b)}

\caption{(a) A set of Feynman diagrams in open closed string field theory 
that can be part of the leading two 
instanton contribution to $1\to 1$ closed string amplitude.  
The symbol 1,2 at the 
$\times$ denotes that the vertex may describe either type 1 D-instanton or type 2
D-instanton.
(b) The same  diagram represented in the closed string
effective field theory, with $\bullet \hskip -.1in {\Large\times}$ representing a
two instanton interaction vertex in the closed string effective field theory. 
As argued in the text, the contribution to these diagrams is suppressed in the
small $a$ limit.
\label{fig88}
}
\end{figure}

\begin{figure}
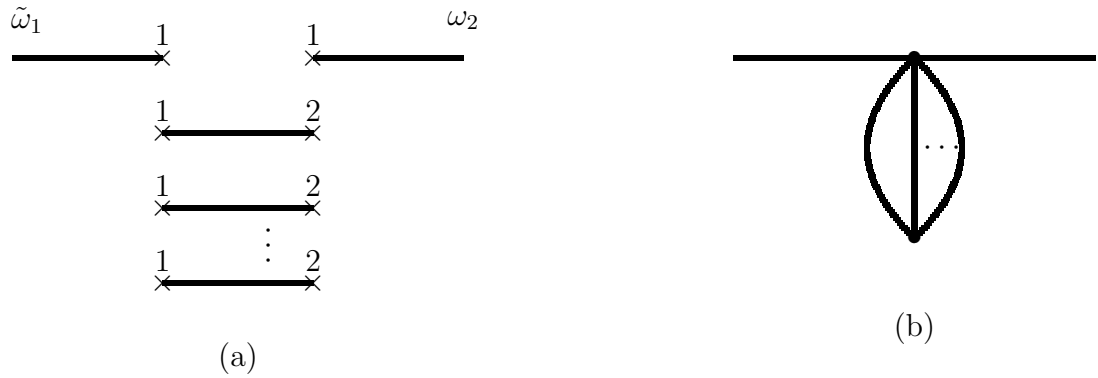


\vskip .2in

\begin{center}
\hbox{\fignew \hskip -2in\figKnew}
\end{center}

\vskip -.8in

\caption{(a) Examples of Feynman diagrams of open closed string field
theory giving part of the
leading 2-instanton contribution to the closed string 2-point amplitude.
We also have a similar diagram where the pair of 1's in the top line are replaced
by a pair of 2's. (b) Representation of the same diagram in the
closed string effective 
field theory. 
\label{figK}
}
\end{figure}

Let us now examine other Feynman diagrams that can contribute to this
amplitude.
A second class of Feynman diagrams that can contribute 
at the same order has been shown in Fig.~\ref{fig88}. 
The contribution from these diagrams will be given by an integral of the form 
\refb{e6.14}, but the integrand will be multiplied by one factor of
\refb{e6.12new}
for every extra closed string propagator. 
Since \refb{e6.12new} is a real and even function of $\chi$, 
the same argument as for \refb{e6.14}
tells us that the imaginary part will come from the points $\chi=\pm 2\pi$.
Now as we take $a\to 0$, $b\to \infty$ limit, the term inside the square bracket 
in \refb{e6.14} remains
bounded, while \refb{e6.12new} vanishes in this limit.
Therefore we conclude
that the amplitudes shown in Fig.~\ref{fig88} do not have any  imaginary part in the large
$b$ limit. This is consistent with the cutting rules since Fig.~\ref{fig88}(b) 
does not admit any cut. This however was not guaranteed from the start 
since we have not established
the reality of the two instanton interaction vertex that appears in Fig.~\ref{fig88}(b).

\begin{figure}
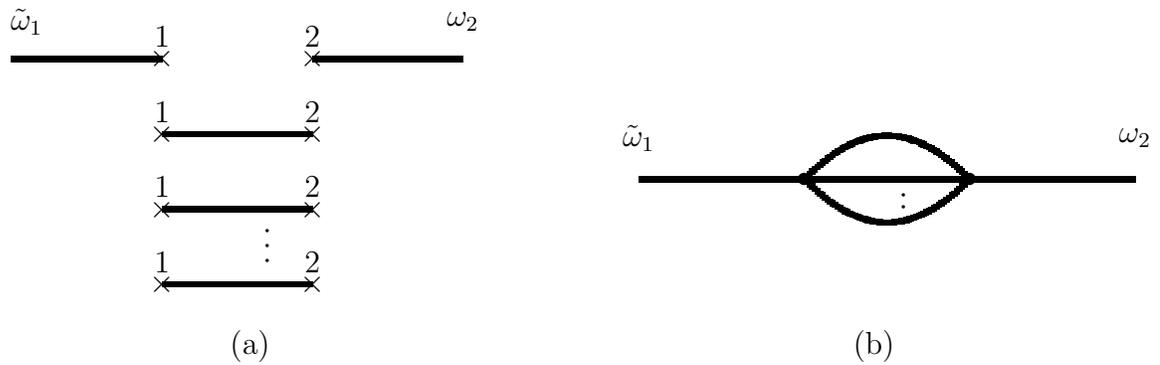

\begin{center}
\hbox{ \figelevennew \qquad \figthirteennew}
\end{center}

\vskip -1.4in

\hbox{\hskip 1.7in (a) \hskip 3in (b)}

\caption{(a) A set of Feynman diagrams in open closed string field theory
that can be part of the leading 2-instanton
contribution to the closed string 2-point function. (b) The same Feynman 
diagram represented in the closed string
effective field theory.
\label{fig18}
}
\end{figure}

Fig.~\ref{fig17} and Fig.~\ref{fig88} 
take into account all the Feynman diagrams that have one or more factor of
$A_{12}$. Now we turn to the Feynman diagrams without this factor.
We shall divide these into two classes of Feynman diagrams, shown in
Fig.~\ref{figK} and Fig.~\ref{fig18}. 
From the representation of these diagrams in the closed string effective field theory,
we see that that they are constructed from two one instanton interaction vertices
in the closed string effective field theory. Since we have taken $\NN$ to be real, the
one instanton interaction vertices do not violate unitarity, and therefore the imaginary
parts of these diagrams will be given by the sum over all cut diagrams. It is easy to
see from Fig.~\ref{figK}(b) that these diagrams do not admit any cut and therefore 
do not have any
imaginary part. 
The cut diagram of Fig.~\ref{fig18} has been shown in Fig.~\ref{fig89}. 
Its contribution,
computed using the usual cutting rules, is given by:
\ben\label{e6.19cutpre}
&&\hskip -.3in  i\, \NN^2 \, e^{-2/g_s} \, 2\pi\delta( \omega_1+\omega_2)
\, 4 \sinh (\pi|\omega_1|) \sinh(\pi|\omega_2|) \,  \sum_{n=1}^\infty {1\over n!}\, 
\int_{P_i>0} dP_1\cdots dP_n \nonumber \\ && \hskip -.3in
\int_{e_i>0} {de_1\over 2\pi} \cdots {de_n\over 2\pi} \, 
2\pi \delta(e_1+\cdots e_n-|\omega_2|) \, \prod_{i=1}^n \left\{ 
2\pi \delta \left({e_i^2\over 4} - P_i^2\right) \, 4\, \sinh^2 (2\pi P_i)\right\}\, . 
\een
Note that we have written down the formula in a way that is valid both for positive
and negative $\omega_2$.
We can simplify this by first carrying out the
integration over the $P_i$'s using the delta functions, and then use
the identity,
\be \label{eidentity}
\sum_{n=1}^\infty {4^n\over n!}\, 
\int_{e_i>0} {de_1\over e_1} \cdots {de_n\over e_n} \, 
2\pi \delta(e_1+\cdots e_n-|\omega_2|) \,
\prod_{i=1}^n  \sinh^2 (\pi e_i)  =4\pi^2 \, \sinh(2\pi |\omega_2|)\, .
\ee
\refb{eidentity} can be proved by multiplying both sides by $e^{-\mu|\omega_2|}$,
integrating over $\omega_2$ from 0 to $\infty$ (or from $-\infty$ to 0)
and showing that the two sides
agree for arbitrary $\mu>2\pi$. This reduces 
\refb{e6.19cutpre} to,
\be \label{e6.19cut}
 i\, \NN^2 \, e^{-2/g_s} \, 2\pi\delta( \omega_1+\omega_2)
\, 4  \sinh^2(\pi\omega_2) \, 4\pi^2 \, \sinh(2\pi |\omega_2|)\, .
\ee
\refb{e6.19cut} 
gives twice the contribution to the imaginary part of the T-matrix. Therefore the
contribution from this cut diagram to the imaginary part of the T-matrix is given by:
\be\label{eim2}
i\, \NN^2 \, e^{-2/g_s} \, 8\pi^2 \, 2\pi\delta(\omega_1+\omega_2)
\, \sinh^2(\pi\omega_2) \,  \sinh(2\pi |\omega_2|)\, .
\ee

Adding \refb{eim1} and \refb{eim2} we get the total contribution to the 
imaginary part of the closed string 2-point amplitude to this order:
\be\label{etotalimaginary}
-i\, \NN^2 \, e^{-2/g_s}\, 8\, \pi^2
\,  \sinh^2(\pi \omega_2)\, 
2\pi\delta(\omega_1+\omega_2) \, \left(1 + e^{-2\pi|\omega_2|}\right)\, .
\ee
This agrees with \refb{e6.7}. Note however that if we were using the ordinary
cutting rules of a field theory with real effective action, then we shall only have
the contribution from the cut diagram of Fig.~\ref{fig89}. Its contribution to the imaginary
part of the T-matrix, given by \refb{eim2}, differs from the full answer \refb{e6.7}. 
Therefore the usual cutting rules fail.
The difference
can be accounted for by the imaginary part of Fig.~\ref{fig17} given in \refb{eim1}, 
which in turn, 
originates from the violation of reality condition by the 2-instanton interaction vertex
of the closed string effective field theory. This signals the violation of unitarity
within the closed string sector. 

In conclusion we see that even if we had taken $\NN$ to be real, unitarity is
violated. This can be traced to the appearance of the open string
tachyons below a critical separation of the two D-instantons, and our
use of Lorentzian prescription for dealing with the associated singularity. 
We also note that
even though the contribution from the individual Feynman diagrams depends
on the string field theory parameter $a$ (or equivalently $b$), the contribution from
the cut diagram is given by the product of physical amplitudes in the single instanton
sector, and is independent of the string field theory parameter. On the other hand
the imaginary part of the original 2-point amplitude, given in \refb{e6.7}, is also
independent of the string field theory parameters. Therefore the violation of cutting rules
is a physical effect that cannot be changed by changing the string field theory parameters.

\subsection{Unitarity restoration via unitary prescription} \label{sunires}

We shall now examine what happens if we use the unitary prescription
for dealing
with the singularities associated with the 
critical separation of the D-instantons where a tachyon
appears. The leading 2-instanton contribution to the 2-point function of closed
string tachyons, computed with this prescription, 
is given in \refb{e6.7prime}. It is easy to see that this agrees with the 
prediction of the cutting rules, given in \refb{eim2}, for the imaginary part of the
amplitude.

This can be also seen at the level of the effective action. It follows from the
analysis in \S\ref{stwo} that the
only source of imaginary term in the effective action is in the part of the amplitude
given in \refb{eim1}, which will be absent if we used the unitary prescription instead
of the  Lorentzian prescription in \refb{e6.14}. Therefore  
the resulting effective action will be real.

\subsection{Connection to the matrix model} \label{smattwo}

In this section we shall compare our results to that in \cite{1912.07170}. 
During the analysis of $k$ instanton amplitudes for $k\ge 2$, \cite{1912.07170}
used the prescription of integrating over the zero modes directly in the Lorentzian
space without any factor of $i$. We shall first show why this is consistent with our
procedure of defining the integral in the Euclidean space and picking up appropriate 
factors of $i$ if we need to rotate the integral, as was done while arriving at \refb{e6.7}.
For this we note that in our convention, for the
general case of $k$ instantons we shall have an
overall normalization factor $\NN^k = (-i{\cal N})^k$,
where ${\cal N} = i\NN = -1/(8\pi^2)$ in the notation of \cite{1912.07170}. 
On the other hand while computing
the amplitude we have to integrate over 
$(k-1)$ euclidean variables labelling the differences in the Euclidean time coordinates
of the D-instantons -- integration over  the variable labelling the average time coordinate 
produces the energy conserving delta function and has already been discussed in
\S\ref{seuclid}. 
After rotation to Lorentzian variables, this gives a factor of $i^{k-1}$.
To get the S-matrix from the T-matrix we have an extra factor of $i$.
Therefore the
net normalization, in the convention where we integrate the zero modes along
the Lorentzian direction, is given by:
\be
(-i{\cal N})^k i^{k-1} i = {\cal N}^k\, .
\ee
This
agrees with the prescription of \cite{1912.07170} 
of integrating over the zero modes in Lorentzian space
and using normalization factor ${\cal N}^k$ for $k$ instanton amplitude.

We can test this by comparing the results for the leading
2-instanton contribution to the
two point function of closed string tachyons. 
We use the relation $\NN=-i\NNN$ to
rewrite \refb{e6.7} as:
\be
S_{\rm 2-inst} = i T_{\rm 2-inst}=  - {\cal N}^2 \, e^{-2/g_s}\, 
 8\pi^2 \,  \sinh^2(\pi \omega_2)\, 2\pi\delta(\omega_1+\omega_2) \,
  \left(1 + e^{-2\pi|\omega_2|}\right)\, .
 \ee
This agrees
with the result of \cite{1912.07170} for the  leading two instanton contribution to the
closed string 2-point function, and agrees with the matrix model results after adding
the contribution from the (2,1) D-instanton.

This is the result using Lorentzian prescription for the integration contour of the
zero modes. As we have 
argued in \S\ref{smultireal}, 
this prescription leads to non-unitary amplitudes even when we expect
the theory to be unitary, {\it e.g.} for D-instanton anti-D-instanton contribution to the 
amplitudes in type IIB string theory. The unitary prescription gives unitary amplitudes
in these theories. Even though in two dimensional bosonic string theory we do not expect
a unitary amplitude, and indeed unitarity is already violated by the one instanton
amplitudes by having an imaginary $\NN$, we shall now explore what
results we get if we use the unitary prescription in two dimensional string theory. We have 
already seen that for the two
instanton contribution to the two point amplitude, we shall have a missing term 
 \refb{eim1} in the unitary prescription. Multiplying this by $i$ to translate this to a 
 contribution to the S-matrix, using $\NN=-i\NNN$, and using the on-shell condition
 $P_i=|\omega_i|/2$ for $i=1,2$, we get the missing term to be,
 \ben \label{ediff}
&& - {\cal N}^2 \, e^{-2/g_s}\, 4\, \pi^2 \, 
\sinh^2(\pi \omega_2)\, 2\pi \delta(\omega_1+\omega_2) 
\left[e^{2\pi  \omega_2}
+ e^{-2\pi  \omega_2} +2\right] \nonumber \\
&=& -{1\over 16\pi^2} \,  e^{-2/g_s}\, 
2\pi \delta(\omega_1+\omega_2)  \, \sinh^2(2\pi\omega_2)\, ,
\een
where we have used $\NNN=-1/(8\pi^2)$.

Since \cite{1912.07170} 
got agreement with the matrix model results, if we want the unitary prescription
to also agree with the matrix model results, we must have some other source of the
contribution \refb{ediff} to compensate for this missing term. To this end we note that 
\cite{1912.07170} parametrized
the contribution to the same S-matrix element from (2,1) ZZ-instanton as:
\be\label{eold}
4\, \NNN_2\,  e^{-2/g_s}\, 
2\pi \delta(\omega_1+\omega_2)  \, \sinh^2(2\pi\omega_2)\, ,
\ee
for some constant $\NNN_2$. Comparing \refb{ediff} and 
\refb{eold} we see that that the missing contribution \refb{ediff} may be
compensated by a shift in $\NNN_2$ of the form $\NNN_2\to\NNN_2-1/(64\pi^2)$.
Now the result of \cite{1912.07170} for $\NNN_2$ was $3/(64\pi^2)$. 
Therefore our new 
result for $\NNN_2$ with the unitary prescription is:
\be\label{enewn2}
\NNN_2 = {1\over 32\pi^2}\, .
\ee

We shall now verify that the same shift in $\NNN_2$ also correctly reproduces the
2-instanton contribution to the $n$-point amplitude. The difference between the 
Lorentzian and unitary prescription for the $n$-point amplitude comes from different
treatment of the
analog of the Feynman diagram shown in Fig.~\ref{fig17}(a), except that instead of having
just two external closed strings, we now have $n$ external closed strings, each of which
can connect to a type 1 D-instanton vertex or a type 2 D-instanton vertex. In the Lorentzian
prescription the result of this diagram can be written down in a manner analogous to
\refb{e6.14}:
\be \label{e6.14new}
{1\over 2}\, \NN^2 \, e^{-2/g_s}\, 4\, \pi^2 \, \left\{\prod_{i=1}^n 2 \sinh(2\pi P_i)\right\}\, 
 2\pi \delta\left(\sum_{i=1}^n\omega_i\right) 
\int \, {d\chi\over \chi^2-4\pi^2+i\eps}\, f(\chi) 
\sum_{S\subset\{1,\cdots\, n\}} e^{-\chi \sum_{i\in S} \omega_i} \, .
\ee
Here $S$ denotes a subset of $\{1,\cdots, n\}$ labelling the collection of external
closed strings that connect to type 1 D-instanton vertices, the rest of the external closed
strings connect to type 2 D-instanton vertices.
The sum over $S$ can be easily performed yielding $\prod_i\{ 2\cosh(\chi\omega_i/2)\}$.
Using on-shell condition $P_i=|\omega_i|/2$, we can rewrite \refb{e6.14new} as:
\be
{1\over 2}\, \NN^2 \, e^{-2/g_s}\, 4\, \pi^2 \, \left\{\prod_{i=1}^n 2\sinh(\pi |\omega_i|)\right\}\, 
 2\pi \delta\left(\sum_{i=1}^n\omega_i\right) 
\int \, {d\chi\over \chi^2-4\pi^2+i\eps}\, f(\chi) 
\prod_i\{ 2\cosh(\chi\omega_i/2)\}\, .
\ee
In the unitary prescription, instead of using $i\eps$ we use the average of $i\eps$
and $-i\eps$ prescription to carry out
the $\chi$ integration. The difference is proportional to $\delta(\chi^2-4\pi^2)$. Using
the result $f(\pm 2\pi)=1$, we can write the extra contribution in the Lorentzian 
prescription as,
\be
- {i} \,
\NN^2 \, e^{-2/g_s}\, \pi^2 \, \left\{\prod_{i=1}^n 2 \sinh(2\pi |\omega_i|)\right\}\, 
 2\pi \delta\left(\sum_{i=1}^n\omega_i\right) \, .
 \ee
Multiplying this by $i$ and using $\NN=-i\NNN=i/(8\pi^2)$, 
we get the extra contribution to the
S-matrix in the Lorentzian prescription:
\be\label{emissing}
- {1\over 64\pi^2} \, e^{-2/g_s}\,  \left\{\prod_{i=1}^n 2 \sinh(2\pi |\omega_i|)\right\}\, 
 2\pi \delta\left(\sum_{i=1}^n\omega_i\right) \, .
 \ee
On the other hand, the contribution to the $n$-instanton amplitude due to the
(2,1) ZZ-instanton was computed in \cite{1912.07170} to be:
\be 
\NNN_2 \, e^{-2/g_s}\,  \left\{\prod_{i=1}^n 2 \sinh(2\pi |\omega_i|)\right\}\, 
 2\pi \delta\left(\sum_{i=1}^n\omega_i\right) \, .
\ee
Therefore the missing contribution \refb{emissing} in the unitary prescription can
be compensated by a shift $\NNN_2\to \NNN_2- 1/(64\pi^2)$. This again leads us
back to \refb{enewn2} as the new value of $\NNN_2$ in the unitary prescription.

\subsection{Three instanton amplitude in the unitary prescription} \label{sthree}

We shall now study the effect of using unitary prescription for the three instanton
amplitude. For this we shall use the expressions for various amplitudes computed 
in \cite{1912.07170}
in the Lorentzian prescription and apply the procedure described at the end of 
\S\ref{s6.1} to get the result in the unitary prescription. 
In order to avoid the proliferation of the absolute
value symbol, we shall write all subsequent formul\ae\ in this  
subsection for an incoming closed string of energy
$\tilde\omega_1=-\omega_1>0$ and an outgoing closed string of energy $\omega_2>0$.

As already emphasized, there are two
differences between 
the notation in \cite{1912.07170} and the ones used in this paper. First of
all \cite{1912.07170} gave the results for the S-matrix elements while the various 
manipulations we have described is for the T-matrix element. As a result, while converting
the result of \cite{1912.07170} to the notations of this paper, we have to multiply by a
factor of $-i$. Second, \cite{1912.07170} gave various results using the normalization
constants $\NNN_k$ for the $(k,1)$ instanton amplitudes, which will be related by the
normalization constants $\NN_k$ in our notation by $\NNN_k = i\NN_k$. Therefore, in
an expression in \cite{1912.07170} containing $r$ factors of different $\NNN_k$'s, there
are $(r-1)$ hidden $i$'s that will remain after we have stripped off the normalization
factors of $\NN$ from the T-matrix elements. Under complex conjugation this will pick
up a factor of $(-1)^{r-1}$.  The procedure discussed at the end
of \S\ref{s6.1} now tells us that in order to convert such an expression from the
Lorentzian prescription to the unitary prescription, we need to
anti-symmetrize the expression under $\omega_i\to -\omega_i$ for even $r$,
and symmetrize the expression under $\omega_i\to -\omega_i$ for odd $r$.
The expressions in \cite{1912.07170} do not have explicit $i$'s in the normalization
factor, but if there were such factors then each factor of $i$ will shift the effective $r$ by 1.

As an example we note that expression \refb{e6.7}, after being converted to the notation
of \cite{1912.07170}, would have two factors of $\NNN$. Therefore here $r=2$ and we have
to antisymmetrize \refb{e6.7} under $\omega_i\to -\omega_i$ 
to get the result in the unitary prescription.
This is what we had concluded in \S\ref{s6.1}.
One point we should remember is that this anti-symmetrization should not involve any
factor coming from Liouville correlators. For the two point functions that we shall
discuss, these factors remain unchanged under $\omega_i\to-\omega_i$, and so we do
not need to be careful about this.

Let us now apply this to compute the leading order $e^{-3/g_s}$ contribution to the
2-point function of closed strings, with the incoming string carrying energy
$\tilde\omega_1=-\omega_1>0$ and the outgoing string carrying energy $\omega_2>0$.
The contribution computed in \cite{1912.07170} come from three sources. First the
three instanton contribution to the S-matrix was given by:
\be
e^{-3/g_s} \, 2\pi\delta(\tilde\omega_1-\omega_2)\, \NNN_1^3\, {64\pi^4\over 3} \, 
\sinh^2(\pi\omega_2) \, \left(1+e^{-2\pi\omega_2}+e^{-4\pi\omega_2}\right)\, ,
\ee
where $\NNN_1$ (called $\NNN$ in \S\ref{smattwo}) is the normalization factor associated
with the one instanton contribution. Since there are 3 factors of $\NNN_k$, we need
to symmetrize the expression under $\omega_i\to -\omega_i$ to get the result
for unitary prescription. This gives:
\be\label{eexp1}
e^{-3/g_s} \, 2\pi\delta(\tilde\omega_1-\omega_2)\, \NNN_1^3\, {64\pi^4\over 3} \, 
\sinh^2(\pi\omega_2) \, {1\over 2}
\left(2+e^{-2\pi\omega_2}+e^{-4\pi\omega_2}
+e^{2\pi\omega_2}+e^{4\pi\omega_2}\right)\, .
\ee
At the same order we have a two instanton contribution with one (2,1) and one (1,1)
instanton. In the Lorentzian prescription this was given by:
\be
-e^{-3/g_s} \, 2\pi\delta(\tilde\omega_1-\omega_2)\, \NNN_1\, \NNN_2\, 
{32\pi^2\over 3} \, 
\sinh^2(\pi\omega_2) \, \left(e^{2\pi\omega_2}+3 + 3 \, e^{-2\pi\omega_2} + 2\, e^{-4\pi
\omega_2}
\right)\, ,
\ee
where $\NNN_2$ is the normalization factor associated with the (2,1) instanton. Since we
have two factors of $\NNN_k$, we need to antisymmetrize the expression under
$\omega_i\to -\omega_i$ to get the corresponding expression in the unitary prescription.
This gives:
\ben\label{eexp2}
&& -e^{-3/g_s} \, 2\pi\delta(\tilde\omega_1-\omega_2)\, \NNN_1\, \NNN_2\, 
{32\pi^2\over 3} \, 
\sinh^2(\pi\omega_2) \, {1\over 2}\,
\left(e^{2\pi\omega_2} + 3 \, e^{-2\pi\omega_2} + 2\, e^{-4\pi
\omega_2} \right. \nonumber \\
&& \hskip 2in  \left. -
e^{-2\pi\omega_2} - 3 \, e^{2\pi\omega_2} - 2\, e^{4\pi
\omega_2}
\right)\, .
\een
The value of $\NNN_2$ in the unitary prescription has been given in \refb{enewn2}.
Finally the contribution from the single (3,1) instanton takes the same form in the 
Lorentzian and the unitary prescription:
\be\label{eexp3}
e^{-3/g_s} 
 \, 2\pi\delta(\tilde\omega_1-\omega_2)\, 4\, \NNN_3\, \sinh^2(3\pi\omega_2)\, .
 \ee
The sum of \refb{eexp1}, \refb{eexp2} and \refb{eexp3} should be compared with the
expected answer from the matrix model\cite{1912.07170}:
\ben\label{eefin4}
&& -e^{-3/g_s} 
 \, 2\pi\delta(\tilde\omega_1-\omega_2)\, {1\over 24\pi^2}\, \sinh^2(\pi\omega_2) \,
 \left[4 + 5\cosh(2\pi\omega_2) + 3\cosh(4\pi\omega_2)\right. \nonumber \\ &&
 \hskip 2in \left. + 2\sinh(2\pi\omega_2) 
 +2\sinh(4\pi\omega_2)\right]\, .
\een
We find that the following choice of $\NNN_3$ satisfies the requirement:
\be
\NNN_3 = -{1\over  96\, \pi^2}\, .
\ee
Related observation has been made independently in \cite{private}. 
In contrast in the Lorentzian prescription the required value of $\NNN_3$ was 
found\cite{1912.07170}
to be $-5/(192\pi^2)$.

\sectiono{Contour prescription from  Picard-Lefschetz theory} \label{spicard}

In this section we shall use Picard-Lefschetz theory to
argue that non-BPS D-instantons in type IIA string theory
and instanton anti-instanton system in type IIB string theory give real contribution to
the closed string effective action.

In Picard-Lefschetz 
theory (see 
\cite{1001.2933,1009.6032,1206.6272,1511.05977,1601.03414,1802.10441}
for review), we consider
a finite dimensional integral of the form,
\be\label{epic1}
I=\int_\CC \left\{\prod_{i=1}^N dx_i\right\}\, e^{S(\{x_i\})}\, ,
\ee
where $S(\{x_i\})$ is a function of $N$ variables $x_1,\cdots, x_N$, carrying an 
overall normalization factor $1/g_s$ for some small parameter $g_s$, and 
$\CC$ denotes some $N$ real dimensional subspace of the $N$ 
complex dimensional space of the $x_i$'s. The integral
is evaluated as follows. We first identify the critical points of $S$ in the complex 
$x_i$ space. Then for each critical point $\sigma$, we associate the Lefschetz thimble
$\JJ_\sigma$, an $N$ real dimensional subspace of the $N$ complex dimensional space
of the $x_i$'s, by solving the gradient flow equations:
\be\label{epic2}
{dx_i\over d\tau} = {\p S^* \over \p x_i^*}, \qquad {dx_i^*\over d\tau} = {\p S \over \p x_i}\, ,
\ee
where $\tau$ is a real parameter, $*$ denotes complex conjugation and the flow lines
approach the critical point $\sigma$ at $\tau\to\infty$. The collection of all such flow lines
define $\JJ_\sigma$. We define
\be\label{epic3}
I_\sigma = \int_{\JJ_\sigma} \left\{\prod_{i=1}^N dx_i\right\}\, e^{S(\{x_i\})}\, .
\ee
The original integration contour in \refb{epic1} is then expressed as a weighted sum 
$\sum_\sigma n_\sigma \JJ_\sigma$
for appropriate choice of the multipliers $n_\sigma$ and
the integral \refb{epic1} is  given by 
$\sum_\sigma n_\sigma I_\sigma$. An important point to keep in mind is that the
integrals $I_\sigma$ do not change under small deformations of $\JJ_\sigma$
as long as the asymptotic boundaries are kept fixed (or kept within the range along 
which $S(\{x_i\})$ approaches $-\infty$ sufficiently rapidly).

In the current context, 
the D-instanton systems represent the critical points in the configuration space of open
and closed string fields, and
the contribution to the amplitude from a given D-instanton system is expressed as
the path integral over the open and closed string fields along the thimble 
corresponding to that particular instanton, obtained by solving \refb{epic2}.
However, since our interest is in computing the closed string effective action by integrating
out the open string modes, we can focus on the integration over the open string modes
only. Furthermore, the massive open string modes can be integrated out first by standard
procedure, leading to an effective action of the massless and tachyonic open string modes
and closed strings. 
We can also drop the integration over the collective coordinates describing the center of
degrees of freedom of the full system of D-instantons, since the dependence on these
modes is trivial. 
Therefore we focus on integration over the rest of the massless modes 
and tachyonic open
string modes only. These are finite in number. Of course the effective action involving these
modes and the closed string modes is not known, and is likely to be very complicated.
However, since the result of integration depends only on the topological class of the
integration contour and not on the details, 
we expect that the essential features of the integration contours can be captured
by working with some prototype actions that capture the essential properties of these
modes. Finally, since we treat the closed strings states perturbatively, expanding the
action in powers of these fields, the integration contours over the open string modes
can be analyzed by examining the part of the action obtained by setting the closed 
string modes to zero, and then using the same integration contour even in the presence
of the closed strings. 
For this reason, we shall drop the closed string fields during
the analysis of the integration contour, 
and consider the integration over the tachyonic and massless open
string modes, weighted by 
the exponential of the
effective action of these modes. 
Our goal will be to express the desired integration 
contour on the open string modes as weighted union of the thimbles associated with
different critical points, thereby resolving some of the ambiguities encountered earlier
in determining the contribution from different instantons.

\begin{figure}
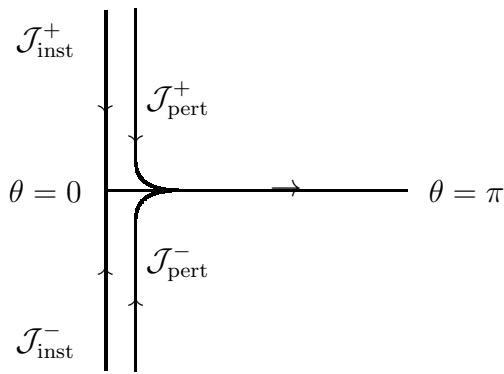

\begin{center}
\figpicard
\end{center}

\caption{The steepest descent contours for the action \refb{eppp1}.
\label{figpicard}
}
\end{figure}

\subsection{Non-BPS D-instanton in type IIA string theory} \label{s8.1}

The non-BPS D-instanton in type IIA string theory has a real tachyonic mode. The
tachyon potential is symmetric under $T\to -T$, with a maximum at $T=0$
representing the D-instanton and a pair of minima at some values representing the
perturbative vacuum\cite{0410103}. 
Furthermore, the two minima are expected to describe the same
perturbative vacuum. For this reason it is more appropriate to regard the tachyon
as an angular variable. We shall denote by $\theta$ the appropriately normalized angular
variable and use the following form the action:
\be\label{eppp1}
S(\theta) = -{1\over \sqrt 2\, g_s} (1+\cos\theta)\, .
\ee
The point $\theta=0$, with action $-\sqrt 2/g_s$, describes 
the D-instanton configuration and
$\theta=\pi$, with action 0, denotes the vacuum without D-instanton. The 
natural choice for the original integration
contour is along the real $\theta$ axis. 
We shall restrict the integration range to 
$0\le\theta\le \pi$. We note however that the qualitative features of our result will not
depend on the details of the potential -- in particular we shall get similiar results also for
the standard double well potential.

For real positive $g_s$, the steepest descent
contour for the saddle point at $\theta=\pi$, representing the perturbative vacuum, 
runs along
the real axis from 0 to $\pi$. Let us call this $\JJ_{\rm pert}$.
On the other hand, there is a pair of 
steepest descent contours associated with the saddle point
at 0, representing the
instanton, running along the imaginary axis from $\pm i\infty$ to 0. 
We shall denote these by $\JJ_{\rm inst}^\pm$.
Therefore the integration cycle contains only the
steepest descent contour 
associated with the perturbative vacuum. In our convention,
this can be summarized by saying 
that the normalization
constant $\NN$, associated with the instanton, vanishes.

If instead of taking $g_s$ to be real, we take this to have a small imaginary part, then the
steepest descent contours have more intricate structure\cite{1511.05977}.
The steepest descent contours 
associated with the instanton are only slightly deformed from their original form, but only
one of them stays inside the range $0\le {\rm Re}(\theta)\le\pi$. Whether it is 
$\JJ_{\rm inst}^+$ or $\JJ_{\rm inst}^-$ depends on the sign of the imaginary part of
$g_s$.
The
steepest descent contour associated with the perturbative vacuum now starts at
$\pm i\infty$, runs towards zero almost along the imaginary axis and then runs almost
along
the real axis till $\theta=\pi$. Whether it begins at $i\infty$ or $-i\infty$  depends 
on the sign of the imaginary part of $g_s$.  This has been shown by the
contours $\JJ^\pm_{\rm pert}$  in Fig.~\ref{figpicard}.
Therefore in this
case the integration cycle will be given by $\JJ^\pm_{\rm pert}-
\JJ^\pm_{\rm inst}$, and whether we use $+$ or $-$ sign depends on the sign of the
imaginary part of $g_s$. 
This shows that in this case there is a contribution from the steepest
descent contour associated with the instanton. However now  
since the steepest descent contour $\JJ^\pm_{\rm pert}$ is deformed into the
complex plane, its contribution is not real, and the imaginary part cancels with the
one coming from 
$\JJ^\pm_{\rm inst}$ in the limit of real $g_s$.
We can avoid this by taking the average
between the limits from the ${\rm Im}(g_s)>0$ and ${\rm Im}(g_s)<0$ sides. In this
case the imaginary contributions to $\JJ^\pm_{\rm pert}$
cancel, but the contributions $\JJ^\pm_{\rm inst}$ 
from the steepest
descent contours associated with the instanton also cancel. Therefore we are again
led to the result $\NN=0$.

Note that since the full result is continuous across the real axis, the limits from the
${\rm Im}(g_s)>0$ and ${\rm Im}(g_s)<0$ sides give the same result. However for this
we need to include the contributions from the steepest descent contours for
the perturbative vacuum and the instanton. Only by taking the average, or
by working strictly with real $g_s$, we get vanishing contribution from the instanton
and real contribution from the perturbative vacuum.

We must also remind the reader that the results described above depend crucially on 
the original choice of the integration contour. For example, if instead of taking the
integration over the tachyon to run along the real axis we had taken it to run from $-i\infty$
to $i\infty$, then it can be expressed as $\JJ^-_{\rm inst}-\JJ^+_{\rm inst}$ and we shall
get imaginary contribution to the integral.

\subsection{Instanton anti-instanton system in type IIB string theory} \label{s8.2}

We now turn to the system containing a D-instanton  anti-D-instanton
pair in type IIB string theory.
In this case the relevant degrees of freedom include the mode 
describing the relative location
between the D-instanton and the anti-D-instanton, and the complex
scalar mode  $T$ that becomes tachyonic
when the separation is less than a critical value. 
At zero separation, the tachyon potential has a maximum at $T=0$ 
representing the instanton
anti-instanton system, and a circle of minima at a non-zero value of $|T|$ representing
the perturbative vacuum without any instantons\cite{0410103}. 
There is strong reason to believe,
however, that the circle of minima, differing by the choice of phase of $T$, describes the 
same vacuum. For this reason, it will be more appropriate to represent the
space of tachyons as a sphere labelled by polar angles $(\theta,\phi)$, with $\theta=0$
denoting the maximum of the potential and $\theta=\pi$ denoting the minimum
of the potential.  
On the other hand the relative location between the D-instantons
is described by 10 real coordinates since type IIB D-instanton has
10 transverse space-time coordinates. However only the distance variable $\chi$ will
be of importance in our analysis, since using the spherical symmetry in ten dimensions,
the integration over the angular variable can be factored out of the integral.
Therefore we consider an integral over $\theta,\phi,\chi$ and model the action by,
\be \label{eppp2}
S(\theta, \phi, \chi ) = -{1\over g_s} \, [(1+\cos\theta) + \chi ^2 \, (1-\cos\theta)]\, .
\ee
Here $\theta=0$ represents the instanton anti-instanton configuration, 
separated by distance $\chi $,  with
action $-2/g_s$. On the other hand, $\theta=\pi$, $\chi =0$ represents
the vacuum without D-instanton with vanishing action. 
Note that we have appropriately rescaled $\chi$ to avoid factors of $2\pi$.
For definiteness, the integration measure over
$(\theta,\phi)$ will be taken to be $\sin\theta \, d\theta\, d\phi$ although full spherical
symmetry in $(\theta,\phi)$ space is not guaranteed, and the integration
measure over $\chi $ will be taken as $\chi^9\, d\chi $. 
Natural choice of the integration contour $\CC$ 
is the real $\theta$-$\phi$-$\chi $ plane and 
the integral is finite and real, except for a possible divergence from the large $\chi $, small
$\theta$ region. However, 
as we have discussed below \refb{eactuala12}, in the construction of the closed
string effective action this integral is cut off
at large $\chi$, and there is no divergence. 

We shall now describe the decomposition of the integration contour in terms of the 
Lefschetz thimbles around the saddle points. Since the integrand does not depend on
$\phi$, we can carry out the $\phi$ integral, picking a factor of $2\pi$. Eventually this
will be canceled when we divide the result by the volume of the rigid $U(1)$ gauge groups
living on the (anti-) D-instanton. Therefore we
shall not discuss this any further. Furthermore, we shall not use the method
of steepest descent to carry out the integration
over $\chi $ -- instead we shall envisage carrying out the integral over $\chi $ at the end
explicitly after carrying out the integration over $\theta$ using the
steepest descent contours for fixed $\chi $. This reduces the problem to the one 
discussed in \S\ref{s8.1}, with $g_s/(1-\chi^2)$ representing the effective coupling. 
Since this blows up at $\chi=1$, the method of steepest descent breaks down, although
the full integral remains finite. For this reason we shall evaluate the integral by deforming
the $\chi$ integration contour into the complex plane around $\chi=1$. 
Since the full
integral is non-singular at $\chi=1$, the result does not depend on how we deform the
contour. However following our discussion in \S\ref{s8.1}, we see that if we want to keep
the contribution from the steepest descent contour associated with the perturbative
vacuum real, we need to average over the contours above and below 1 in the complex
$\chi$ plane, and this is the route we shall follow.

\begin{figure}
\begin{center}
\figpicardtwo
\end{center}

\vskip -.2in

\caption{The steepest descent contours for the action \refb{eppp2} 
in the complex $\theta$ plane for $x^2>1$.
\label{figpicardtwo}
}
\end{figure}

For the steepest descent contours associated with
the saddle point at $\theta=0$, the integration over $\theta$ now produces the factor
proportional to
$e^{-2/g_s} (\chi^2-1)^{-1}$ that appeared {\it e.g.} in \refb{eres}, 
and averaging over the contours above and below the pole
at 1, we get the unitary prescription for the contribution from this pair of saddle points.
On the other hand, integration over $\theta$ for the steepest descent contour 
associated with the saddle point at $\theta=\pi$ produces the factor proportional to
$-e^{-2\, \chi^2/g_s} (\chi^2-1)^{-1}$. Taking the average over the contours 
$\JJ^+_{\rm pert}$ and $\JJ^-_{\rm pert}$ we get real contribution from the perturbative
vacuum as well. 

If we label the contour near $\chi=1$ as $\chi=1 +\eps\, e^{i\alpha}$ and vary $\alpha$
from $\pi$ to 0 or $-\pi$ to 0 depending on whether the contour passes above or below
$\chi=1$, the role of the two saddle points get exchanged as $\alpha$
crosses $\pm\pi/2$. For Re($\chi$)$>1$, the
saddle point at $\theta=0$ becomes the dominant saddle point and the steepest
descent contours in the complex $\theta$ plane get deformed as 
shown in Fig.~\ref{figpicardtwo}.

\subsection{Instanton in two dimensional bosonic string theory} \label{s8.3}

An instanton in two dimensional bosonic string theory has a real tachyonic mode $T$
with potential that has a maximum at $T=0$ representing the D-instanton and 
a local minimum
at a positive value of $T$ representing the perturbative vacuum. The potential approaches
$\pm\infty$ as $T\to\pm\infty$. We can model this by choosing:
\be\label{eppp3}
S(T) = -{1\over g_s}(1- 3\,  T^2 + 2\, T^3) \, .
\ee
In this case a real integration contour over $T$ does not give a well defined result, since
$S(T)$ approaches $\infty$ as $T\to -\infty$. Instead, the allowed asymptotic regions
in which the contour must lie consists of three cones, each of opening angle
$\pi/3$, around the lines
${\rm Arg}[T]=0$ and ${\rm Arg}[T]=\pm 2\pi i/3$ (see {\it e.g.} the discussion on Airy
function in \cite{1001.2933,1206.6272}). 
Therefore we
have to choose a contour in the complex $T$ plane. This makes the result complex. 
For physical reasons, we would like
the integral, representing the contribution to the closed string effective action, 
to have positive imaginary part so that it represents a dissipative system. This can be
achieved by choosing the
integration contour so that asymptotically it lies within the
cones around ${\rm Arg}[T]=0$ and ${\rm Arg}[T]=- 2\pi i/3$. This has been shown in
Fig.~\ref{figpicardthree} by the thick line. Alternatively we could also choose the 
integration contour to begin at infinity within the cone around ${\rm Arg}[T]=- 2\pi i/3$
and end at infinity within the cone around ${\rm Arg}[T]=2\pi i/3$.

\begin{figure}
\begin{center}
\figpicardthree
\end{center}

\vskip -.2in

\caption{The steepest descent contours for the action \refb{eppp3} 
in the complex $T$ plane.
\label{figpicardthree}
}
\end{figure}

The saddle point at $T=0$ with action $-1/g_s$ represents the unstable D-instanton,
while the saddle point at $T=1$ with vanishing action represents the vacuum without
any D-instanton. The steepest descent contours
for $T=1$ run along the real axis from
0 to 1 and $\infty$ to 1. We shall denote these by $\JJ_{\rm pert}^-$ and
$\JJ_{\rm pert}^+$ respectively.  
The steepest descent contours  for the saddle point at $T=0$ run from
$e^{\pm 2\pi i/3} \infty$ to $0$. We shall denote these by
$\JJ_{\rm inst}^\pm$ respectively. Then we have, from Fig.~\ref{figpicardthree},
\be 
\CC = \JJ_{\rm inst}^- +\JJ_{\rm pert}^- - \JJ_{\rm pert}^+ \, .
\ee
The first terms gives complex contribution while the last two terms give real
contribution. This provides a qualitative understanding of why in this case
the D-instanton
contribution to the closed string effective action may be complex.

\subsection{Final remarks}

Although the analysis described above
gives a qualitative understanding of the ambiguities involved in
computing D-instanton contribution to the closed string effective action, we need to keep
in mind that the full problem is more involved and new features may appear there.
For example, from the point of view of open string theory, the saddle point representing
the perturbative vacuum is the dominant saddle, and its contribution should dominate
the closed string effective action. However, it is known that the boundary state of the
D-instanton vanishes at that saddle\cite{0810.1737}. 
Therefore we would expect that all closed string
amplitudes on Riemann surfaces with boundaries ending on the D-instanton, on which
open strings have condensed to the tachyon vacuum, will actually vanish. 
This suggests that the perturbative contribution to the closed string effective action from the
dominant saddle point actually vanishes (unless we include the purely closed string 
contribution $S_0$ as part of the contribution from this saddle point).
Nevertheless we would expect a non-perturbative 
contribution to  survive at least for complex $g_s$, 
since in the
${\rm Im}(g_s)\to 0^\pm$ limit, these contributions are supposed to cancel the imaginary
parts of the contribution from the Lefschetz thimble associated with  the D-instanton.
It will certainly be important to understand this better.

The general lesson that is likely to survive in the full theory is that when the tachyon 
potential is bounded from below, as in superstring theory, we can choose real integration
contour over the open string modes on the D-instanton, and therefore integration over
these open string modes will give real contribution to the closed string effective action.

\sectiono{Conclusions}

The main results of this paper may be summarized as follows:
\begin{enumerate}
\item We have described a systematic procedure for constructing a novel 
gauge invariant effective action obtained by integrating out the open string modes
on the D-instanton. This differs from the usual method of integrating out a subset of
the degrees of freedom, in that we need to sum over connected and 
disconnected diagrams in defining the effective action, and furthermore we have an
unknown normalization constant $\NN\,e^{-C/g_s}$ appearing in the expression for
the effective action. Nevertheless, the effective action constructed this
way satisfies the BV master equation. 
\item We have shown that cutting rules and unitarity of the amplitudes follow as long as the
effective action constructed this way is real. For multiple D-instantons and multiple
anti-D-instantons in type IIB string theory, the reality condition holds and their contribution 
to the amplitude is unitary.
\item For D-instanton systems with tachyonic modes, {\it e.g.} instanton anti-instanton 
system in type IIB string theory and the non-BPS D-instanton in type IIA string theory,
the reality condition could fail due to two effects:
the overall normalization $\NN$ could be complex and the contour prescription for 
integrating over the relative positions of the instantons and anti-instantons may lead to
complex effective action.
However, 
using insights from Picard-Lefschetz theory we have argued that when the tachyon
potential is bounded from below, as in the case of critical superstring theories, the 
effective action is actually real.
\item Non-unitarity of two dimensional string theory identified 
in \cite{1907.07688,1912.07170} can be traced to
both the sources mentioned above. In particular the overall normalization $\NN$, found
in \cite{1907.07688} 
by comparison with matrix model results, is purely imaginary. Furthermore, in
the multi-instanton sector, \cite{1912.07170} 
used the `Lorentzian' prescription to integrate over the
relative locations of the instantons, leading to complex contribution to the effective action.
However, if we do not attempt comparison with the matrix model results, we can in
principle recover unitary D-instanton amplitudes in two dimensional bosonic string theory
by taking $\NN$ to be real and using unitary contour prescription for integrating over the
relative locations of the instantons.
\end{enumerate}

\bigskip

\noindent {\bf Acknowledgement:} I wish to thank Bruno Balthazar,  
Victor Rodriguez and Xi Yin for many useful discussions and for informing me of many of
their unpublished results. 
I also wish to thank
Michael Green, Michael Gutperle
 and Barton Zwiebach for useful discussions.
This work was
supported in part by the 
J. C. Bose fellowship of 
the Department of Science and Technology, India and the Infosys chair professorship.

\end{document}